\documentclass{jfm}
\usepackage[section]{placeins}
\usepackage{graphicx}
\usepackage{epstopdf, epsfig}
\usepackage{subfig}
\usepackage{floatrow}
\usepackage{enumitem, amsmath,bm}
\usepackage{bbold}
\usepackage{dsfont}
\usepackage{comment}
\usepackage[dvipsnames]{xcolor}
\usepackage{mathtools}
\usepackage[normalem]{ulem}

\usepackage{scalerel,stackengine}
\stackMath
\newcommand\reallywidehat[1]{%
\savestack{\tmpbox}{\stretchto{%
  \scaleto{%
    \scalerel*[\widthof{\ensuremath{#1}}]{\kern.1pt\mathchar"0362\kern.1pt}%
    {\rule{0ex}{\textheight}}%WIDTH-LIMITED CIRCUMFLEX
  }{\textheight}% 
}{2.4ex}}%
\stackon[-6.9pt]{#1}{\tmpbox}%
}

\newcommand{\strm}{\text{strm}}
\newcommand{\spwise}{\text{span}}
\newcommand\Llam{L_{\rm lam}}
\newcommand\Lturb{L_{\rm turb}}

\newcommand\utot{\boldsymbol{u}}
\newcommand\ubar{\overline{\boldsymbol{u}}}
\newcommand\uprime{\boldsymbol{u}^\prime}
\newcommand\uhat{\boldsymbol{\hat{u}}}

\newcommand\utilde{\tilde{\boldsymbol{u}}}

\newcommand\ujphatc{\widehat{u_j^\prime}^{*}}

\newcommand\ujmhatc{\widehat{\overline{u}}_j^*}

\newcommand\tke{\widehat{K}}

\newcommand\Prods{\widehat{\Pi}} %Spectral production
\newcommand\Dissips{\widehat{D}}
\newcommand\Prodms{\widehat{\overline{\Pi}}} %Spectral prod in the mean balance
\newcommand\Dissipms{\widehat{\overline{D}}}
\newcommand\Prodsy{\Prods_y}
\newcommand\Dissipsy{\Dissips_y}

\newcommand\Advs{\widehat{A}}
\newcommand\Advms{\widehat{\overline{A}}}
\newcommand\Advsy{\widehat{A}_y}

\newcommand\Transnls{\widehat{T}_{nl}}
\newcommand\Transnlsy{\widehat{T}_{nl, y}}

\newcommand\lambdamax{\widetilde{\lambda}_{\max}}
\newcommand\lambdamaxMBU{\widehat{\lambda}_{\max}}
\newcommand\ksmax{k_{\text{rolls}}} % Scale of 
\newcommand\kLS{k_{\text{LS}}}
\newcommand\kSS{k_{\text{SS}}} 
\newcommand\dz{\Delta z}

\newcommand\Repg{Re_{\rm pg}}
\newcommand\Regu{Re_{\rm gu}}

\newcommand\resub[1]{{\color{black}#1}}

\shorttitle{Pattern emergence and optimal wavelength in transitional shear turbulence}
\shortauthor{S. Gom\'e, L.S. Tuckerman and D. Barkley}

\title{Patterns in transitional shear turbulence. 
\\Part 2: Emergence and optimal wavelength}

\author{S\'ebastien Gom\'e\aff{1},
  Laurette S. Tuckerman\aff{1}
    \corresp{\email{laurette.tuckerman@espci.fr}}
 \and Dwight Barkley\aff{2}}

\affiliation{\aff{1}Laboratoire de Physique et M\'ecanique des Milieux H\'et\'erog\`enes, CNRS, ESPCI Paris, PSL Research
University, Sorbonne Universit\'e, Universit\'e Paris-Cit\'e, Paris 75005, France
\aff{2}Mathematics Institute, University of Warwick, Coventry CV4 7AL, United Kingdom}

\begin{document}

\maketitle

\begin{abstract}
Low Reynolds number turbulence in wall-bounded shear flows \emph{en route} to laminar flow takes the form of oblique, spatially-intermittent turbulent structures. \resub{In plane Couette flow}, these emerge from uniform turbulence via a spatiotemporal intermittent process in which localised quasi-laminar gaps randomly nucleate and disappear.  
For slightly lower Reynolds numbers, spatially periodic and approximately stationary turbulent-laminar patterns predominate. 
The statistics of quasi-laminar regions, including the distributions of space and time scales and their Reynolds number dependence, are analysed.
A smooth, but marked transition is observed between uniform turbulence and flow with intermittent quasi-laminar gaps, whereas the transition from gaps to regular patterns is more gradual.
Wavelength selection in these patterns is analysed via numerical simulations in oblique domains of various sizes. 
\resub{Via lifetime measurements in minimal domains, and a wavelet-based analysis of wavelength predominance in a large domain, we quantify the existence and non-linear stability of a pattern as a function of wavelength and Reynolds number.}
We report that
the preferred wavelength
maximises the energy and dissipation of the large-scale flow along laminar-turbulent interfaces. 
This optimal behaviour is primarily due to the advective nature of the large-scale flow, 
with turbulent fluctuations playing only a secondary role.

\end{abstract}

\begin{keywords}
Turbulence, transition, pattern formation
%Authors should not enter keywords on the manuscript, as these must be chosen by the author during the online submission process and will then be added during the typesetting process (see http://journals.cambridge.org/data/\linebreak[3]relatedlink/jfm-\linebreak[3]keywords.pdf for the full list)
\end{keywords}

\section{Introduction}

\resub{Turbulence in wall-bounded shear flows in the transitional regime is
characterised by coexisting turbulent and laminar regions, with the turbulent fraction increasing with Reynolds number.} This
phenomenon was first described by \cite{coles1966progress} and by \cite{andereck1986flow} in
Taylor-Couette flow. Later, by constructing Taylor-Couette and plane
Couette experiments with very large aspect ratios, \cite{prigent2002large,prigent2003long}
showed that these coexisting turbulent and laminar regions,
\resub{called \emph{bands} and \emph{gaps} respectively},
spontaneously formed regular patterns with a selected
wavelength and orientation that depend systematically on $Re$.
These patterns have been simulated numerically and studied
intensively in plane Couette flow
\citep{barkley2005computational,barkley2007mean,duguet2010formation, rolland2011ginzburg, tuckerman2011patterns},
plane Poiseuille flow
\citep{tsukahara2005dns,tuckerman2014turbulent,ShimizuPRF2019,kashyap2021subcritical},
and Taylor-Couette flow \citep{meseguer2009instability,dong2009evidence,wang_2022}.

In pipe flow, the other canonical wall-bounded shear flow, only the streamwise direction is long, and  transitional turbulence takes the form of \emph{puffs}, \resub{also called} \emph{flashes} \citep{Reynolds:1883, wygnanski1973transition}, which are the one-dimensional
%equivalent 
analog of turbulent bands. In contrast to bands in planar shear flows, experiments and direct numerical simulations show that puffs 
do not spontaneously form %regular 
spatially periodic patterns \citep{moxey2010distinct, avila2013nature}. Instead, the spacing between them is dictated by short-range interactions \citep[]{hof2010eliminating, samanta2011experimental}. 
Puffs have been extensively studied, especially in the context of the model derived by \cite{barkley2011a, barkley2011b,barkley2016theoretical} from the viewpoint of \emph{excitable media}.  In this framework, fluctuations from uniform turbulence trigger quasi-laminar gaps (\resub{i.e.\ low-turbulent-energy holes within the flow}) at random instants and locations, as has been seen in direct numerical simulations (DNS) of pipe flow. The bifurcation scenario giving rise to localised gaps has been investigated by \citet[]{frishman2022dynamical}, who called them \emph{anti-puffs}. Interestingly, spatially periodic solutions like those observed in planar shear flows are produced in a centro-symmetric version of the Barkley model \citep{barkley2011b} \resub{although the mechanism for their formation has not yet been clarified.}

In this paper, we will show that in plane Couette flow, as in pipe flow, short-lived localised gaps emerge randomly from uniform turbulence at the highest Reynolds numbers in the transitional range, which we will see is $Re\simeq 470$ in the domain which we will study.  The first purpose of this paper is to investigate these gaps.
The emblematic regular oblique large-scale bands appear at slightly lower Reynolds numbers, which we will see is $Re\simeq 430$.

If the localised gaps are disregarded, it is natural to associate the bands with a \emph{pattern-forming instability} of the
uniform turbulent flow.
This was first suggested by \citet[]{prigent2003long} and
later investigated by \cite{rolland2011ginzburg}.
\citet[]{manneville2012turbulent} and \citet[]{kashyap2021subcritical}
proposed a Turing mechanism to account for the appearance of patterns
by constructing a reaction-diffusion model based on an extension of the \citet[]{waleffe1997self} model of the streak-roll self-sustaining process. 
\citet{reetz2019exact} discovered a sequence of bifurcations leading to a large-scale steady state that resembles a skeleton for the banded pattern, arising from tiled copies of the exact \citet{nagata1990three} solutions of plane Couette flow.
%and not from uniform turbulence. 
The relationship between these pattern-forming frameworks and local nucleation of gaps is unclear.

The adaptation of classic stability concepts to turbulent flows is currently a major research topic. 
%see, e.g., \citet[]{markeviciute2022improved}.  
At the simplest level, it is always
formally possible to carry out linear stability analysis of a mean
flow, as was done by \citet[]{barkley2006linear} for a limit cycle in the cylinder wake. The mean
flow of uniformly turbulent plane Couette flow has been found to be
linearly stable \citep[]{tuckerman2010instability}. However, this procedure makes the drastic simplification of neglecting the Reynolds stress entirely in the stability problem and hence its interpretation is uncertain \citep[e.g.,][]{bengana2021frequency}. The next level of complexity and accuracy is to represent the Reynolds stress via a closure model. 
\resub{However, classic closure models for homogeneous turbulence (e.g.\ $(K,\Omega)$) have yielded predictions that are completely incompatible with results from full numerical simulation or experiment \citep[]{tuckerman2010instability}.}
Another turbulent configuration in which large, 
spatially periodic scales emerge are zonal jets, characteristic of geophysical turbulence. For zonal jets, a closure model provided by a cumulant expansion \citep[]{ srinivasan2012zonostrophic, tobias2013direct} has led to a plausible stability analysis \citep[]{parker2013zonal}. Other strategies are possible for turbulent flows in general: \citet[]{kashyap2022linear} examined the averaged time-dependent response of uniform turbulence to large-wavelength perturbations and provided evidence for a linear instability in plane channel flow. They computed a dispersion relation which is in good agreement with the natural spacing and angle of patterns.

Classic analyses for non-turbulent pattern-forming flows, such as
Rayleigh-B\'enard convection or Taylor-Couette flow, yield not only a threshold 
\resub{and a preferred wavelength, but also
existence and
stability ranges for other wavelengths} through the Eckhaus instability
\citep{busse1981transition,ahlers1986wavenumber, riecke1986stability,
  tuckerman1990bifurcation, cross2009pattern}. As the control
parameter is varied, this instability causes spatially periodic
states to make transitions to other periodic states whose wavelength
is preferred.  
Eckhaus instability is also invoked in turbulent zonal jets \citep[]{parker2013zonal}.
The second goal of this paper is to study the regular patterns
of transitional plane Couette flow and to 
determine the wavelengths at which they can exist and thrive.
At low enough Reynolds numbers, patterns will be shown to destabilise and to acquire a  
different wavelength.

Pattern formation is sometimes associated with maximisation principles obeyed by the preferred wavelength, as in the canonical Rayleigh-B\'enard convection. 
\resub{Such principles, like maximal dissipation, also have a long history for turbulent solutions.
\citet[]{malkus1954heat} and
\citet[]{busse1981transition} 
%(and references therein) 
proposed a principle of maximal heat transport, or equivalently maximal dissipation, obeyed by convective turbulent states.
The maximal dissipation principle, as formulated by \citet[]{malkus1956outline} in shear flows,
occurs in other systems such as von K\'arm\'an flow \citep[]{ozawa2001thermodynamics, mihelich2017turbulence}.
\resub{(This principle has been somewhat controversial and  was challenged by \citet[]{reynolds1967stability} within the context of stability theory. See a modern revisit of Malkus stability theory with statistical closures by \citet[]{markeviciute2022improved}.)}}
Using the energy analysis formulated in our companion paper \citet[]{gome1}, we will associate the selected wavelength to a maximal dissipation observed for the large-scale flow along the bands.

\section{Numerical setup}
\label{sec:numerics}

\begin{figure}
    \centering
  %  \subfloat[]{ \includegraphics[width=0.5\columnwidth]{mf_v_arrows_R360.pdf}}
\includegraphics[width=0.8\columnwidth]{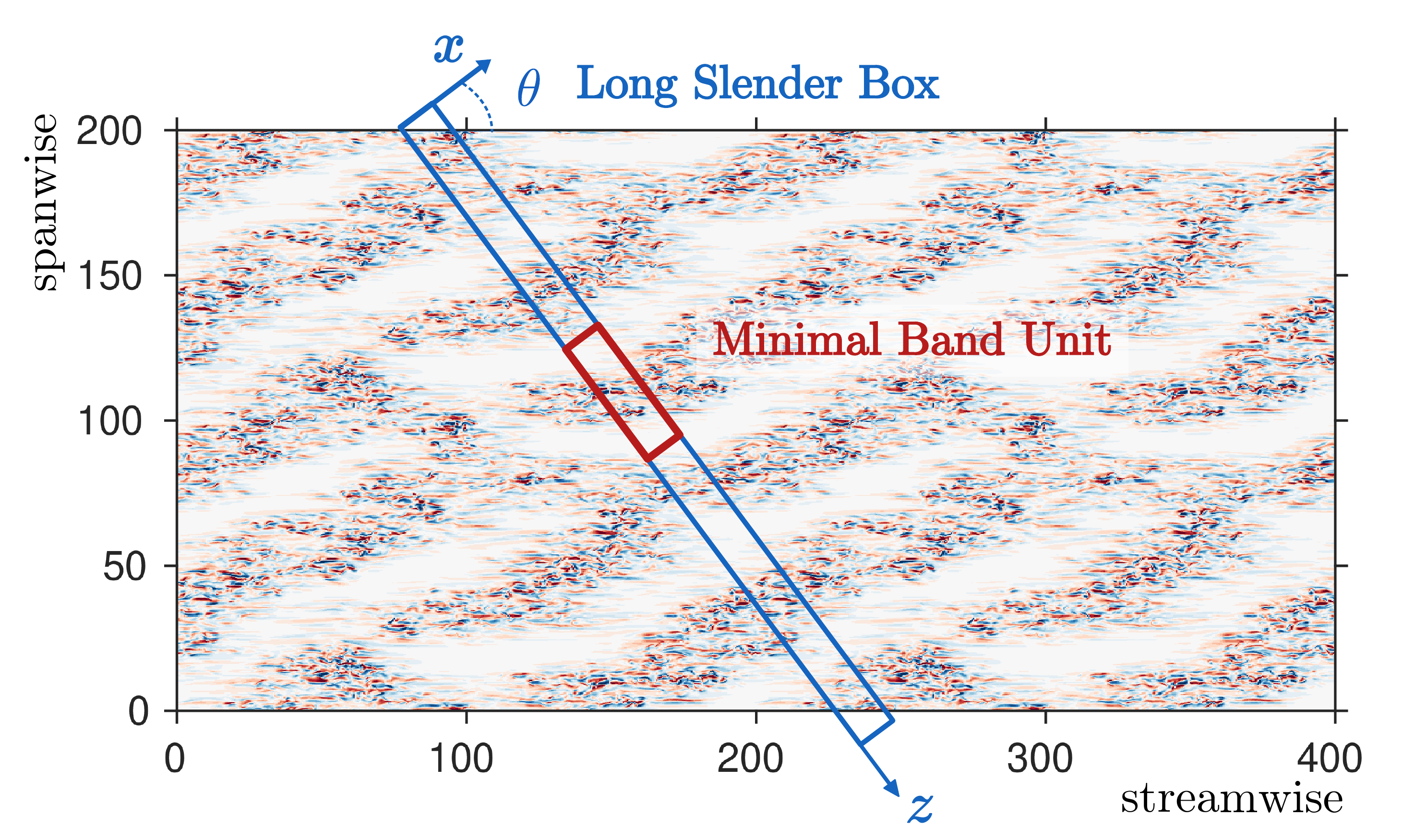}
\caption{Spatial visualization of our numerical domains at $Re=360$. Colors show the wall-normal velocity $v$ at the midplane $y=0$ (blue: $-0.2$, white: 0, red: 0.2) in a domain of size $L_{\strm}=400$, $L_{\spwise}=200$. Red and blue boxes respectively show a Minimal Band Unit and a Long Slender Box. }
    \label{fig:domain}
\end{figure}

Plane Couette flow consists of two parallel rigid plates moving at different velocities, here equal and opposite velocities $\pm U_\text{wall}$. Lengths are nondimensionalised by the half-gap $h$ between the plates and velocities by $U_\text{wall}$. The Reynolds number is defined to be $Re \equiv U_\text{wall}h/\nu$. We will require one further dimensional quantity that appears in the friction coefficient -- the mean horizontal shear at the walls, which we denote by $U^\prime_\text{wall}$. We will use non-dimensional variables throughout except when specified. 
We simulate the incompressible Navier-Stokes equations
\begin{subequations}
    \begin{align}
    \frac{\partial \utot}{\partial t}
    + \left(\utot \cdot \nabla\right) \utot
    &= -\nabla p  + \frac{1}{Re} \nabla^2 \utot, \\
    \nabla \cdot \utot &= 0,
    \end{align}
    \label{eq:NS}
\end{subequations}
using the pseudo-spectral parallel code {\tt Channelflow} \citep{channelflow}.
Since the bands are found to be oriented obliquely with respect to the streamwise direction, we use a doubly periodic numerical domain which is tilted with respect to the streamwise direction of the flow, shown as the oblique rectangle in figure \ref{fig:domain}. This choice was introduced by \cite{barkley2005computational}
and has become common in studying turbulent bands \citep{shi,lemoult2016directed, paranjape2020oblique,tuckerman2020patterns}.
The $x$ direction is chosen to be aligned with a typical turbulent band and the $z$ coordinate to be orthogonal to the band. The relationship between streamwise-spanwise coordinates and tilted band-oriented coordinates is:
\begin{subequations}
\label{tilted}
\begin{align}
\mathbf{e}_{\strm} &= \quad\cos{\theta} \, \mathbf{e}_x + \sin{\theta} \, \mathbf{e}_z \\
\mathbf{e}_{\spwise} &= -\sin{\theta } \, \mathbf{e}_x + \cos{\theta} \, \mathbf{e}_z \quad 
\end{align}
\end{subequations}
The usual wall-normal coordinate is denoted by $y$ and the corresponding velocity by $v$.
%Thus the boundary conditions are  $\utot(y=\pm1)=\pm \mathbf{e}_{\strm}$ in $y$ and no-flux in $x$ and $z$.
Thus the boundary conditions are  $\utot(y=\pm1)=\pm \mathbf{e}_{\strm}$ in $y$ and periodic in $x$ and $z$, together with a zero-flux constraint on the flow in the $x$ and $z$ directions.
The field visualised in figure \ref{fig:domain} comes from an additional simulation we carried out in a domain of size ($L_{\strm},L_y,L_{\spwise}) = (200,2,100)$ aligned with the streamwise-spanwise directions.
Exploiting the periodic boundary conditions of the simulation, the visualisation shows four copies of the instantaneous field. 

The tilted box effectively reduces the dimensionality of the system by disallowing large-scale variation along the short $x$ direction. The flow in this direction is considered to be statistically homogeneous as it is only dictated by small turbulent scales. In a large non-tilted domain, bands with opposite orientations coexist
\citep{prigent2003long, duguet2010formation, klotz2022phase}, but only one orientation is permitted in the tilted box.

We will use two types of
numerical domains, with different lengths $L_z$.
Both have fixed resolution $\dz = L_z/N_z = 0.08$, along with fixed $L_x=10$ ($N_x=120$), \resub{$L_y=2$ $(N_y=33)$} and $\theta = 24^{\circ}$.
These domains are shown in figure \ref{fig:domain}.

\begin{enumerate}[leftmargin=*,labelindent=8mm,labelsep=3mm, itemindent=0mm, label=(\arabic*)]
    \item \textbf{Minimal Band Units}, an example of which is shown as the dark red box in figure \ref{fig:domain}. These domains accommodate a single band-gap pair and so are used to study strictly periodic pattern of imposed wavelength $\lambda=L_z$. \resub{($L_z$ must typically be below $\simeq 65$ to contain a unique band.)} 
    \item \textbf{Long Slender Boxes}, which have a large $L_z$ direction that can accommodate a large and variable 
number of gaps and bands in the system. The blue box in figure \ref{fig:domain} is an example of such a domain size with $L_z = 240$, but larger sizes ($L_z=400$ or $L_z=800$) will be used in our study.
\end{enumerate}

\section{Nucleation of laminar gaps and pattern emergence}
\label{sec:intermittency}

\begin{figure}
    \centering
    \subfloat[]{  \includegraphics[width=0.5\columnwidth]{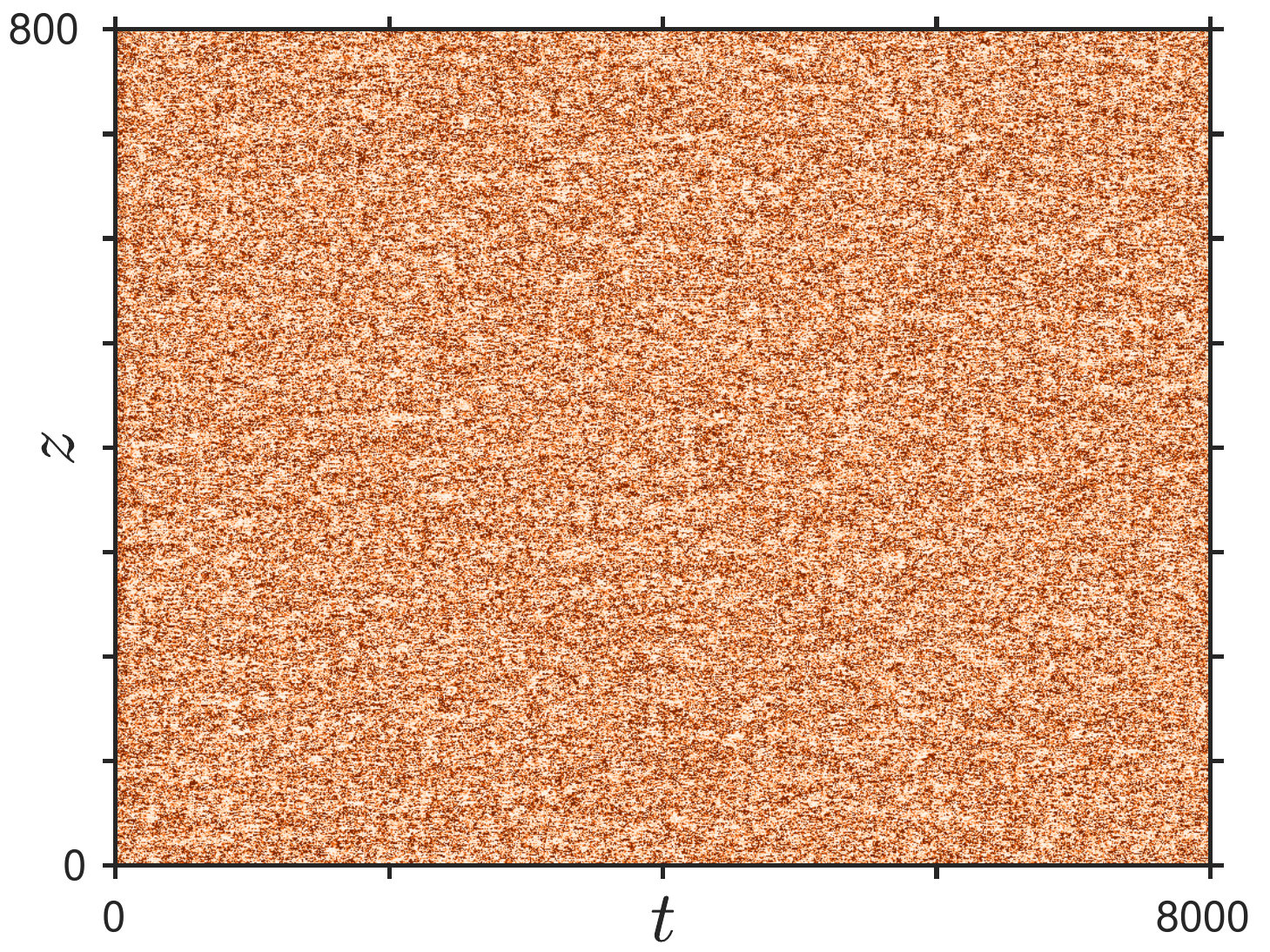} \label{fig:probes_q_R500}} ~
%\subfloat[]{  \includegraphics[width=0.5\columnwidth]{probes_q_R480_or.pdf}  \label{fig:probes_q_R480}} ~
\subfloat[]{  \includegraphics[width=0.5\columnwidth]{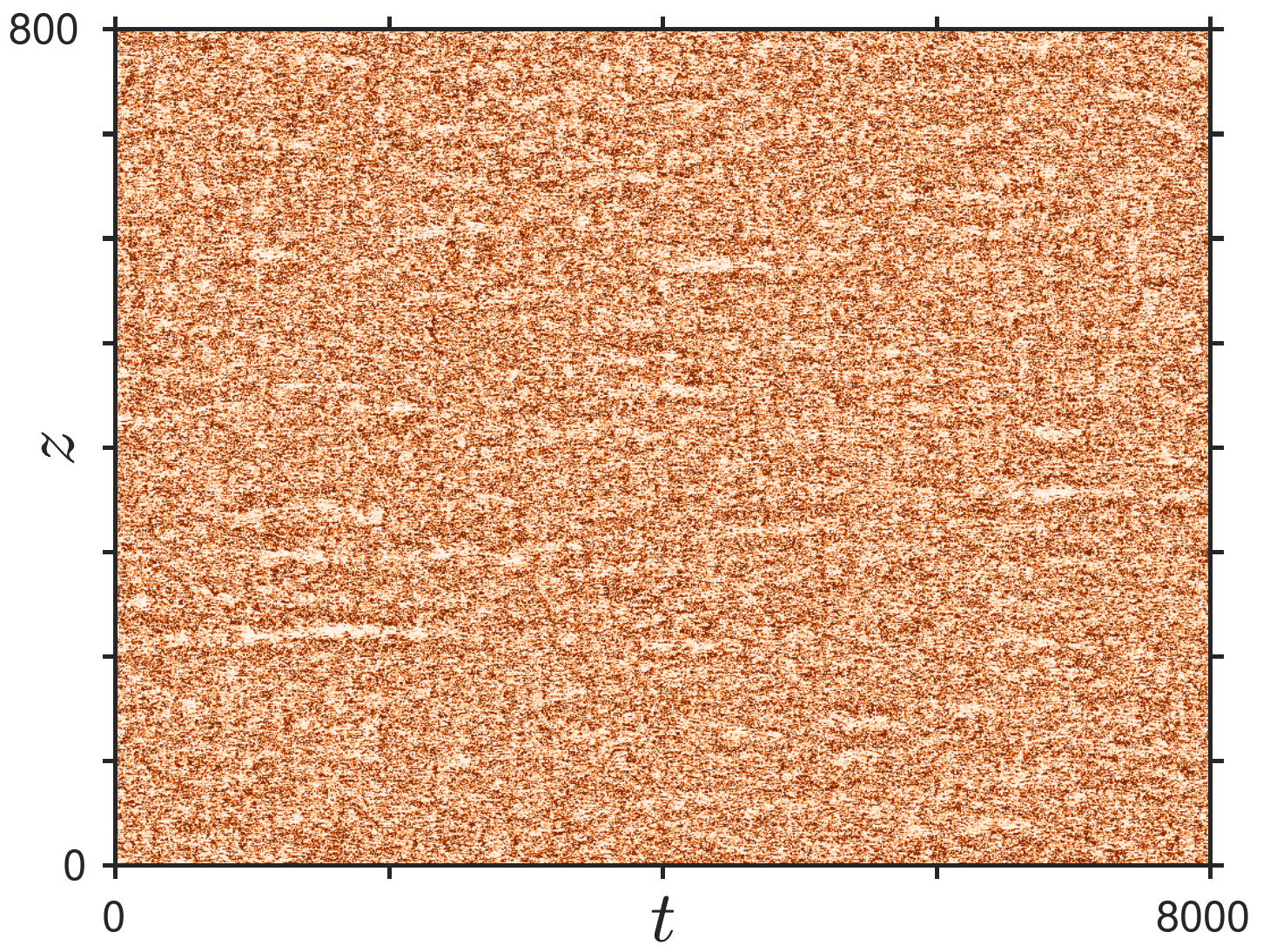} \label{fig:probes_q_R460}
} \\
\subfloat[]{  \includegraphics[width=0.5\columnwidth]{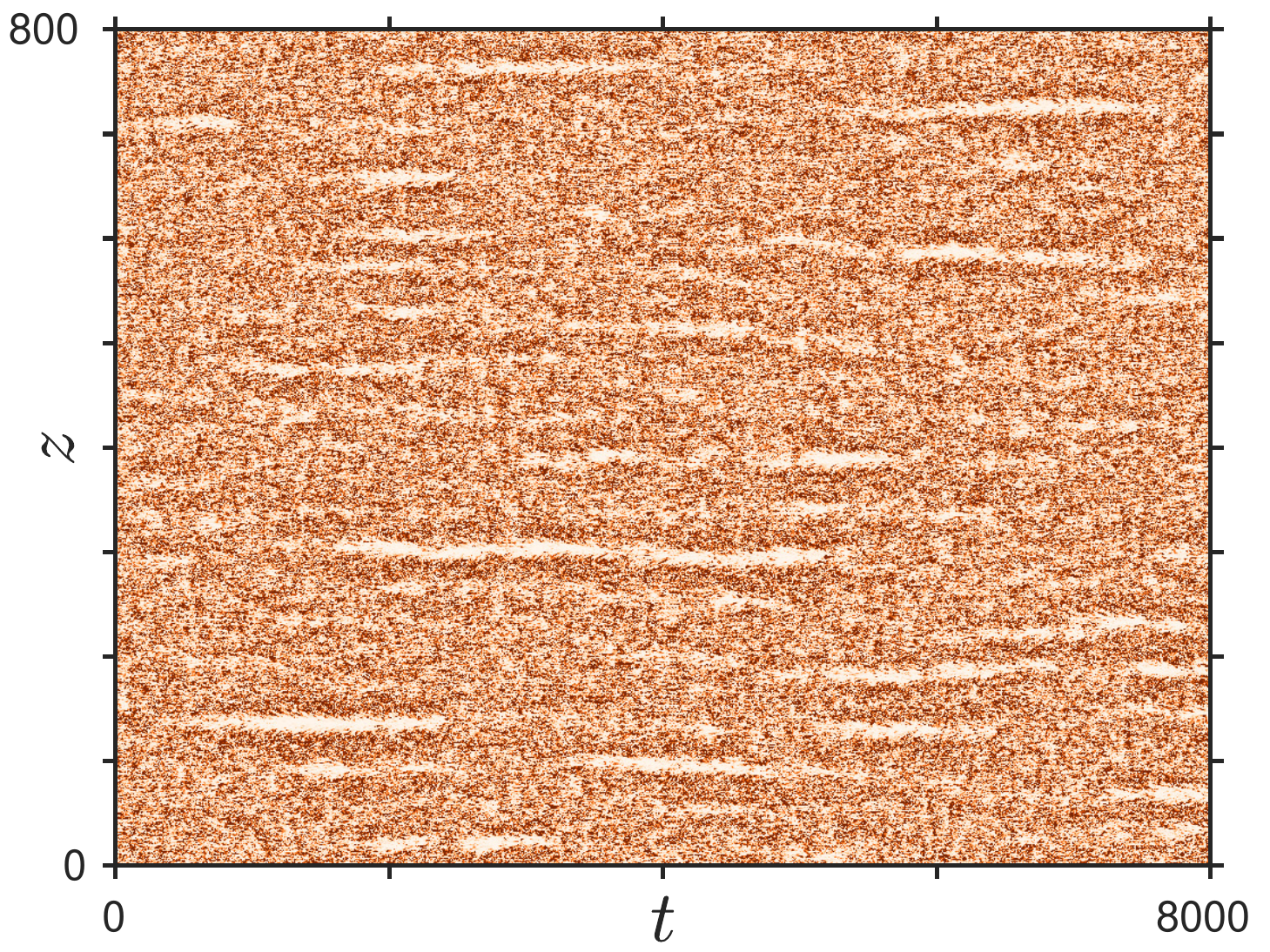}  \label{fig:probes_q_R440}} ~
\subfloat[]{  \includegraphics[width=0.5\columnwidth]{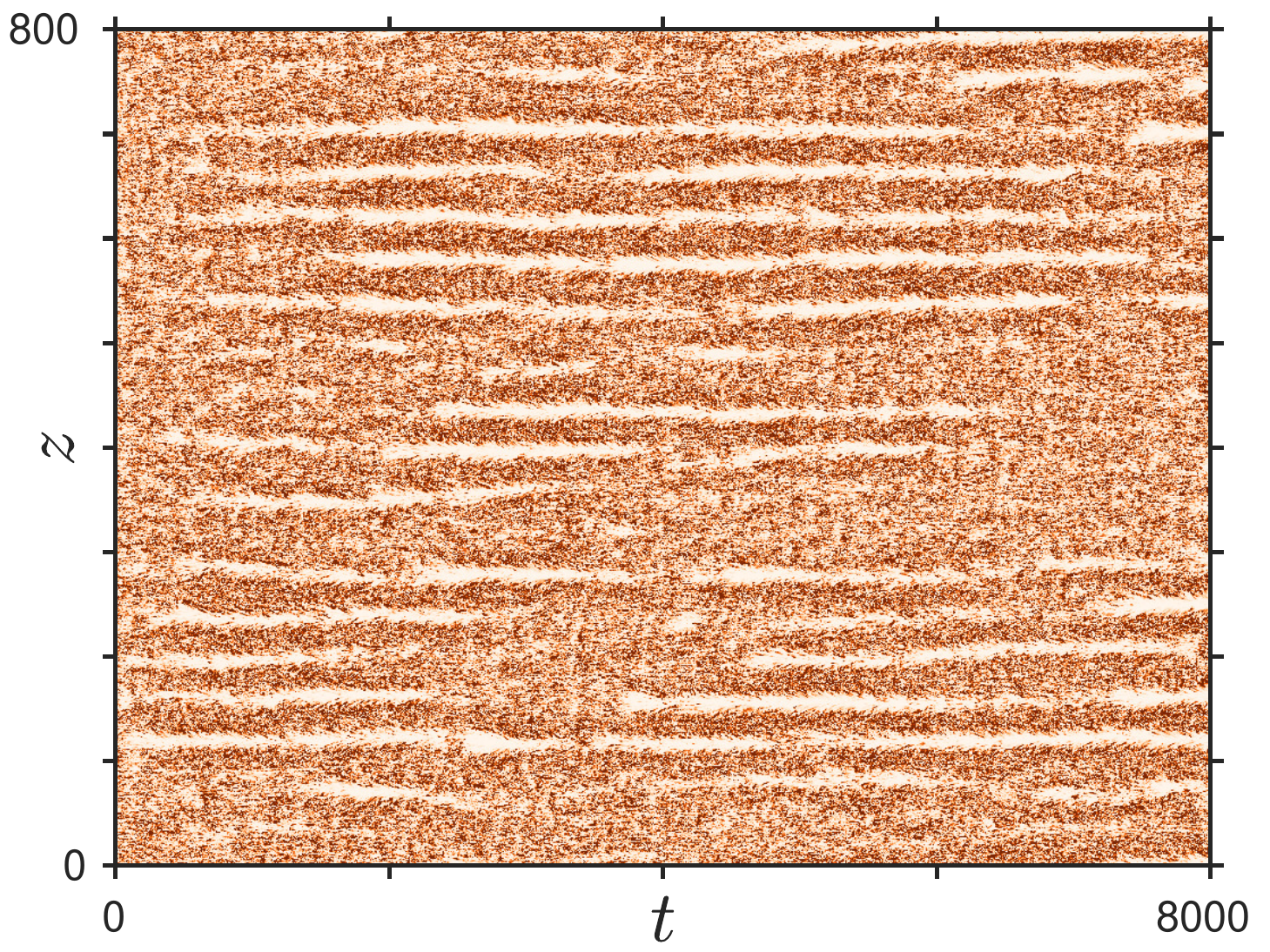} \label{fig:probes_q_R420}} \\
\subfloat[]{  \includegraphics[width=0.5\columnwidth]{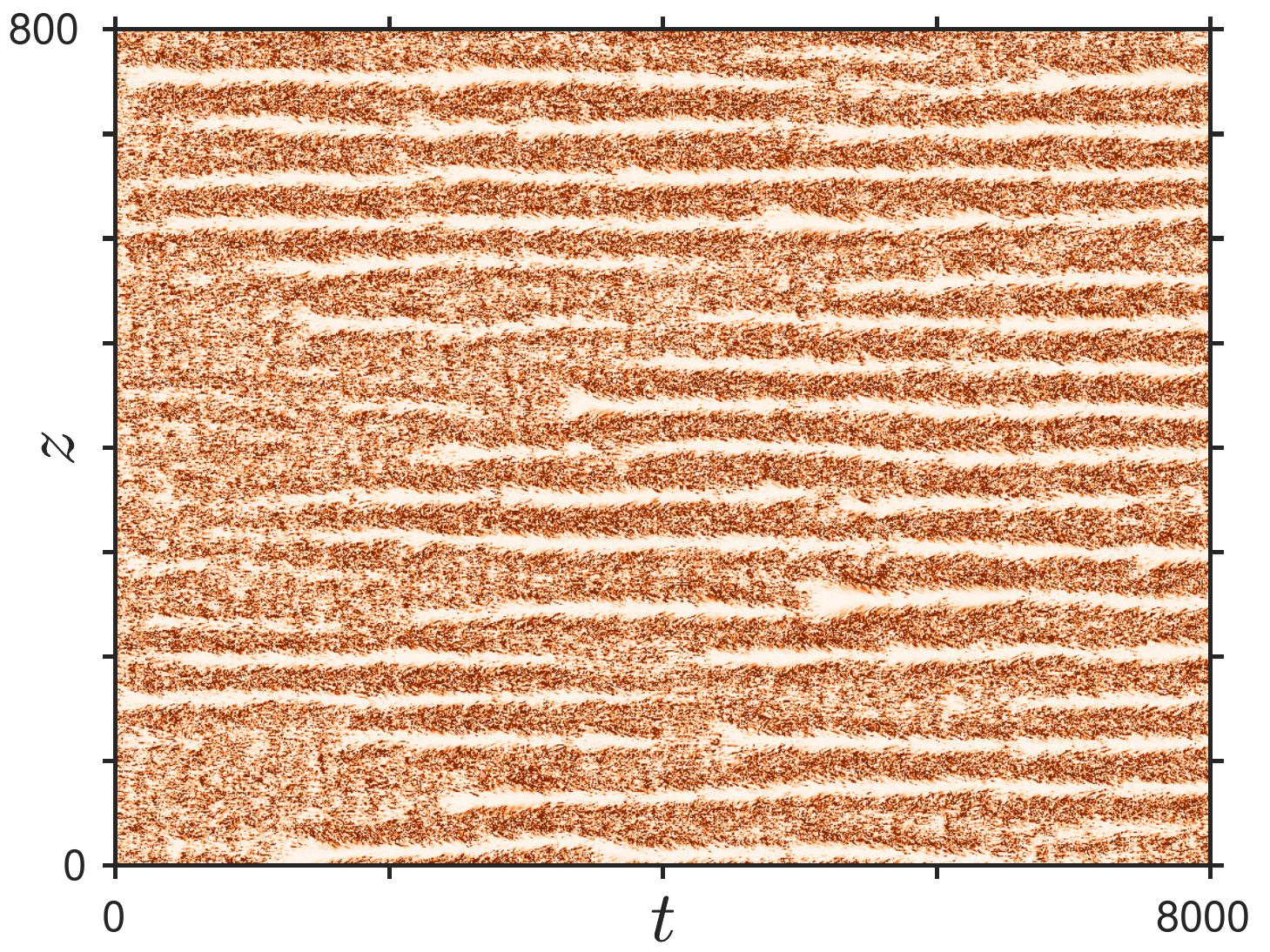} \label{fig:probes_q_R400}} ~
\subfloat[]{  \includegraphics[width=0.5\columnwidth]{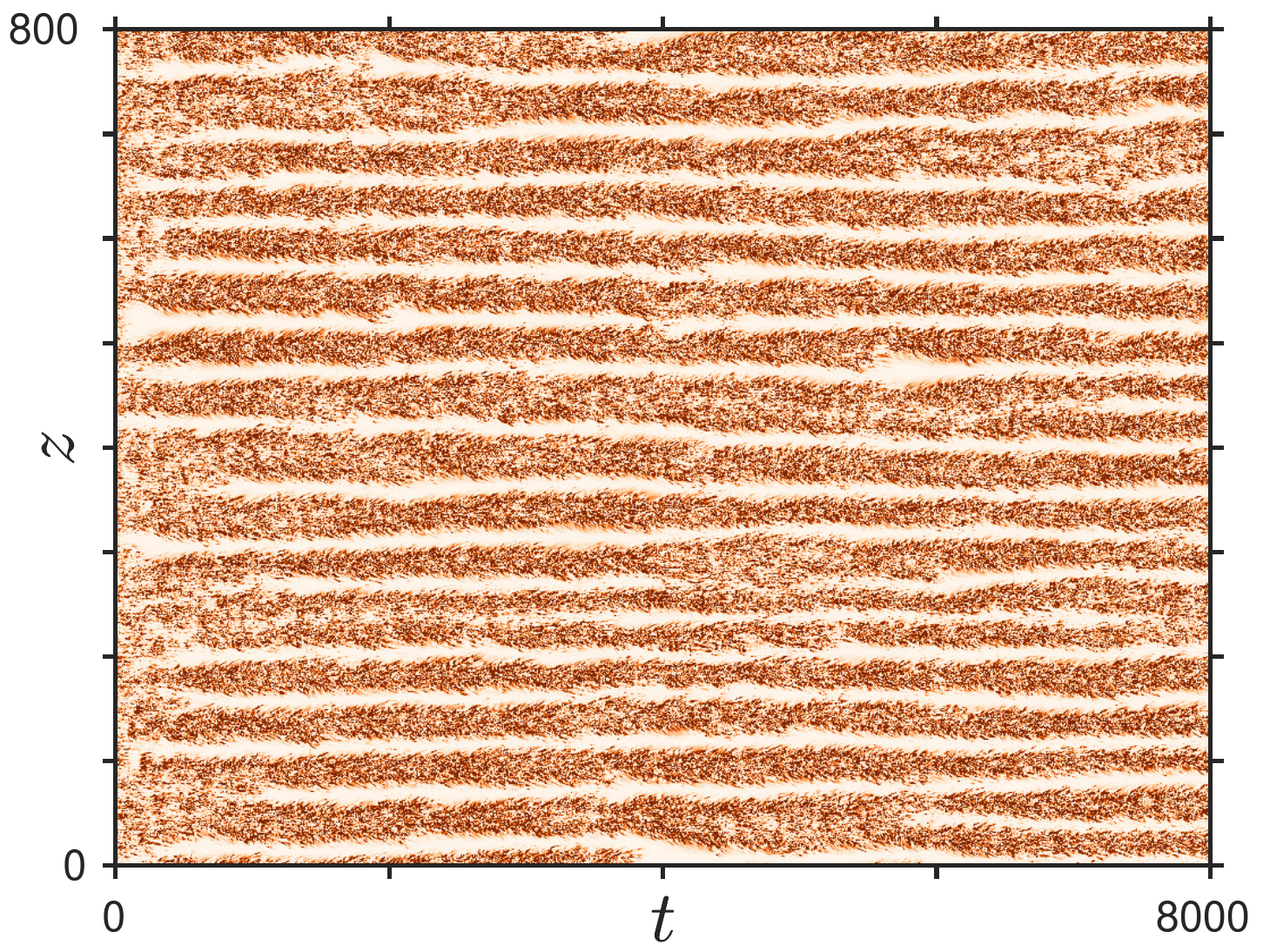} \label{fig:probes_q_R380}} 
\caption{Spatio-temporal visualization of pattern formation with $L_z=800$, for (a) $Re=500$, (b) $Re=460$, (c) $440$, (d) $420$, (e) $400$ and (f) $Re=380$. Flow at $t=0$ is initiated from uniform turbulence at $Re=500$. Color shows cross-flow energy $(v^2 + u_{\spwise}^2)/2$ at $x=L_x/2$, $y=0$ (white: 0, red: 0.02). 
At high $Re$, weak local gaps appear sparsely. 
When $Re$ is decreased, spatio-temporally intermittent patterns of finite spatial extent emerge. These consist of turbulent cores (dark red) and quasi-laminar gaps (white). For still lower $Re$,
quasi-laminar regions live longer, and patterns are more regular and steady.}
\label{fig:probes_Lz800}
\end{figure}

%\subsection{Gaps emergence and scalings}

We have carried out simulations in a Long Slender Box of size $L_z=800$ for various $Re$ with the uniform turbulent state from a simulation at $Re=500$ as an initial condition, a protocol called a quench.
Figure~\ref{fig:probes_Lz800}, \resub{an extension of figure 1 of \citet[Part 1]{gome1}}, displays the resulting spatio-temporal dynamics at six Reynolds numbers. Plotted is the $(z,t)$ dependence of the cross-flow energy $(v^2 + u_{\spwise}^2)/2$ at $(x=L_x/2, y=0)$. 
The cross-flow energy is a useful diagnostic because it is zero for laminar flow \resub{and is therefore a proxy for turbulent kinetic energy}. The choice $x=L_x/2$ is arbitrary since there is no large-scale variation of the flow field in the short $x$ direction of the simulation. 

%Figure~\ref{fig:probes_Lz800} demonstrates strong space-time intermittency and encapsulates the main results of this section: the emergence of patterns out of uniform turbulence is a gradual process. 

Figure~\ref{fig:probes_Lz800} encapsulates the main message of this section: the emergence of patterns out of uniform turbulence is a gradual process involving spatio-temporal intermittency of turbulent and quasi-laminar flow. 
At $Re=500$, barely discernible low-turbulent-energy regions appear randomly within the turbulent background. At $Re=460$ these regions are more pronounced and begin to constitute 
localised, short-lived 
quasi-laminar gaps within the turbulent flow.
%These gaps appear sparsely and are not long lived.  
%At $Re=440$, clearly demarcated, spatially localised quasi-laminar gaps are seen. 
As $Re$ is further decreased, these 
%quasi-laminar 
gaps are more probable
%appear more frequently  
and last for longer times. Eventually, the gaps self-organise into persistent, albeit fluctuating, patterns. The remainder of the section will quantify the evolution of states seen in figure~\ref{fig:probes_Lz800}.

%the emergence of patterns out of uniform turbulence is a gradual process, with laminar gaps appearing randomly and locally in space and time at high $Re$, as seen as the white spots in figure \ref{fig:probes_q_R500} at $Re=500$ and figure \ref{fig:probes_q_R460} at $Re=460$.
%
%and persistently spanning the entire domain when reducing $Re$. 
%Short-lived gaps, shown as white spots, already emerge randomly at $Re=460$ (figure \ref{fig:probes_Lz800}a). 
%When decreasing $Re$, gaps emerge at a higher rate and with longer lifetimes as shown at $Re=440$ in figure \ref{fig:probes_q_R440}. 
%When $Re\leq420$, as pictured in figure \ref{fig:probes_q_R420}e, f and g, the gaps self-organise into persistent patterns. The remainder of the section will quantify this transition to patterns.

\resub{\subsection{Statistics of laminar and turbulent zones}}

\begin{figure}
    \centering
\subfloat[]{\includegraphics[width=0.5\columnwidth]{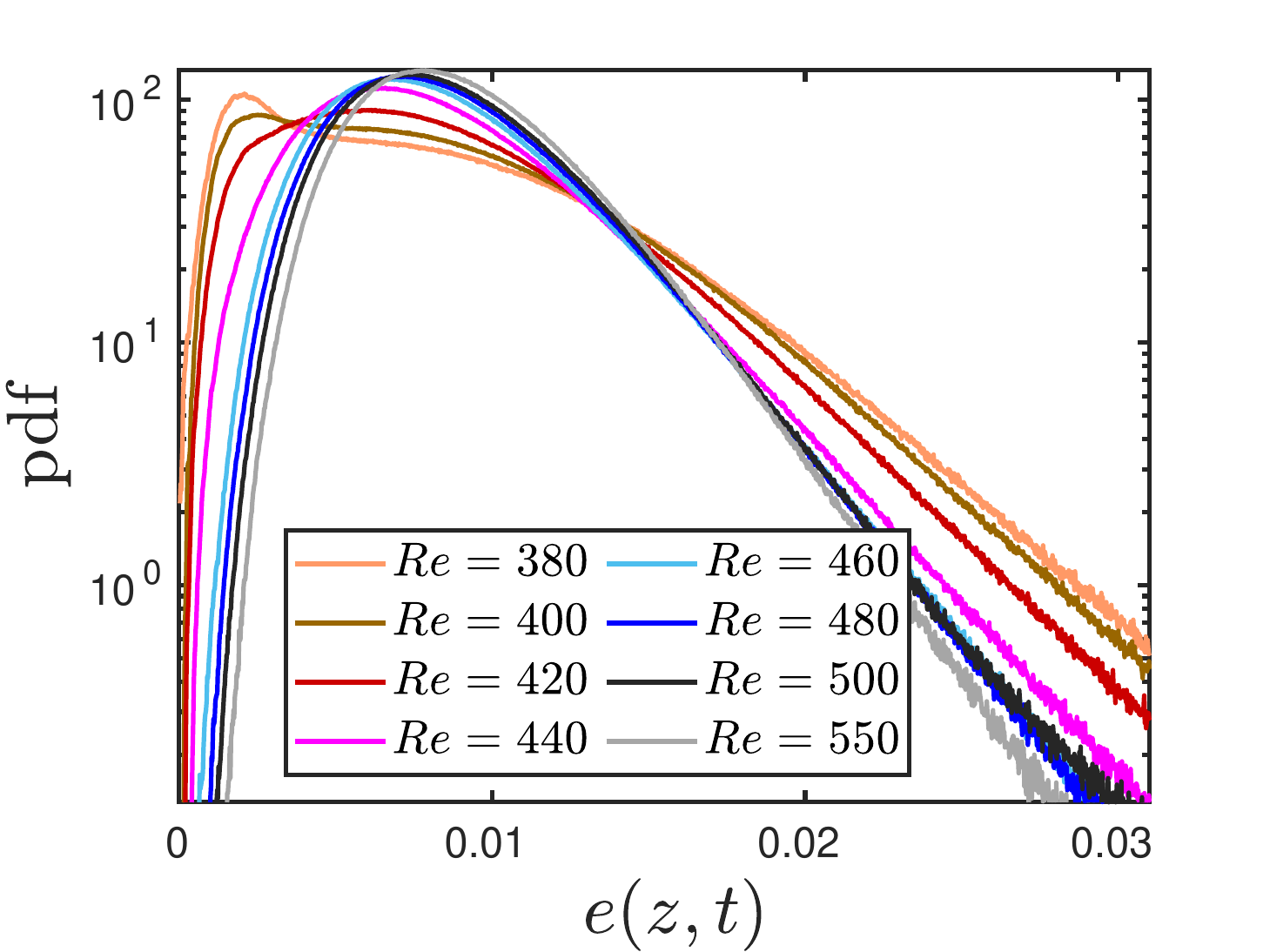} 
\label{fig:pdf_e}}
\subfloat[]{\includegraphics[width=0.5\columnwidth]{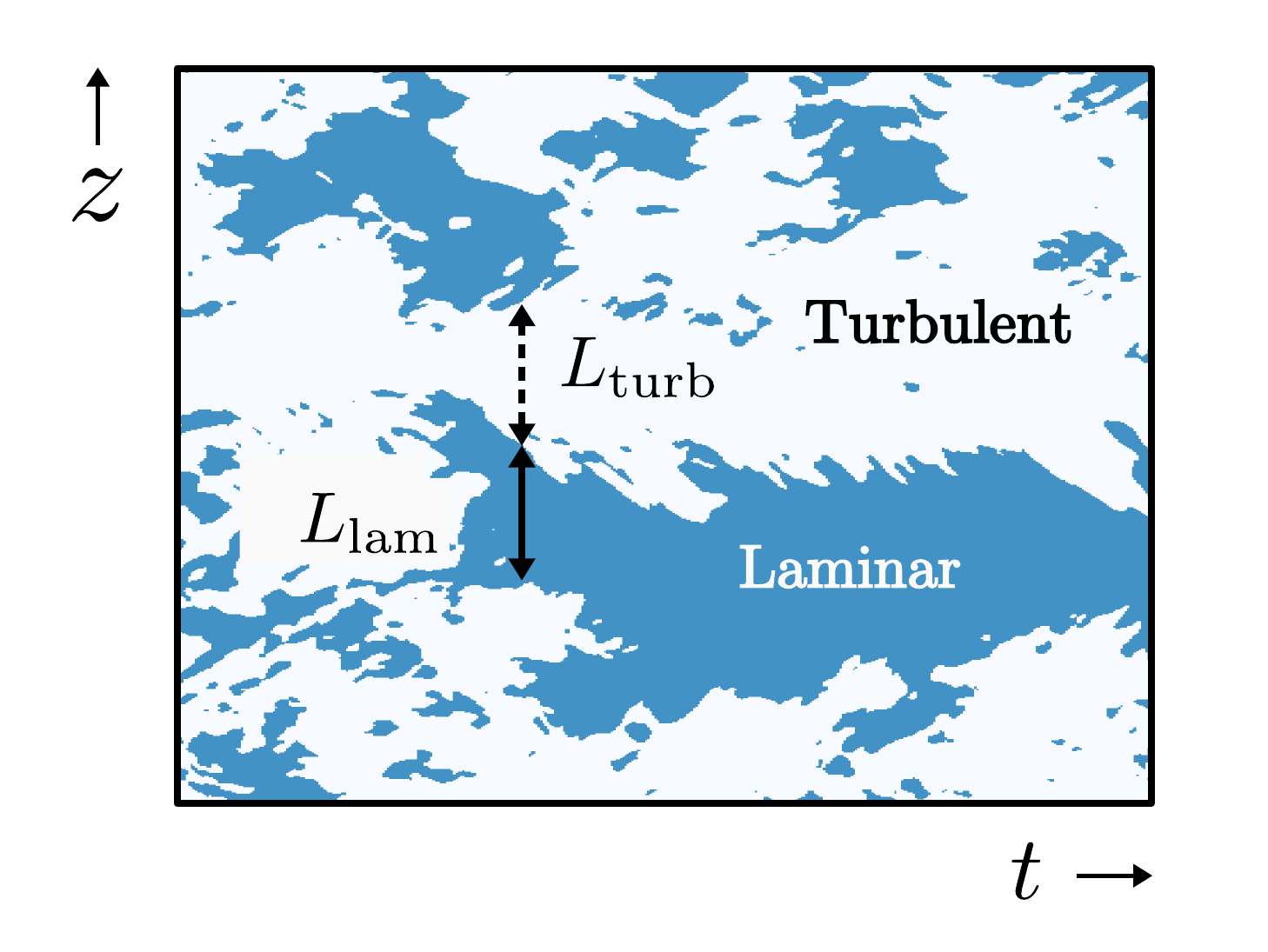} 
\label{fig:sketch_LT}}\\
\subfloat[]{\includegraphics[width=0.5\columnwidth]{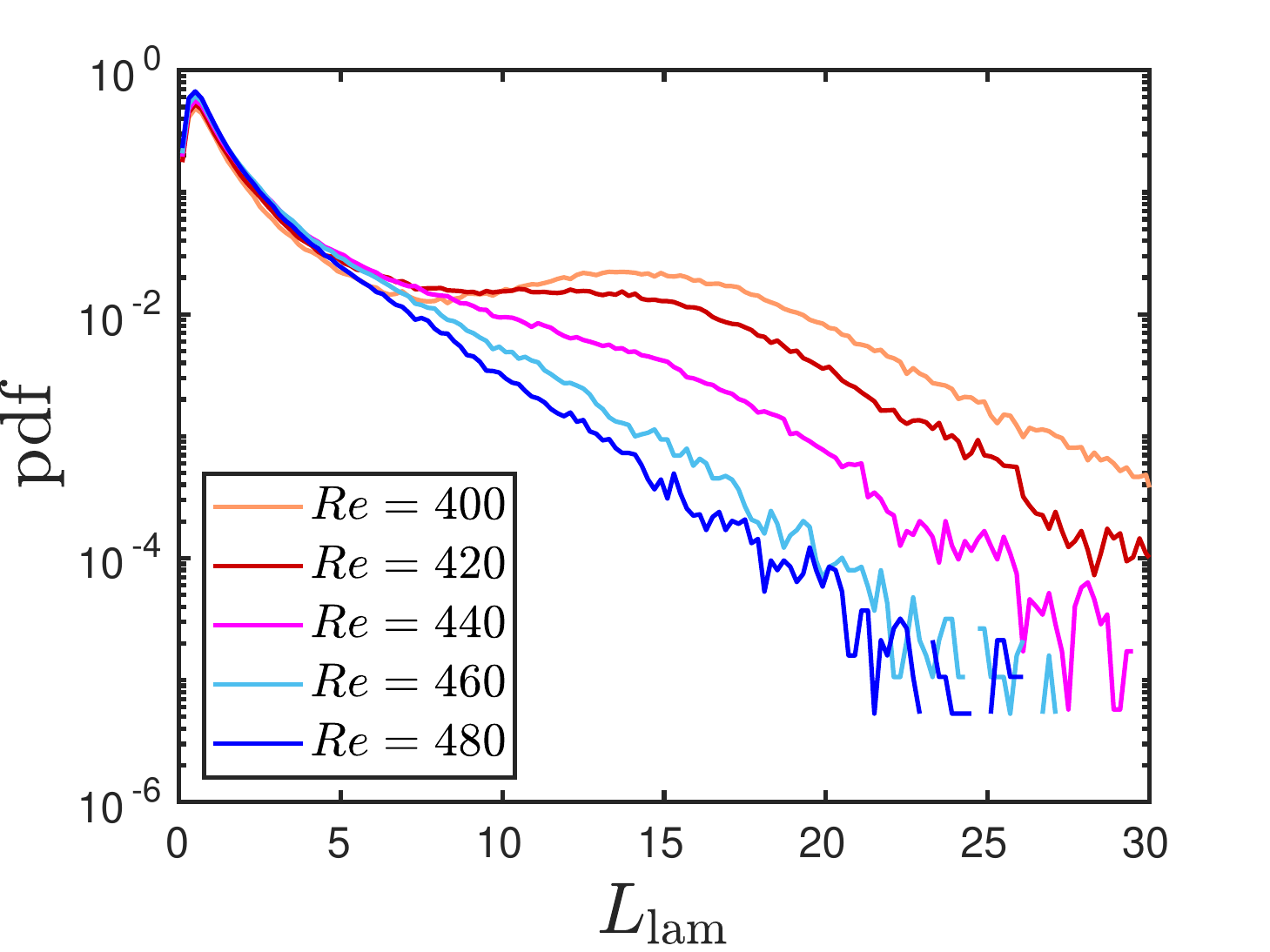} 
\label{fig:pdf_llam}} ~
\subfloat[]{\includegraphics[width=0.5\columnwidth]{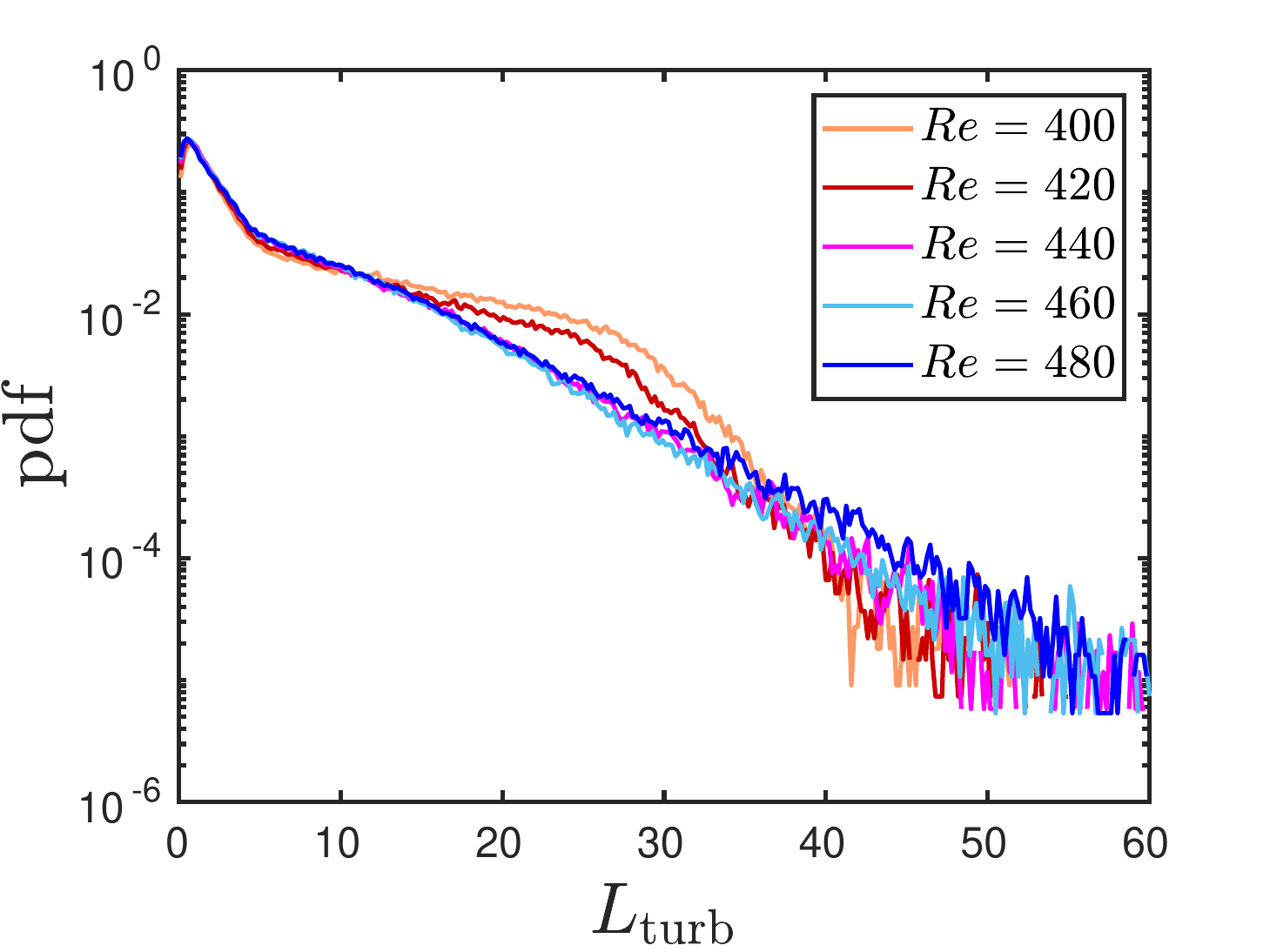} 
\label{fig:pdf_lturb}} ~
\caption{(a) PDFs of local cross-flow energy $e(z,t)$ defined in \eqref{eq:e}. Maximum at $e\simeq 0.002$ appears for $Re\leq 420$.  (b) Illustration of the thresholding $e(z,t) < e_{\rm turb}$, of a turbulent-laminar field at $Re=440$ with turbulent regions, $e(z,t) > e_{\rm turb}$ in white and quasi-laminar regions in blue. 
Definitions of $L_{\rm lam}$ and $L_{\rm turb}$, the lengths of quasi-laminar and turbulent regions, are illustrated.
(c) PDFs of laminar gap widths $L_{\rm lam}$ showing plateaux near 15 appearing for $Re\leq 440$. (d) PDFs of widths of turbulent regions $L_{\rm turb}$ showing local increase near 20 for $Re\leq 420$.
    }
\end{figure}

We consider the $x, y$-averaged cross-flow energy
\begin{equation}
\label{eq:e}
 e(z,t) \equiv \frac{1}{L_x L_y} \int_{-1}^1 \int_0^{L_x} \frac{1}{2}( v^2 + u_\spwise^{2})(x,y,z,t) ~ \mathrm{d}x \,\mathrm{d}y   
\end{equation}
as a useful diagnostic of  quasi-laminar and turbulent zones.
The probability density functions (PDFs) of $e(z,t)$ are shown in figure~\ref{fig:pdf_e} for various values of $Re$. 
%For $Re\geq 460$, the PDFs are maximal around $e \simeq 0.007$, and show strong asymmetry between their right and left tails.
%
The right tails, corresponding to high-energy events, are broad and exponential for all $Re$. The left, low-energy portions of the PDFs vary qualitatively with $Re$, unsurprisingly since these portions correspond to the weak turbulent events and hence include the gaps. 
%
%The steep left side of the distribution corresponds to events of weak turbulence or short-lived gaps, such as those visible at $Re=460$ in figure~\ref{fig:probes_Lz800}a. In comparison, the right tail is much broader and exponential. 
%
For large $Re$, the PDFs are maximal around $e \simeq 0.007$. As $Re$ is decreased, a low-energy peak emerges at $e\simeq 0.002$, corresponding to the emergence of long-lived quasi-laminar gaps seen in figure~\ref{fig:probes_Lz800}. The peak at $e\simeq 0.007$ flattens and gradually disappears. 
%An interesting feature is that the right portion of the PDF increases with decreasing $Re$. 
An interesting feature is that the distributions broaden with decreasing $Re$ with both low-energy and high-energy events becoming more likely.
This reflects a spatial redistribution of energy that accompanies the formation of gaps, with turbulent bands extracting energy from quasi-laminar regions and consequently becoming more intense.
(See figure 6 of \citet[Part 1]{gome1}.)
%This is presumably an effect of turbulent bands extracting energy from the laminar gaps. (See figure 6 of \cite{gome1}.)
%This is presumably the effect of turbulent bands extracting energy from the quasi-laminar regions and becoming more intense. \LT{ Merge two preceding sentences? "Quasi-laminar gaps", "formation of" repeated too much.} 

%with increased cross-flow energy within bands compared with uniformly turbulent flow \cite{barkley2005computational}

An intuitive way 
%to characterise the intermittent creation of gaps is 
to define turbulent and quasi-laminar regions is by thresholding the values of $e(z,t)$. In the following, a region will be called quasi-laminar if $e(z, t) < e_{\rm turb}$ and turbulent if $e(z, t) \geq e_{\rm turb}$. 
As the PDF of $e(z,t)$ evolves with $Re$, we define a $Re$-dependent threshold as a fraction of its average value, $e_{\rm turb} = 0.75 ~\overline{e}$. 
The thresholding is illustrated in figure \ref{fig:sketch_LT}, which is an enlargement of the flow at $Re=440$ that shows turbulent and quasi-laminar zones as white and blue areas, respectively.
Thresholding within a fluctuating turbulent environment can conflate long-lived gaps with tiny, short-lived regions in which the energy fluctuates below the threshold $e_{\rm turb}$. These are seen as the numerous small blue spots in figure \ref{fig:sketch_LT} that differ from the wider and longer-lived gaps.
This deficiency is addressed by examining the statistics of the spatial and temporal sizes of quasi-laminar gaps.

We present the length distributions of laminar $\Llam$ and turbulent zones $\Lturb$ in figures~\ref{fig:pdf_llam} and \ref{fig:pdf_lturb} at various Reynolds numbers. These distributions have their maxima at very small lengths, reflecting the large number of small-scale, low-turbulent-energy regions that arise due to thresholding the fluctuating turbulent field. 
As $Re$ is decreased, the PDF for $\Llam$ begins to develop a 
%peak near $\Llam\simeq 15$, 
\resub{plateau around $\Llam\simeq 15$},
corresponding to the scale of the gaps visible in figure \ref{fig:probes_Lz800}. 
%The right tails of the distribution shift upwards, but remain exponential. 
The right tails of the distribution are exponential and shift upwards with decreasing $Re$. 
The PDF of $\Lturb$ also varies with $Re$, but in a somewhat different way. As $Re$ decreases, the likelihood of a turbulent length in the range $15 \lesssim \Lturb \lesssim 35$ increases above the exponential background, but at least over the range of $Re$ considered, a maximum does not develop.

The laminar-length distributions show the emergence of structure at $Re$ higher than 
%prior to 
the turbulent-length distributions. This is  visible at $Re=440$, where the distribution of $\Lturb$ is indistinguishable from those at higher $Re$, while the distribution of $\Llam$ is substantially altered. This is entirely consistent with the impression from the visualisation in figure~\ref{fig:probes_q_R440} that quasi-laminar gaps emerge from a uniform turbulent background. 
Although the distributions of $\Llam$ and $\Lturb$ behave differently, the length scale emerging as $Re$ decreases are within a factor of two. This aspect is not present in the pipe flow results of \citet{avila2013nature}. (See Appendix \ref{app:comp_pipe} for a more detailed comparison.)\\

\subsection{\resub{Gap lifetimes and transition to patterns}}

\begin{figure}
\centering
\subfloat[]{\includegraphics[width=0.5\columnwidth]{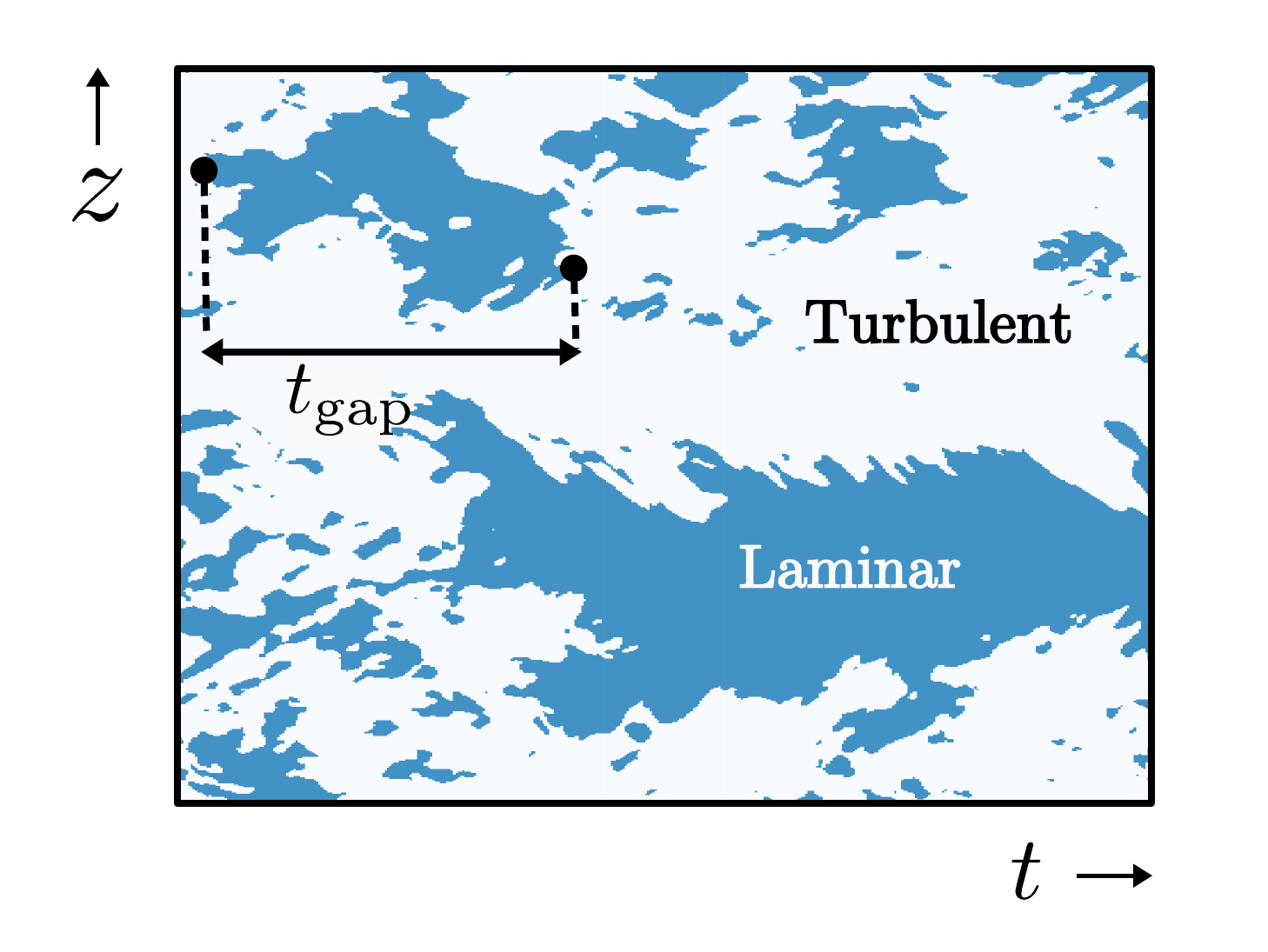} 
\label{fig:sketch_tgap}}
\subfloat[]{ \includegraphics[width=0.5\columnwidth]{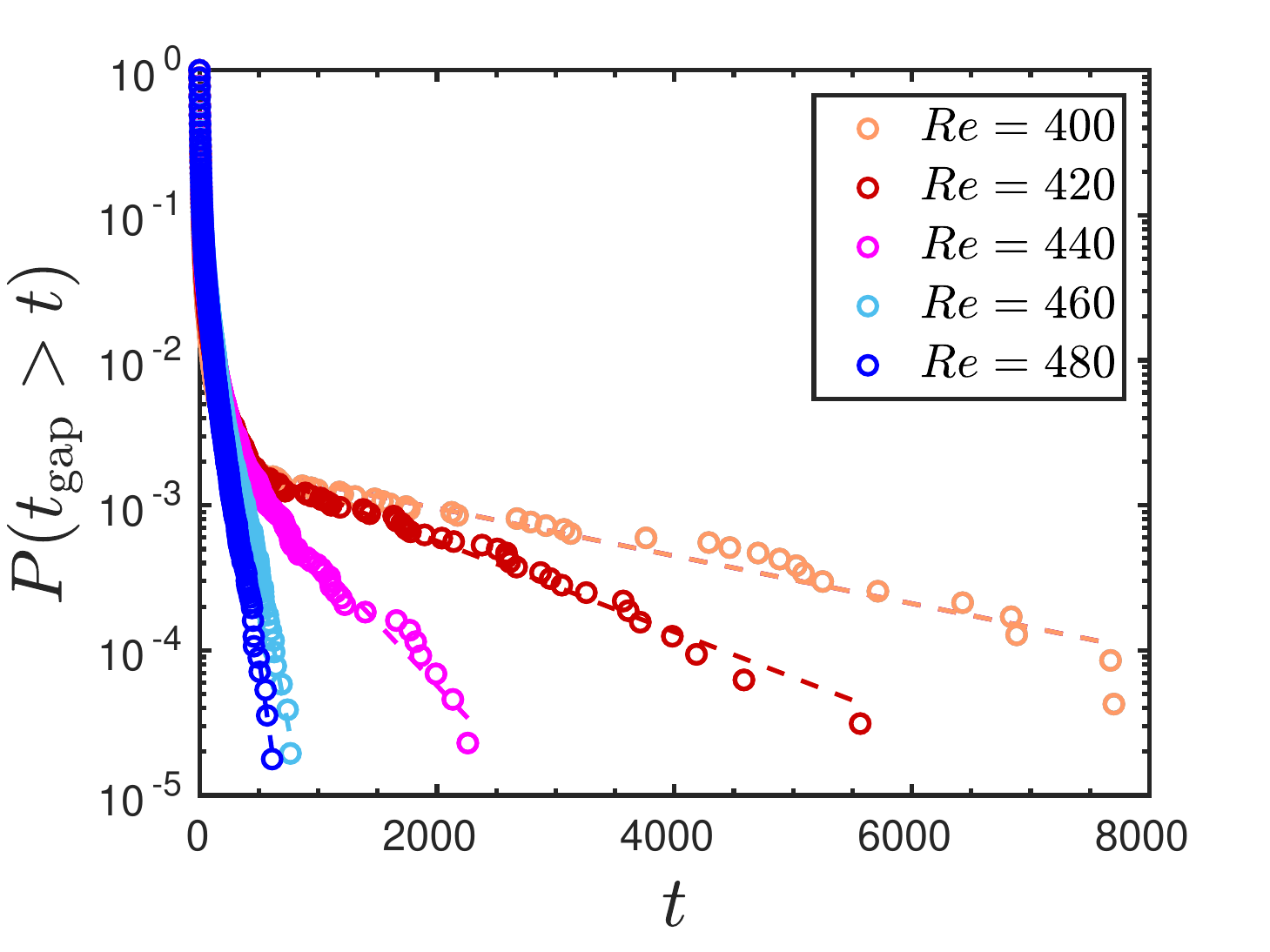} 
 \label{fig:survival_g}}\\
 \subfloat[]{ \includegraphics[width=0.5\columnwidth]{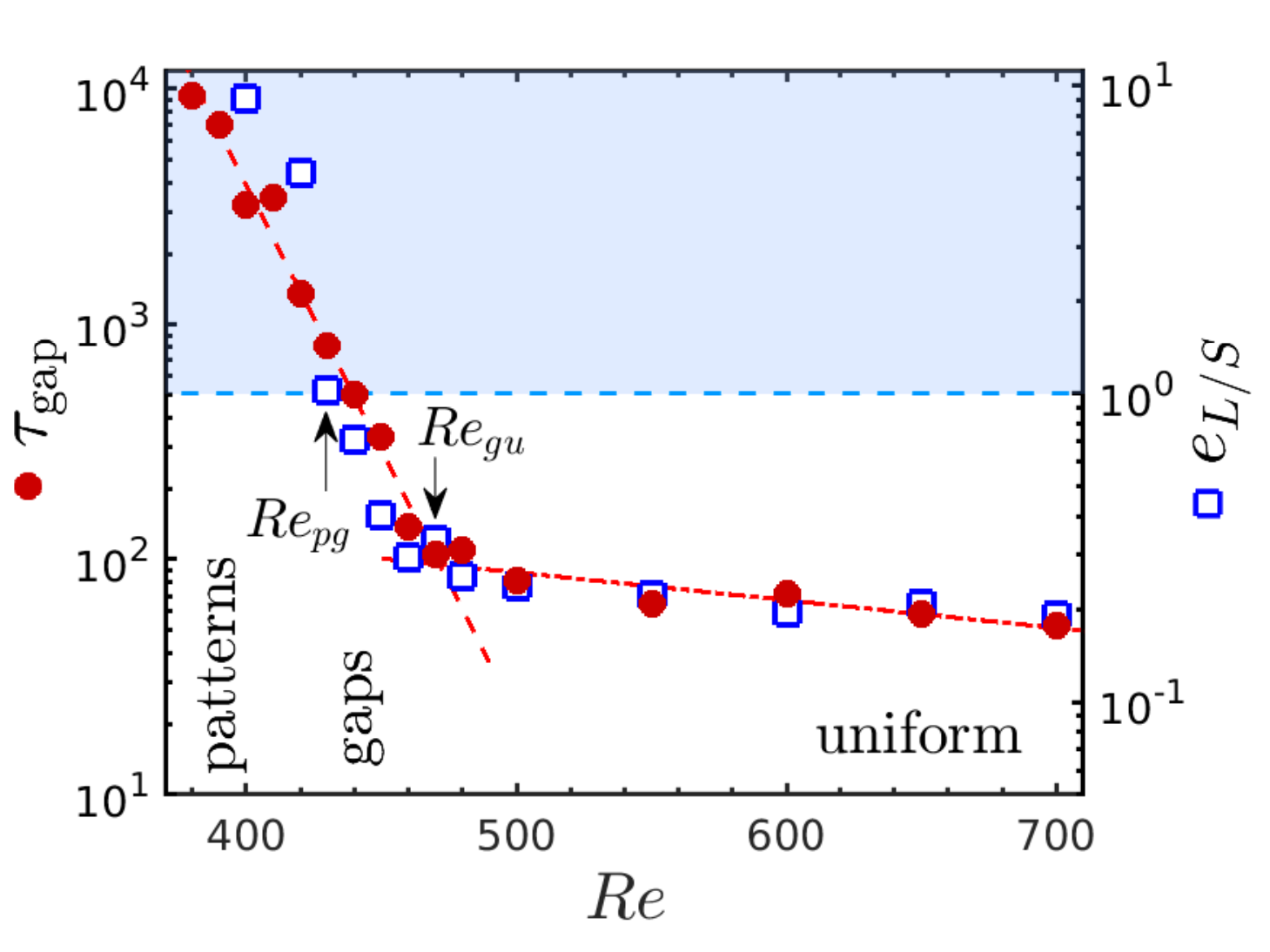}
 \label{fig:taug}}
 \subfloat[]{  \includegraphics[width=0.5\columnwidth]{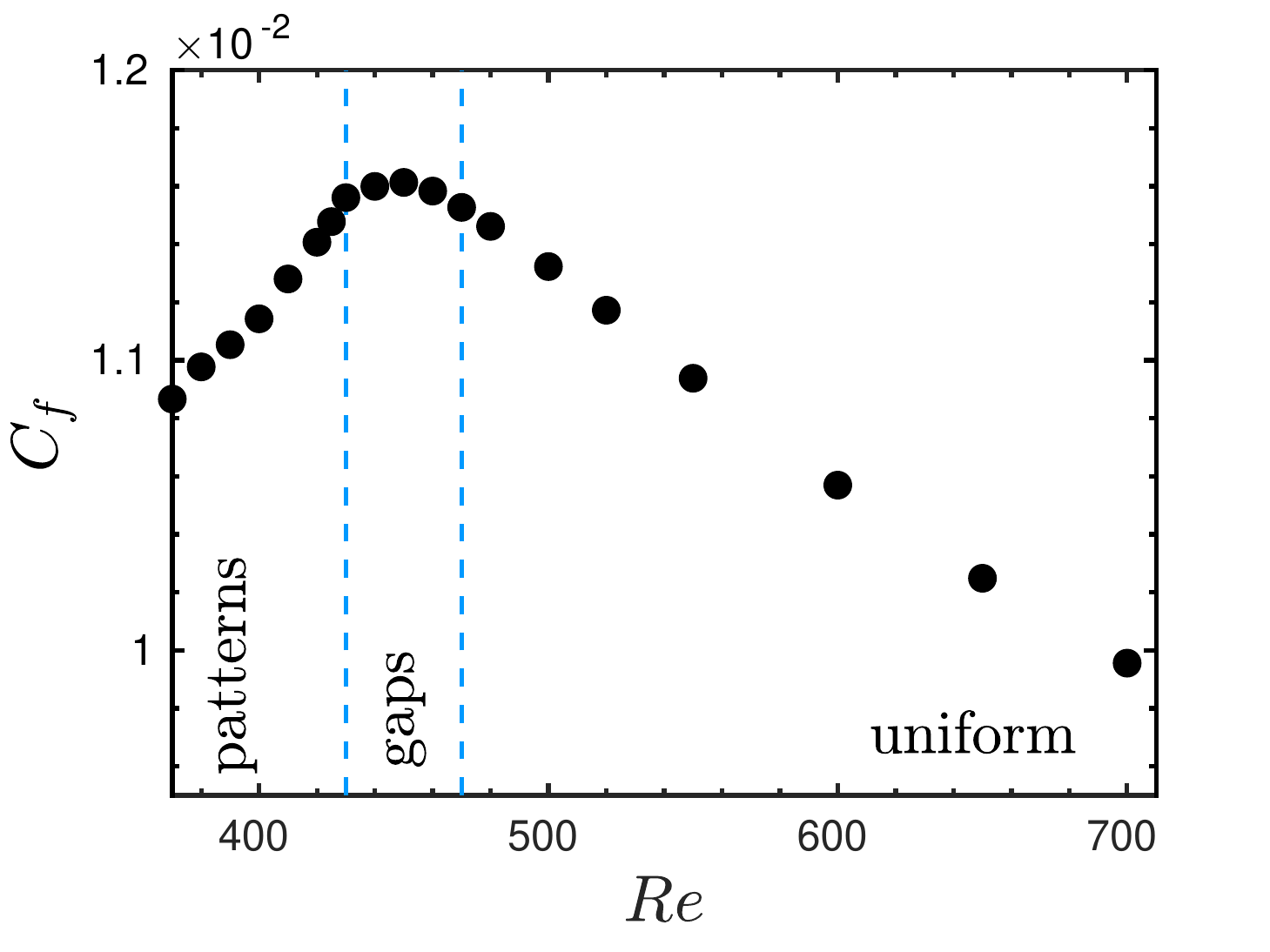} 
 \label{fig:Cf}}
\caption{(a) Same as figure \ref{fig:sketch_LT}, 
%but illustrating the definition of lifetimes of quasi-laminar gaps $t_{\text{gap}}$. 
\resub{
but illustrating the definition of $t_{\text{gap}}$, the lifetime of a quasi-laminar gap.
}
(b) Survival functions of $t_{\text{gap}}$. After initial steep portions, slopes yield the characteristic times.
(c) Evolution with $Re$ of characteristic time $\tau_{\text{gap}}$ and of ratio of large to small scale energy $e_{L/S}$ defined by \eqref{eq:e_LS}. Both of these quantities present two exponential regimes, with the same slopes and a common crossover at $\Regu$. \resub{The horizontal dashed line delimits the region $e_{L/S} > 1$, defining $\Repg$ below which regular patterns dominate. We estimate $\Repg \simeq 430$ and $\Regu\simeq 470$ (to two significant figures).}
(d) Evolution of friction coefficient $C_f$ with $Re$, with the three regimes 
delimited by $\Repg$ and $ \Regu$, as defined from (c).
}
 \label{fig:Cf_Cf}
\end{figure}

Temporal measurements of the gaps are depicted in figure \ref{fig:Cf_Cf}. Figure \ref{fig:sketch_tgap} shows the procedure by which we define the temporal extents $t_{\text{gap}}$ of quasi-laminar gaps. For each gap, i.e.\ a connected zone in $(z,t)$ satisfying $e(z,t)<e_{\rm turb}$, we locate its latest and earliest times and define $t_{\rm gap}$ as the distance between them.
Here again, we fix the threshold at $e_{\rm turb}=0.75~\overline{e}$. 
Figure~\ref{fig:survival_g} shows the temporal distribution of gaps, via the survival function of their lifetimes. 
In a similar vein to the spatial gap lengths, two characteristic behaviours are observed: for small times, many points are distributed near zero (as a result of frequent fluctuations near the threshold $e_{\rm turb}$), while for large enough times, an exponential regime is seen:
\begin{equation}
    \label{eq:exp_P}
    P(t_{\text{gap}} > t) \propto e^{-t/\tau_{\text{gap}}(Re)} \text{ for } t>t_0,
\end{equation}
\resub{where $t_0=500$ has been used for all $Re$,
although the exponential range begins slightly earlier for larger values of $Re$.}

%
%FRACTION OF GAPS? vs Ft 
%not contaminating

The slope of the exponential tail is extracted at each $Re$ and the resulting characteristic time-scale $\tau_{\text{gap}}$ is shown in figure~\ref{fig:taug}.
The evolution of $\tau_{\rm gap}$ with $Re$ displays two regimes, each with nearly exponential dependence on $Re$, but with very different slopes on the semi-log plot. For $Re\geq 470$, the characteristic lifetimes are $\tau_{\rm gap} = O(10^2)$ and vary weakly with $Re$.
%and lie on a line with a slow slope.
%This corresponds to the short-lived, low-energy events on the left tails of the PDFs for $e(z,t)$ in figure \ref{fig:pdf_e}, seen as small white events in figure \ref{fig:probes_q_R500}.
These short timescales correspond to the small white events visible in figure \ref{fig:probes_q_R500} and are associated with  low-energy values on the left tails of the PDFs for $e(z,t)$ in figure \ref{fig:pdf_e}.
Discounting these events, we refer to such states as uniform turbulence.  
%For $Re<470$, the slope in $\tau_{\rm gap} (Re)$ is steeper, and the abrupt change in slope marks the transition between uniform turbulence and the local gap regime. 
For $Re<470$, $\tau_{\rm gap}$ varies rapidly with $Re$, \resub{increasing by two orders of magnitude between $Re = 470$ and $Re=380$.}
\resub{The abrupt change in slope seen in figure~\ref{fig:taug}, which we denote by $Re_{\rm gu}$, 
\resub{marks the transition between gaps and uniform turbulence;}
%marks the transition between uniform turbulence and the emergence of local gaps as $Re$ is decreased; 
%We denote by $\Regu=470$ the Reynolds number at which this transition occurs.
we estimate $\Regu = 470$ (to two significant figures).} 
%The abrupt change in slope seen in figure~\r0ef{fig:taug} marks the transition between uniform turbulence and the emergence of local gaps as $Re$ is decreased. We denote by $\Regu=470$ the Reynolds number at which this transition occurs.
%and estimate $\Regu = 470$. 
%
%While this abrupt change in behaviour of $\tau_{\rm gap}$ provides a useful criterion for defining distinguishing uniform turbulence and flow with 
We stress that as far as we have been able to determine, there is no critical phenomenon associated with this change of behaviour. That is, the transition is smooth and lacks a true critical point. It is nevertheless evident that the dynamics of quasi-laminar gaps changes significantly in the region of $Re = 470$ and therefore it is useful to define a reference Reynolds number marking this change in behaviour.

Note that typical lifetimes of laminar gaps must become infinite by the threshold $Re \simeq 325$ below which turbulence is no longer sustained \citep{lemoult2016directed}.
(We believe this to be true even for $Re\lesssim 380$ when the permanent banded regime is attained, although this is not shown here.) For this reason, we have restricted our study of gap lifetimes to $Re\gtrsim 380$ and we have limited our maximal simulation time to $\sim 10^4$. 

To quantify the distinction between localized gaps and patterns,
we introduce a variable $e_{L/S}$ as follows. Using the Fourier transform in $z$,
\begin{equation}
\uhat(x,y,k_z,t) =  
\frac{1}{L_z}\int_{0}^{L_z}  \boldsymbol{u}(x,y,z,t) e^{- i k_z z } \,\text{d}z \:,
\label{eq:fourier}
\end{equation}
we compute the averaged spectral energy 
\begin{align}
\widehat{E}(y,k_z)\equiv\frac{1}{2}\overline{\uhat\cdot\uhat^\ast}, \qquad\qquad
\widehat{E}(k_z)\equiv \langle\widehat{E}(y,k_z)\rangle_y
\end{align}
where the overbar designates an average in $x$ and $t$.
This spectral energy is described in figure 3a of our companion paper \citet[Part 1]{gome1}.
We are interested in the ratio of $\widehat{E}(k_z)$ at large scales (pattern scale) to small scales (roll-streak scale), as it evolves with $Re$. For this purpose, we define the ratio of large-scale to small-scale maximal energy:
\begin{equation}
    \label{eq:e_LS}
   e_{L/S} = \frac{ \underset{ k_z < 0.5 }{ \max}\widehat{E} (k_z) }{\underset{ k_z \geq 0.5 }{ \max}\widehat{E} (k_z)}
\end{equation}
\resub{The choice of wavenumber $k_z=0.5$ to delimit large and small scales comes from the change in sign of non-linear transfers, as established in \citet[Part 1]{gome1}.}
This quantity is shown as blue squares in figure \ref{fig:taug} and is highly correlated to $\tau_{\rm gap}$. This correlation is in itself a surprising observation for which we have no explanation.

For $Re \gtrsim 430$, we have $e_{L/S} < 1$, signaling that the dominant peak in the energy spectrum is at the roll-streak scale, while for $Re\lesssim 430$, the large-scale pattern begins to dominate the streaks and rolls, as indicated by $e_{L/S} > 1$ (dashed blue line on figure \ref{fig:taug}). 
Note that $Re=430$ is also the demarcation between unimodal and bimodal PDFs of $e(z,t)$ in figure \ref{fig:pdf_e}. 
The transition from gaps to patterns is smooth.
In fact, we do not even observe a qualitative feature sharply distinguishing gaps and patterns.
We nevertheless find it useful to define a reference Reynolds number associated to patterns starting to dominate the energy spectrum.
This choice has the advantage of yielding a quantitative criterion, 
which we estimate as $\Repg \simeq 430$ \resub{(to two significant figures)}.
\resub{We find a similar estimation of the value of $Re$ below which patterns start to dominate via a wavelet-based measurement, see Appendix \ref{app:wavelet}.}

%Typical lifetimes of laminar gaps should become infinite for $Re< 380$ and certainly by the threshold $Re=325$ for maintenance of any kind of turbulence.
%
%The exponential behaviour in figure \ref{fig:taug} at $Re<470$ cannot be extrapolated to lower $Re$, since the typical lifetimes of laminar gaps become infinite for  $Re\lesssim 380$, when bands no longer coalesce
%LST -- do you want to say the phrase below? ("no longer" instead of "cannot")
%LST Is "coalesce" the right word here?

%Along with the lifetime measurement, 
In addition to the previous quantitative measures, we also extract the friction coefficient. This is defined as the ratio of the mean wall shear stress $\mu U^\prime_\text{wall}$ to the dynamic pressure $\rho U_\text{wall}^2/2$, which we write in physical units and then in non-dimensional variables as:
\begin{align}
C_f \equiv \frac{\mu U^\prime_\text{wall}}{\frac{1}{2}\rho U_\text{wall}^2} 
%= \frac{2\nu U^\prime_\text{wall}}{U_\text{wall}^2} 
= \frac{2\nu}{h U_\text{wall}} \frac{U^\prime_\text{wall}}{U_\text{wall}/h} 
= \frac{2}{Re} 
\frac{\partial \left<u_\strm \right>_{x,z,t}}{ \partial y} \bigg \rvert_{\rm wall}
\label{eq:cfdefine}
\end{align}
In \eqref{eq:cfdefine}, the dimensional quantities $h$, $\rho$, $\mu$, and $\nu$ are the half-height, the density, and dynamic and kinematic viscosities, and $U_{\rm wall}$ and $U^\prime_{\rm wall}$ are the velocity and mean velocity gradient at the wall.
%shear stress at the wall. 
%
We note that the behavior of $C_f$ in the transitional region has been investigated \resub{in plane channel flow} by \cite{ShimizuPRF2019} and \cite{kashyap2020flow}.
Our measurements of $C_f$ are shown in figure \ref{fig:Cf}. 
%We distinguish three regimes. In the uniform regime $Re>\Regu=470$, $C_f$ increases with decreasing $Re$. In the patterned regime $Re < \Repg=430$, $C_f$ decreases with decreasing $Re$. Between the two, in the localised-gap regime $\Repg < Re < \Regu$, $C_f$ is approximately constant. 
%for $Re< 430$.
\resub{We distinguish different trends within each of the three regimes defined earlier in figure \ref{fig:taug}. 
In the uniform regime $Re>\Regu=470$, $C_f$ increases with decreasing $Re$. In the patterned regime $Re < \Repg=430$, $C_f$ decreases with decreasing $Re$. The localised-gap regime $\Repg < Re < \Regu$ connects these two tendencies, with $C_f$ reaching a maximum at $Re=450$.}

\resub{The presence of a region of maximal $C_f$ (or equivalently maximal total dissipation) echoes the results on the energy balance presented in 
\citet[Part 1]{gome1}: the uniform regime dissipates more energy as $Re$ decreases, up to a point where this is mitigated by the many laminar gaps nucleated. This is presumably due to the mean flow in the turbulent region needing energy influx from gaps to compensate for its increasing dissipation.}

\subsection{\resub{Laminar-turbulent correlation function}}

\begin{figure}
    \centering
    \subfloat[]{\includegraphics[width=0.5\columnwidth]{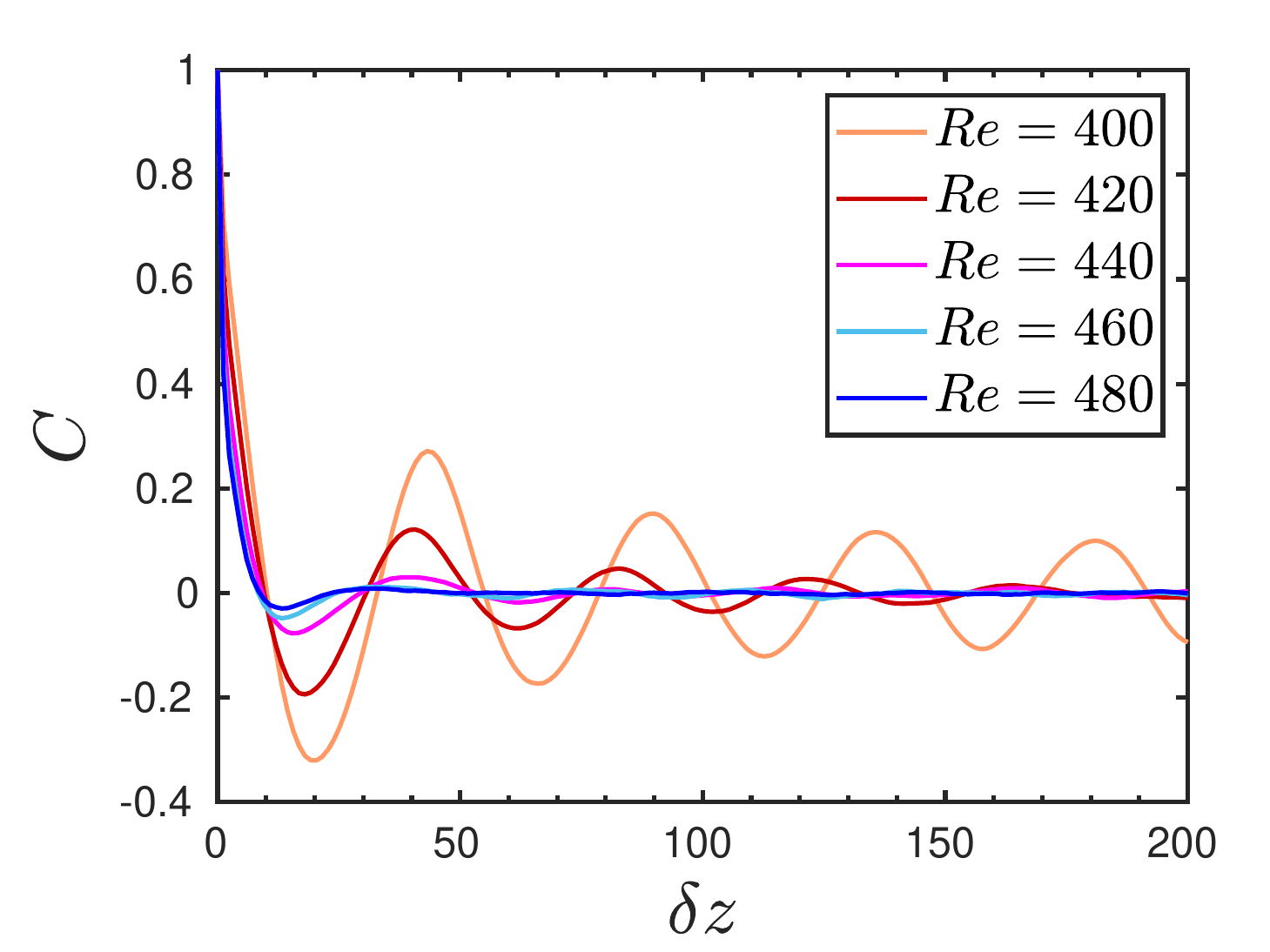}
    \label{fig:Corr}}~
    \subfloat[]{\includegraphics[width=0.5\columnwidth]{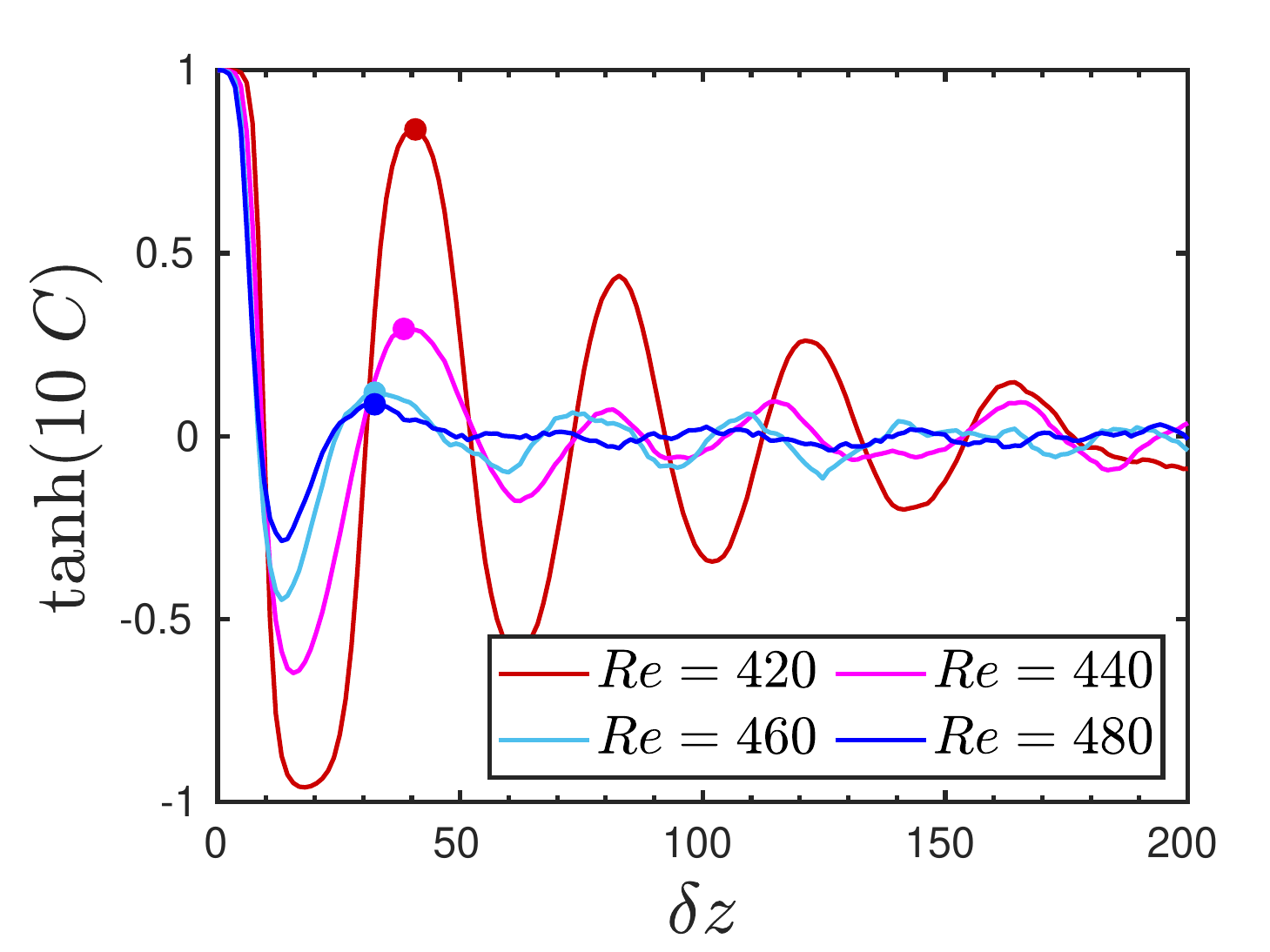}
    \label{fig:Corr_th}}\\
 \caption{(a) Gap-to-gap correlation function $C (\delta z)$ defined by \eqref{eq:Corr} for various values of $Re$. (b) \resub{For $Re\gtrsim 440$ the weak variation and short-ranged maxima are enhanced by plotting $\tanh(10 ~C (\delta z))$}.
 %The oscillations at $Re=420$ are weak at $Re=460$ and disappear at $Re=480$. 
 The dots correspond to the first local maximum, indicating the selection of a finite distance between two local gaps, including at the highest $Re$. \resub{Large-scale modulations smoothly leave room to weak short-range interaction as $Re$ increases and the flow visits patterned, local-gap and uniform regimes.}}
\end{figure}

The changes in regimes and the distinction between local gaps and patterns can be further studied
by measuring the spatial correlation between quasi-laminar regions within the flow. We define 
\begin{equation}
    \Theta(z,t) = \begin{dcases} 
1 ~~\text{   if }
e(z,t) < e_{\rm turb} \text{ (laminar) }\\
0 ~~\text{   otherwise (turbulent) }
\end{dcases}
    \label{eq:Theta}
\end{equation}
(this is the quantity shown in blue and white in figures \ref{fig:sketch_LT} and \ref{fig:sketch_tgap}).
We then compute its spatial correlation function:
\begin{equation}
    C (\delta z)  = \frac{\left< {\Theta}(z) {\Theta}(z+\delta z) 
\right>_{z, t} - \left< {\Theta}(z)\right>^2_{z, t}}{\left< {\Theta}(z)^2\right>_{z, t} - \left<{\Theta}(z)\right>^2_{z, t}}.
    \label{eq:Corr}
\end{equation}
Along with $(z,t)$ averaging, $C$ is also averaged over multiple realisations of quench experiments. As $\Theta$ is a Heaviside function, $C$ can be understood as the average probability of finding %quasi-laminar flow 
a gap at a distance $\delta z$ from a gap
%quasi-laminar flow 
at position $z$.
%for the flow at $z+\delta z$ to be laminar, conditioned on the the flow at $z$ being laminar.
The results are presented in figure \ref{fig:Corr}. 
\resub{The comparative behaviour of $C$ at near-zero values}
is enhanced by plotting $\tanh(10 ~C)$ in figure \ref{fig:Corr_th}. At long range, $C$ approaches zero with small fluctuations at $Re=480$, a noisy periodicity at $Re=460$, and a nearly periodic behaviour for $Re\leq420$.

In all cases, $C$ initially decreases from one and reaches a first minimum 
\resub{at $\delta z \simeq 20$},
due to the minimal possible size of a turbulent zone that suppresses the creation of neighbouring laminar gaps. 
%in the range $\delta z \lesssim 30$. 
$C$ has a prominent local maximum $\delta z_{\rm max}$ right after its initial decrease, at $\delta z_{\rm max} \simeq 32$ at $Re=480$, which increases to $\delta z_{\rm max} \simeq 41$ at $Re=420$. These maxima, shown as coloured circles in figure \ref{fig:Corr_th}, indicate that gap nucleation is preferred at distance $\delta z_{\rm max}$ from an existing gap. The increase in $\delta z_{\rm max}$ and in the subsequent extrema as $Re$ is lowered agrees with the trend of increasing wavelength of turbulent bands as $Re$ is decreased in the fully banded regime at lower $Re$ \citep{prigent2003long,barkley2005computational}.

%

%This first part aimed at observing and characterizing how gaps emerge from uniform turbulence in plane Couette flow in a Long Slender Box, whose long direction allows for the formation of multiple turbulent and laminar structures.
% We showed that gaps are themselves intermittent and characterised by exponential-tailed distributions, and by a characteristic time $\tau_{\text{gap}}$ that behaves exponentially with $Re$, when $Re$ is restricted to $[380-480]$. We also see that gaps randomly nucleate, without creating neighbouring gaps when $Re\geq 440$. When $Re<440$, gaps organise into patterns. This is due to an increased probability of gap formation, and to the finite-size of gaps and bands. 
%Our observations argue against a linear instability of the uniform flow as a 
%Our observations confirm the absence of large-scale modulation in the uniform regime $Re>470$ (as defined in figure \ref{fig:taug}), but emphasise the presence of (weak) gap interaction at a finite distance in this regime. This preference is stronger as $Re$ decreases and multiple gaps appear close to one another.

\resub{The smooth transition from patterns to uniform flow is confirmed in the behaviour of the correlation function. Large-scale modulations characteristic of the patterned regime gradually disappear with increasing $Re$, as gaps become more and more isolated. Only a weak, finite-length interaction subsists in the local-gap and uniform regimes, and will further disappear with increasing $Re$.}
%even in the uniform regime.
%However, local gaps can still slightly infer the presence of a neighbouring gap. This effect suggests a preference for a finite distance at which gaps interact.
%, even local in space,
%SG:
%due to the gaps interacting at a finite distance.
%The selection of a finite gap spacing will be investigated in \S \ref{sec:exist_stab} and \S \ref{sec:spec_optim}.
\resub{This is the selection of this finite gap spacing that we will investigate in \S \ref{sec:exist_stab} and \S \ref{sec:spec_optim}.
}

%\section{Existence and stability bounds for patterns}
\section{Wavelength selection for turbulent-laminar patterns}
\label{sec:exist_stab}

In this section, we investigate the existence of a preferred pattern wavelength by using as a control parameter the length $L_z$ of the Minimal Band Unit.
In a Minimal Band Unit, the system is constrained and the distinction between local gaps and patterns is lost; \resub{see section 3 of our companion paper \citet[Part 1]{gome1}}.
%(MBU). 
%In a Minimal Band Unit, 
%$L_z$ is chosen such as to accommodate a single turbulent band and a single gap, which can be viewed as one period of a perfectly periodic pattern.
$L_z$ is chosen such as to accommodate at most a single turbulent zone and a single quasi-laminar zone, which due to imposed periodicity, can be viewed as one period of a perfectly periodic pattern.
By varying $L_z$, we can verify whether a regular pattern of given wavelength $L_z$ can emerge from uniform turbulence, disregarding the effect of scales larger than $L_z$ or of competition with wavelengths close to $L_z$. We refer to these simulations in Minimal Band Units as \emph{existence} experiments. Indeed, one of the main advantages of the Minimal Band Unit is the ability to create patterns of a given angle and wavelength which may not be stable in a larger domain.
%SG: last sentence repetitive?

In contrast, in a Long Slender Box, $L_z$ is large enough to accommodate multiple bands and possibly even patterns of different wavelengths.
An initial condition consisting of a regular pattern of wavelength $\lambda$ can be constructed by concatenating bands produced from a Minimal Band Unit of size $\lambda$. The \emph{stability} of such a pattern is studied by allowing this initial state to evolve via the non-linear Navier-Stokes equations.
Both existence and stability studies can be understood in the framework of the Eckhaus instability \citep[]{kramer1985eckhaus, ahlers1986wavenumber, tuckerman1990bifurcation, cross2009pattern}.

%LST I do not understand this passage. I do not feel we characterised how the patterns appeared, we just studied the patterns (once they had appeared). Or perhaps you mean appear as Re is lowered rather than appearing in time. I wouldn't say that trig modes are not favored in uniform turbulent state -- they are not present by definition of uniform turbulent state!
In previous studies of transitional regimes, \citet{barkley2005computational} studied the evolution of patterns as $L_z$ was increased.
%Another way is to compute statistics of single-points velocity \citep{moxey2010distinct}, which is deeply affected by the presence of quasi-laminar gaps within the turbulent flow. This measurement can be useful to quantify the change in statistics accompanying the emergence of patterns, but might however be hindered in characterizing the transition, because it is hard to distinguish statistics coming from turbulent fluctuations, that wiggle with a most probable value near 0, and the impact of laminar gaps that also have almost null velocities.
In Section \ref{sec:existence}, we extend 
%the approach of \citet[]{barkley2007mean} 
this approach 
to multiple sizes of the Minimal Band Unit
by comparing lifetimes of patterns that naturally arise in this constrained geometry. The stability of regular patterns of various wavelengths will be studied in Long Slender Domains ($L_z=400$) in Section \ref{sec:stability}.

\subsection{Temporal intermittency of regular patterns in a short-$L_z$ box}
\label{sec:existence}

\begin{figure}
    \centering
    \subfloat[]{
\includegraphics[width=\columnwidth]{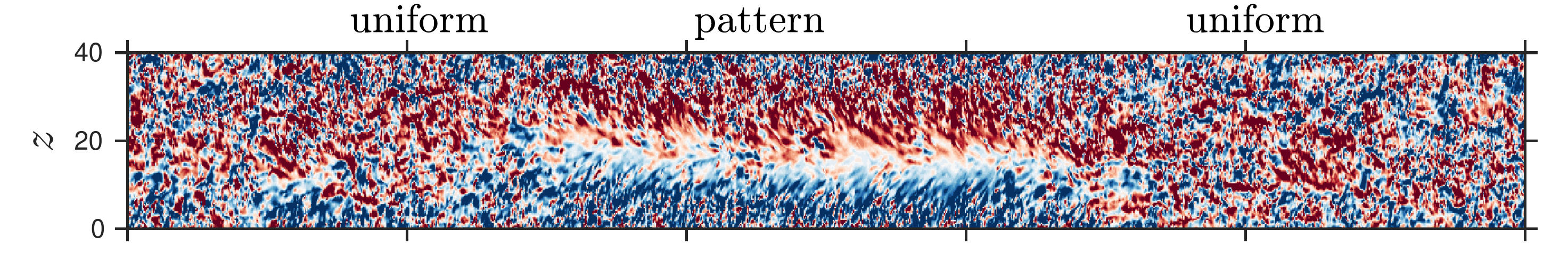} \label{fig:probes_MBU}} \\
\vspace{-1em}
 \subfloat[]{
\includegraphics[width=\columnwidth]{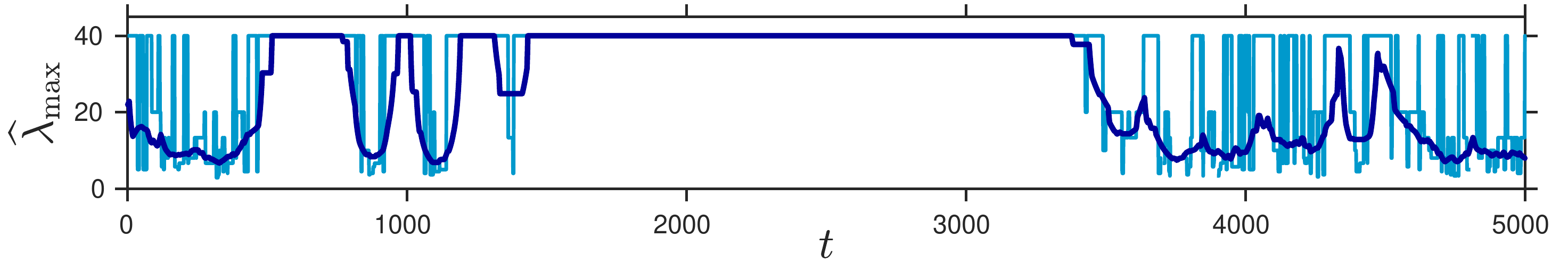} \label{fig:lambda_MBU}}\\
\vspace{-1em}
\subfloat[]{ \includegraphics[width=0.5\columnwidth]{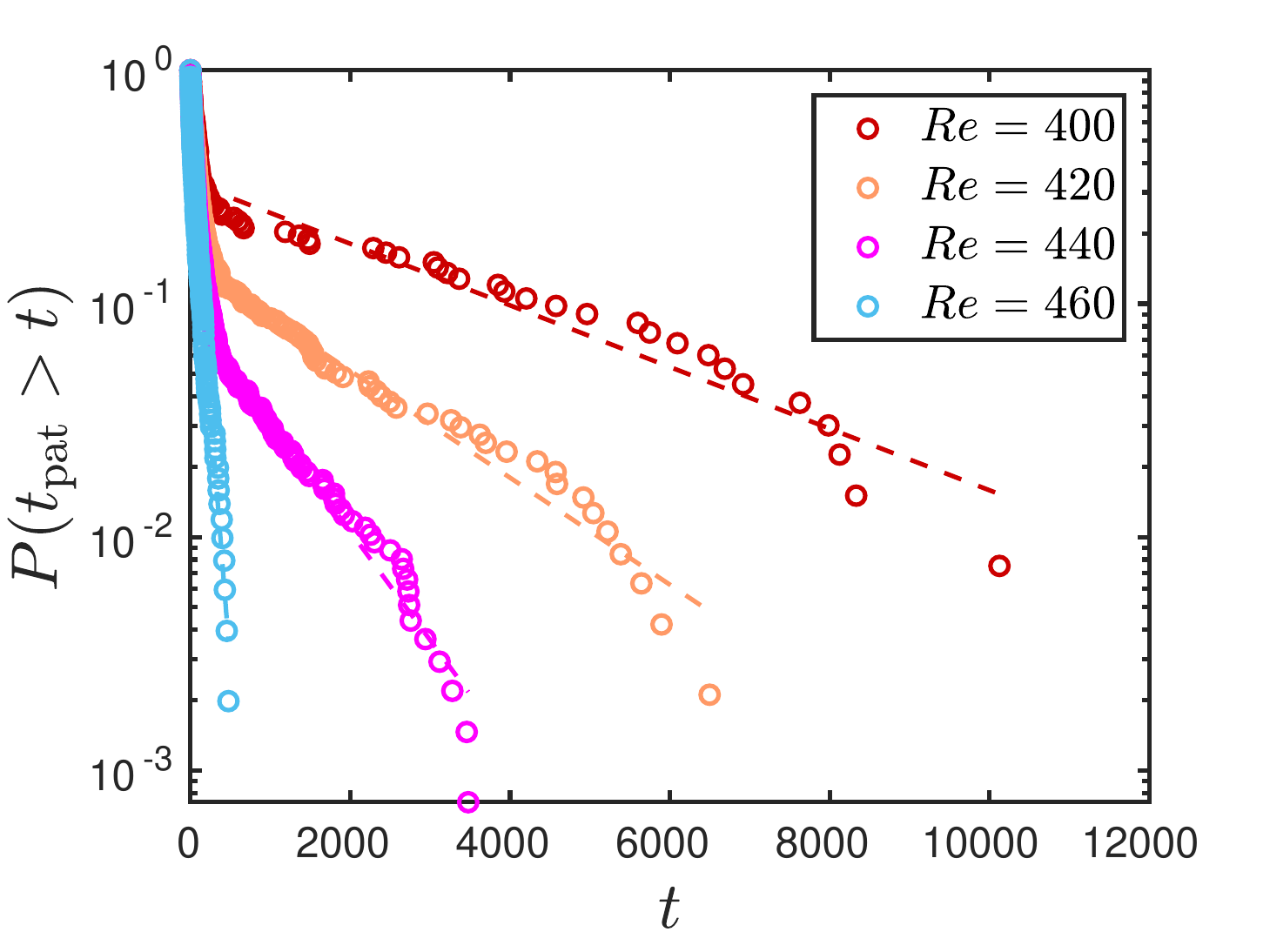} \label{fig:survival_pattern}}~
 \subfloat[]{\includegraphics[width=0.5\columnwidth]{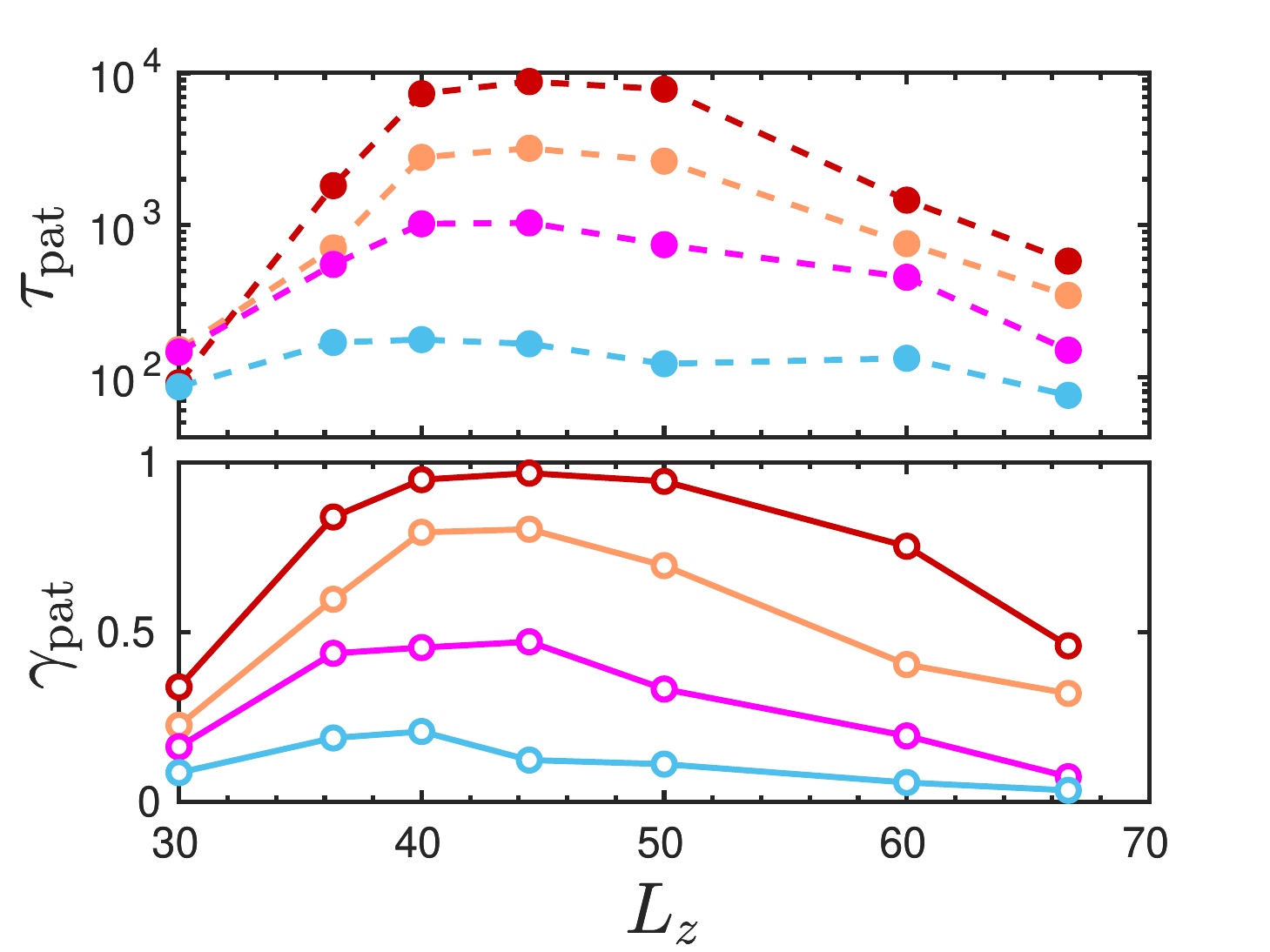} \label{fig:lifetimes_pattern}} 
\caption{Pattern lifetimes. (a) Space-time visualization of a metastable pattern in a Minimal Band Unit with $L_z=40$ at $Re=440$. Colors show spanwise velocity (blue: $-0.1$, white: 0, red: 0.1). (b) Values of the dominant wavelength $\lambdamaxMBU$ (light blue curve) and of its short-time average $\langle\lambdamaxMBU\rangle_{t_a}$ (dark blue curve) are shown; see \eqref{eq:lambdamaxMBU}. A state is defined to be patterned if $\lambdamaxMBU = L_z$.
(c) Survival function of lifetimes of turbulent-laminar patterns in a Minimal Band Unit with $L_z=40$ for various $Re$. The pattern lifetimes $t_{\rm pat}$ are the lengths of the time intervals
during which $\lambdamaxMBU = L_z$. (d) Above: characteristic times $\tau_{\rm pat}$ extracted from survival functions as a function of $L_z$ and $Re$. 
Below: intermittency factor for the patterned state $\gamma_{\rm pat}$, which is the fraction of time spent in the patterned state.
%
%Below: fraction of time patterned state vs. uniform state $\gamma_{\rm pat}$.
}
\end{figure}

Figure \ref{fig:probes_MBU} shows the formation of a typical pattern in a Minimal Band Unit of size $L_z=40$ and at $Re=440$. 
While the system cannot exhibit the spatial intermittency seen in figure \ref{fig:probes_q_R440}, temporal intermittency is possible and is seen as alternation between uniform turbulence and a pattern. 
We plot the spanwise velocity at $y=0$ and $x=L_x/2$.
This is a particularly useful measure of the large-scale flow associated with patterns, 
seen as red and blue zones surrounding a white 
quasi-laminar region, 
i.e.\ a gap.
The patterned state spontaneously emerges from uniform turbulence and remains from $t \simeq 1500$ to $t \simeq 3400$. At $t\simeq500$, a short-lived 
%quasi-laminar zone 
gap 
appears at $z=10$, which can be seen as an attempt to form a pattern.

%The pattern is characterised quantitatively by computing the wavenumber that instantaneously maximises the energy of the Fourier mode $k_z$:
%\begin{equation}
 % \lambdamaxMBU(t) = \frac{2\pi}{\underset{k_z>0}{\text{argmax}}\: |{\langle\widehat{\boldsymbol{u}}(y=0,k_z,t)}\rangle_x|^2}, 
%\label{eq:lambdamaxMBU}
%\end{equation}
%where $\langle\widehat{\boldsymbol{u}}(y=0,k_z,t)\rangle_x$ denotes the $x$ average of the $z$ Fourier transform of the mid-plane velocity.
%
\resub{We characterise the pattern quantitatively as follows.} 
%For each time $t$, we take the $z$ Fourier transform of the mid-plane velocity, then average over $x$, and finally take the square modulus. This yields $|\langle\widehat{\boldsymbol{u}}(y=0,k_z,t)\rangle_x|^2$, which is the instantaneous energy contained in wavenumber $k_z$.
\resub{For each time $t$, we compute $|\langle\widehat{\boldsymbol{u}}(y=0,k_z,t)\rangle_x|^2$, which is the instantaneous energy contained in wavenumber $k_z$ at the mid-plane.
We then determine the wavenumber that maximises this energy and compute the corresponding wavelength.} That is, we define
\begin{equation}
  \lambdamaxMBU(t) \equiv \frac{2\pi}{\underset{k_z>0}{\text{argmax}}\: |{\langle\widehat{\boldsymbol{u}}(y=0,k_z,t)}\rangle_x|^2}.
\label{eq:lambdamaxMBU}
\end{equation}
The possible values of $\lambdamaxMBU$ are integer divisors of $L_z$, here 40, 20, 10, etc. 
Figure \ref{fig:lambda_MBU} presents $\lambdamaxMBU$ and its short-time average $\langle\lambdamaxMBU\rangle_{t_a}$ with $t_a=30$ as light and dark blue curves, respectively. 
When turbulence is uniform, $\lambdamaxMBU$ varies rapidly between its discrete allowed values, while $\langle\lambdamaxMBU\rangle_{t_a}$ fluctuates more gently around 10. The flow state is deemed to be patterned when its dominant mode is $\langle\lambdamaxMBU\rangle_{t_a} =L_z$.
%
%Both the pattern for $1500 \leq t \leq 3400$ and the short-lived gap for $500\leq t \leq 750$ seen in the visualisation of figure \ref{fig:probes_MBU} are manifested as plateaus in $\langle\lambdamaxMBU\rangle_{t_a}$. Note that $L_z$ is a dominant wavelength for other periods shorter than these two noticeable events. 
%
The long-lived pattern occurring for $1500 \leq t \leq 3400$ in figure~\ref{fig:probes_MBU} is seen as a plateau of $\langle\lambdamaxMBU\rangle_{t_a}$ in figure~\ref{fig:lambda_MBU}. There are other shorter-lived plateaus, notably at for $500\leq t \leq 750$.
A similar analysis was carried out by \citet{barkley2005computational,tuckerman2011patterns} using the Fourier component corresponding to wavelength $L_z$ of the spanwise mid-gap velocity. 

Figure~\ref{fig:survival_pattern} shows the survival function $t_{\rm pat}$ of the pattern lifetimes 
%defined 
obtained from
$\langle\lambdamaxMBU\rangle_{t_a}$
% seen in figure \ref{fig:lambda_MBU}
over long simulation times for various $Re$.
This measurement differs from figure \ref{fig:survival_g}, which showed lifetimes of \emph{gaps} in a Long Slender Box and not regular \emph{patterns} obtained in a Minimal Band Unit. %Here, the spatio-temporal intermittency reduces to a temporal problem, since we consider the flow in the Minimal Band Unit to either contain a pattern or not.
%Nevertheless the picture is qualitatively similar.
%: 
The results are however qualitatively similar, with two characteristic zones in the distribution, as in in figure \ref{fig:survival_g}:
% the distributions show two characteristic zones: many short-lived patterns populate the 
at short times, many patterns appear due to fluctuations; while after $t\simeq 200$, the survival functions enter an approximately exponential regime, from which we extract the characteristic times $\tau_{\rm pat}$ by taking the inverse of the slope.

We then vary $L_z$, staying within the Minimal Box regime $L_z \lesssim 65$ in which only one band can fit.  Figure \ref{fig:lifetimes_pattern} (top) shows that $\tau_{\rm pat}$ presents a broad maximum in $L_z$ whose strength and position depend on $Re$: $L_z\simeq 42$ at $Re=440$ and $L_z\simeq 44$ at $Re=400$.
This wavelength corresponds approximately to the natural spacing observed in a Large Slender Box (figure \ref{fig:probes_Lz800}).
Figure \ref{fig:lifetimes_pattern} (bottom) presents the fraction of time that is spent in a patterned state, denoted $\gamma_{\rm pat}$, to reflect that this should be thought of as the intermittency factor for the patterned state.
The dependence of $\gamma_{\rm pat}$ on $L_z$ follows the same trend as $\tau_{\rm pat}$, but less strongly (the scale of the inset is linear, while that for $\tau_{\rm pat}$ is logarithmic).
%and with the most-energetic large scale measured on Fig.~\ref{fig:spec_Lz800_Re}, $\lambda \simeq 42$. 
%For $Re=480$, the survival function is nearly the same as for 460 and $\tau_{\rm pat}$ and $\gamma_{\rm pat}$ are nearly independent of $L_z$; this is the situation for uniform turbulence.

%These results complement the 

The results shown in figure \ref{fig:lifetimes_pattern} complement the Ginzburg-Landau description proposed by \citet[]{prigent2003long} and \citet[]{rolland2011ginzburg}. To quantify the bifurcation from featureless to pattern turbulence, these authors defined an order parameter and showed that it has a quadratic maximum at an optimal wavenumber.
This is consistent with the approximate quadratic maxima that we observe in 
%the logarithmic plot of pattern lifetimes, and in the linear plot of $\gamma_{\rm pat}$ with regard to $L_z$. 
$\tau_{\rm pat}$ and in $\gamma_{\rm pat}$ with regard to $L_z$.
%These important primary results are here generalised to the bistability observed between patterns and uniform turbulence.
\resub{Note that the scale of the pattern can be roughly set from the force balance in the laminar flow regions \citep{barkley2007mean}, $\lambda\simeq Re\sin\theta/\pi $, which 
yields a wavelength of 52 at $Re=400$ (close to the value of 44 found in figure \ref{fig:lifetimes_pattern}).}
%
%Furthermore,

\subsection{Pattern stability in a large domain}
\label{sec:stability}

%% ARXIV VERSION

\begin{figure}
  \centering
\includegraphics[width=\columnwidth]{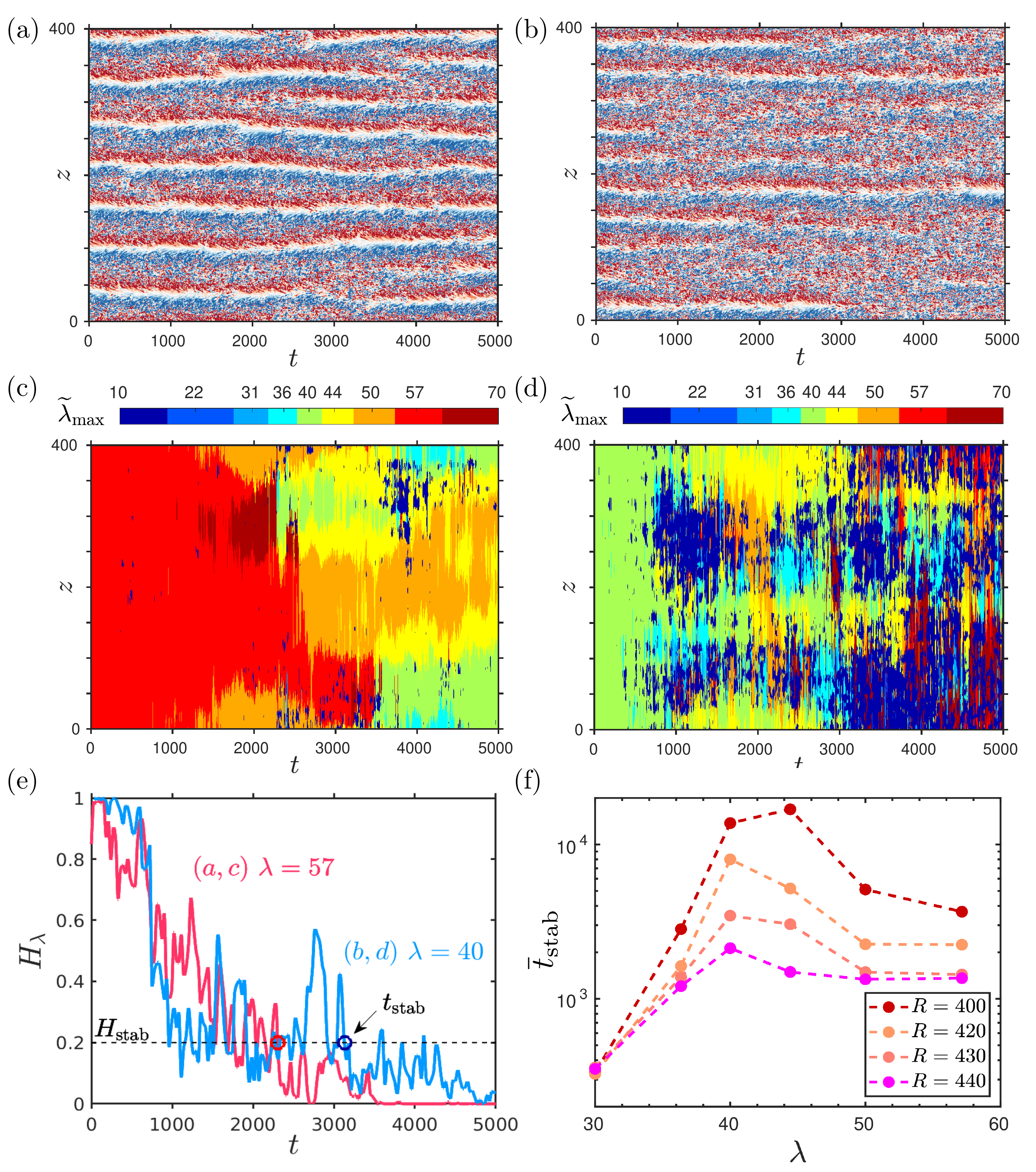}\label{fig:stab} 
  \caption{Simulation in a Long Slender Box from a noise-perturbed periodic pattern with 
(a) initial $\lambda = 57$ at $Re=400$ and (b) initial $\lambda = 40$ at $Re=430$.
Colors show spanwise velocity
(red: 0.1, white: 0, blue: $-0.1$).
(c) and (d) show the local dominant wavelength $\tilde{\lambda}_{\rm max}(z,t)$ determined by wavelet analysis (see Appendix \ref{app:wavelet}) corresponding to the simulations shown in (a) and (b). Color at $t=0$ shows the wavelength $\lambda$ of the initial condition.
(e) shows the wavelet-defined $H_\lambda(t)$ defined in \eqref{eq:H}, which quantifies the proportion of the domain that retains initial wavelength $\lambda$ as a function of time for cases (a) and (b). Circles indicate the times for (a) and (b) after which $H_\lambda$ is below the threshold value $H_{\rm stab}$ for a sufficiently long time. (f) Ensemble-averaged $\bar{t}_\text{stab}$ of the decay time of an imposed pattern of wavelength $\lambda$ for various values of $Re$. The relative stability of a wavelength, whether localised or not, is measured by $\bar{t}_\text{stab}$ via the wavelet analysis. 
}
\label{fig:stab}
\end{figure}

To study the stability of a pattern of wavelength $\lambda$, we prepare an initial condition for a Long Slender Box by concatenating 
repetitions of a single band produced in a Minimal Band Unit. 
We add small-amplitude noise to this initial pattern so that the repeated bands do not all evolve identically.
Figures~\ref{fig:stab}a and \ref{fig:stab}b show two examples of such simulations. Depending on the value of $Re$ and of the initial wavelength $\lambda$, the pattern destabilises to either another periodic pattern
(figure \ref{fig:stab}a for $Re=400$) or to localised patterns surrounded by patches of featureless turbulence (figure \ref{fig:stab}b for $Re=430$).

It can be seen that patterns often occupy only part of the domain. 
For this reason, we turn to the wavelet decomposition \citep[]{meneveau1991analysis, farge1992wavelet} 
to quantify patterns locally.
In contrast to a Fourier decomposition, the wavelet decomposition quantifies the signal as a function of space and scale. From this, we are able to define a local dominant wavelength, $\lambdamax(z,t)$, similar in spirit to
$\lambdamaxMBU(t)$ in \eqref{eq:lambdamaxMBU}, but now at each space-time point. (See Appendix \ref{app:wavelet} for details.) 
Figures~\ref{fig:stab}c and \ref{fig:stab}d show $\lambdamax(z,t)$ obtained from wavelet analysis of the simulations visualised in figures \ref{fig:stab}a and \ref{fig:stab}b.

We now use the local wavelength $\lambdamax(z,t)$ to quantify the stability of an initial wavelength. We use a domain of length $L_z=400$ and we concatenate
$n=7$ to 13 repetitions of a single band to produce a pattern with initial wavelength $\lambda(n)\equiv  400/n\simeq57,50,44\ldots31$. (We have rounded $\lambda$ to the nearest integer value here and in what follows.)
After adding low-amplitude noise, we run a simulation lasting 5000 time units, compute the wavelet transform and calculate from it the local wavelengths $\lambdamax(z,t)$.
We define \resub{$\epsilon_\lambda \equiv \min((\lambda(n+1)-\lambda(n))/2, (\lambda(n)-\lambda(n-1))/2)$}
such that $|\lambda - \lambdamax(z, t)|< \epsilon_\lambda$ if 
$\lambdamax$ is closer to $\lambda(n)$ than to its two neighboring values  . Finally, in order to measure the proportion of $L_z$ in the dominant mode $\lambdamax$ is $\lambda$, we  compute
\begin{equation}
\label{eq:H}
H_\lambda(t) = \left\langle\frac{1}{L_z}\int_0^{L_z} \Theta\left(\epsilon_\lambda - |\lambda - \lambdamax(z, t)|\right)~\text{d}z \right\rangle_{t_a}
\end{equation}
where $\Theta$ is the Heaviside function and the short-time average $\left<\cdot\right>_{t_a}$ is taken over time $t_a=30$ as before.
In practice, because patterns in a Long Slender Box still fluctuate in width, a steady pattern 
may have $H_\lambda$ 
somewhat less than 1. 
If $H_\lambda \ll 1$, a pattern of wavelength $\lambda$ is present in only a very small part of the flow. 
      
Figure~\ref{fig:stab}e shows how wavelet analysis via the Heaviside-like function $H_\lambda(t)$
quantifies the relative stability of the pattern in the examples shown in figures \ref{fig:stab}a and \ref{fig:stab}b.
The flow in figure \ref{fig:stab}a at $Re=400$ begins with $\lambda=57$, i.e.\ 7 bands. 
%LST added -- maybe not?
%SG: we already talk about it later! 
%LST but not so explicitly and I think we should talk about it before talking about e which is derived from c via (4.2)
% I really find this part already extremeley wordy
Figure \ref{fig:stab}c retains the red color corresponding to $\lambda=57$ over all of the domain for $t\lesssim 1200$ and over most of it until $t\lesssim 2300$.
The red curve in figure \ref{fig:stab}e shows $H_\lambda$ decaying quickly and roughly monotonically. One additional gap appears at around $t=2300$ and starting from then, $H_\lambda$ remains low. This corresponds to the initial wavelength $\lambda=57$ losing its dominance to $\lambda=40$, 44 and 50 in the visualisation of $\lambdamax(z,t)$ in figure \ref{fig:stab}c.
 By $t=5000$, the flow shows 9 bands with a local wavenumber $\lambda$ between 40 and 50.

The flow in figure \ref{fig:stab}b at $Re=430$ begins with $\lambda=40$, i.e.\ 10 bands. Figure \ref{fig:stab}d shows that the initial light green color corresponding to 40 is retained until $t\lesssim 800$. 
The blue curve in figure \ref{fig:stab}e representing $H_\lambda$ initially decreases and drops precipitously around $t\simeq 1000$ as several gaps disappear in figure \ref{fig:stab}b.
$H_\lambda$ then fluctuates around a finite value, which is correlated to the presence of gaps whose local wavelength is the same as the initial $\lambda$, visible as zones where $\lambdamax = 40$ in figure \ref{fig:stab}d. The rest of the flow can be mostly seen as locally featureless turbulence, where the dominant wavelength is small ($\lambdamax \le 10$).
The local patterns fluctuate in width and strength, and $H_\lambda$ evolves correspondingly after $t=1000$.
The final state reached in figure \ref{fig:stab}a at $Re=430$ is characterised by the presence of intermittent local gaps.

The lifetime of an initially imposed pattern wavelength $\lambda$ is denoted $t_\text{stab}$ and is defined as follows:
We first define a threshold $H_{\rm stab} \equiv 0.2$ (marked by a horizontal dashed line on figure \ref{fig:stab}e). If $H_\lambda (t)$ is statistically below $H_{\rm stab}$, the imposed pattern will be considered as unstable. 
Following this principle, $t_\text{stab}$ is defined as the first time $H_\lambda$ is below $H_{\rm stab}$, 
\resub{with a further condition to dampen the effect of short-term fluctuations: $t_\text{stab}$ must obey $\left<H_\lambda (t)\right>_{t\in [t_\text{stab}, ~ t_\text{stab} + 2000]} < H_{\rm stab}$, so as to ensure that the final state is on average below $H_{\rm stab}$.}
%with two further conditions to dampen the effect of short-term fluctuations. First, $H_\lambda (t)$ must be below $H_{\rm stab}$ for a period of $\Delta t_1=100$ after $t_\text{stab}$. This avoids selecting a local minimum of little importance. Second, $t_\text{stab}$ must obey $\left<H_\lambda (t)\right>_{t\in [t_\text{stab}, ~ t_\text{stab} + \Delta T_2]} < H_{\rm stab}$, with $\Delta t_2=2000$, so as to ensure that the final state is on average below $H_{\rm stab}$.
The corresponding times in case (a) and (b) are marked respectively by a red and a blue circle in figure \ref{fig:stab}e.

Repeating this experiment over multiple realisations of the initial pattern (i.e.\ different noise realisations) yields an ensemble-averaged $\bar{t}_\text{stab}$. This procedure estimates the time for an initially regular and dominant wavelength to disappear from the flow domain, regardless of the way in which it does so and of the final state approached.
Figure~\ref{fig:stab}f presents the dependence of $\overline{t}_\text{stab}$ on $\lambda$ for different values of $Re$.
\resub{Although our procedure relies on highly fluctuating signals (like those presented on figure \ref{fig:stab}e) and on a number of arbitrary choices ($H_{\rm stab}$, 
 $\epsilon_{\lambda}$, etc.) that alter the exact values of stability times, we find that the trends visualised in figure \ref{fig:stab}f are robust.
 (The sensitivity of $\overline{t}_\text{stab}$ with $H_{\rm stab}$ is shown in figure \ref{fig:stab_Hstab} of Appendix \ref{app:wavelet}.)
 
 A most-stable wavelength ranging between 40 and 44 dominates the stability times for all the values of $Re$ under study.
 This is similar to the results from the \emph{existence} study on figure \ref{fig:lifetimes_pattern}, which showed a preferred wavelength emerging from the uniform state at around $42$ at $Re=440$.
 } 
%We note that a most-stable wavelength emerges from the uniform state, at around $\lambda\simeq 40$ at $Re=440$, similarly to the results from the \emph{existence} study on figure \ref{fig:lifetimes_pattern}, which showed a preferred wavelength of around $42$ at $Re=440$.
Consistently with 
what was observed in Minimal Band Units of different sizes, the most stable wavelength grows with decreasing $Re$. 
%Finite lifetime even at the preferred wavenumber??
% finite lifetimes due to fluctuation between possible wavenumbers 

\subsection{Discussion}

\begin{figure} 
    \centering
 \includegraphics[width=0.6\columnwidth]{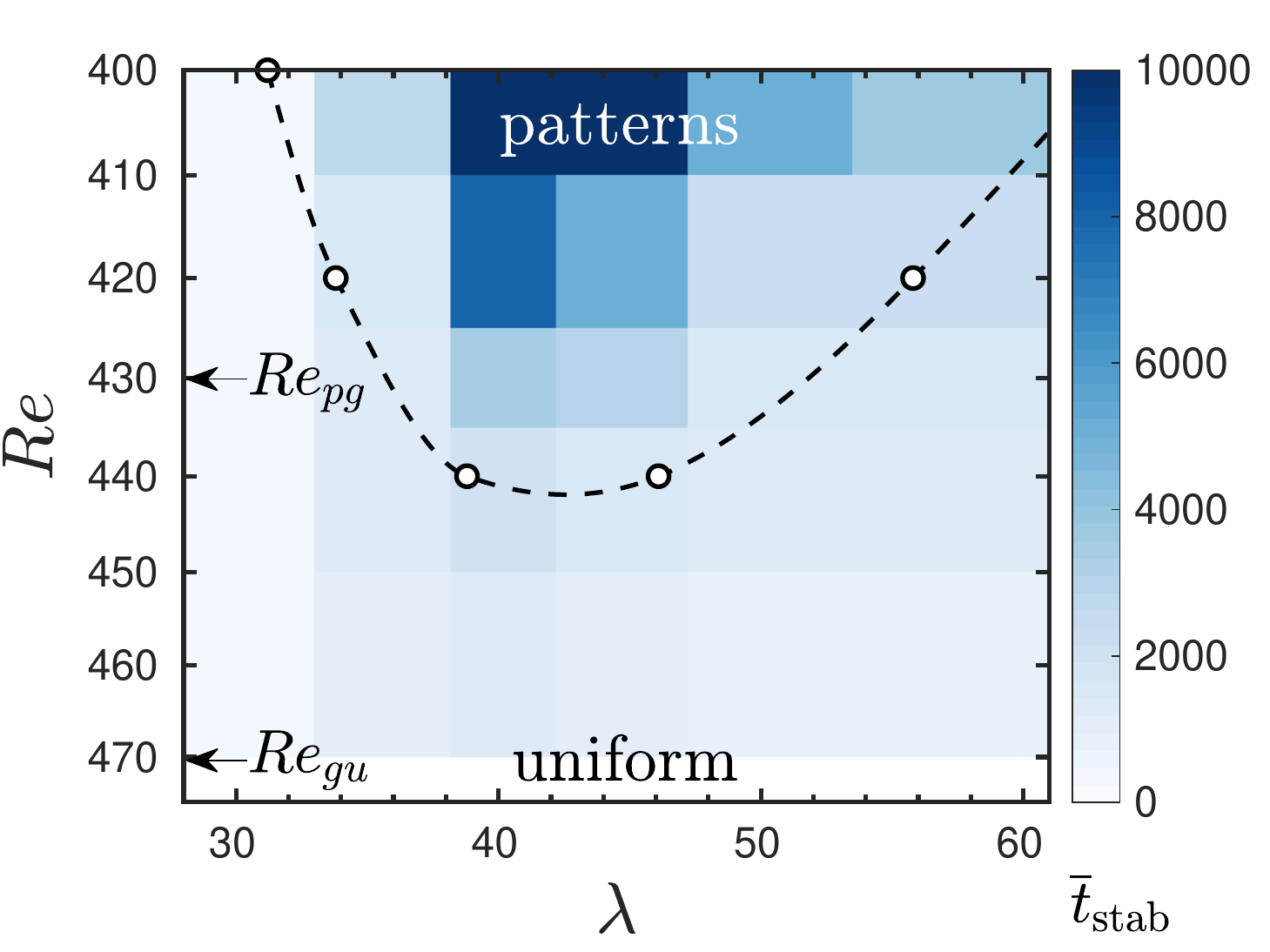} 
    \caption{Visualisation of the pattern selection in the phase space $(\lambda, Re)$.
    \resub{Colours show the stability times $\overline{t}_\text{stab}$, while
    open circles are points $\gamma_{\rm pat}(\lambda, Re)=0.45$. The dashed line is an illustrative fit of these points.} 
    }
    \label{fig:Eckhaus}
\end{figure}

Our study of the \emph{existence} and \emph{stability} of large-scale modulations of the turbulent flow is summarised in figure \ref{fig:Eckhaus}. This figure resembles the existence and stability diagrams presented for usual (non-turbulent) hydrodynamic instabilities such as Rayleigh-B\'enard convection and Taylor-Couette flow \citep[]{busse1981transition, ahlers1986wavenumber, cross2009pattern}. 
In classic systems, instabilities appear with increasing control parameter, while here gaps and bands emerge from uniform turbulent flow as $Re$ is lowered. Therefore, we plot the vertical axis in figure \ref{fig:Eckhaus} with decreasing upwards Reynolds.

%However, since quasi-laminar gaps and bands emerge from uniform turbulent flow as $Re$ is lowered, the vertical axis in figure \ref{fig:Eckhaus} corresponds to decreasing Reynolds number.

We recall that the existence study of $\S$\ref{sec:existence} culminated in the measurement of $\gamma_{\rm pat}(\lambda, Re)$, the fraction of simulation time that is spent in a patterned state, plotted in figure \ref{fig:lifetimes_pattern}.
The parameter values at which $\gamma_{\rm pat}(\lambda, Re) = 0.45$ (an arbitrary threshold that covers most of our data range) are shown as black circles in figure \ref{fig:Eckhaus}. The dashed curve is an interpolation of the iso-$\gamma_{\rm pat}$ points and separates two regions, with patterns more likely to exist above the curve than below. 
The minimum of this curve is estimated to be $\lambda\simeq42$. 
This is a preferred wavelength at which patterns first statistically emerge as $Re$ is decreased from large values. 

The final result of the stability study in section \S\ref{sec:stability}, shown in figure \ref{fig:stab}f, was $\overline{t}_\text{stab} (Re,\lambda)$, 
a typical duration over which a pattern initialised with wavelength $\lambda$ would persist. The colours in figure \ref{fig:Eckhaus} show $\overline{t}_\text{stab}$.
\resub{The peak in $\overline{t}_\text{stab}$ is first discernible at $Re \simeq 440$ and occurs at $\lambda \simeq 40$.}
The pattern existence and stability zones are similar in shape and in their lack of symmetry with respect to line $\lambda=42$. The transition seen in figures \ref{fig:stab}a and 
\ref{fig:stab}c from $\lambda=57$ to $\lambda=44$ at $Re=400$ corresponds to motion from a light blue to a dark blue area in the top row of figure \ref{fig:Eckhaus}. 
This change in pattern wavelength resembles 
the Eckhaus instability which, in classic hydrodynamics, leads to transitions from unstable wavelengths outside a stability band to stable wavelengths inside. 

%transitions from unstable to stable wavelengths. 

%In this section, we
%adopted a pattern-formation framework to 
%evaluated the presence of a most-sustained wavelength in the patterned regime. 

The presence of a most-probable wavelength confirms the initial results of \citet[]{prigent2003long} and those of \citet[]{rolland2011ginzburg}. This is also consistent with the instability study of \citet[]{kashyap2022linear} in plane Poiseuille flow.
However, contrary to classic pattern-forming instabilities, 
the turbulent-laminar pattern does not emerge from an exactly uniform state, but instead from a state in which local gaps are intermittent, as established in
 Section \ref{sec:intermittency}.
%Nucleation and patterning are therefore not contradictory pictures.
In Section \ref{sec:spec_optim}, we will emphasise the importance of the mean flow in the wavelength selection that we have described.

\section{Optimisation of the large-scale flow}
%\section{Optimisation of the mean flow along  }
\label{sec:spec_optim}

This section is devoted to the dependence of various energetic features of the patterned flow on the domain length $L_z$ of a Minimal Band Unit. We fix the Reynolds number at $Re=400$. In the existence study of \S \ref{sec:exist_stab}, the wavelength $\lambda \simeq 44$ was found to be selected by patterns. (Recall the uppermost curves corresponding to $Re=400$ in figure \ref{fig:lifetimes_pattern}.) We will show that this wavelength also extremises quantities in the energy balances of the flow.

\subsection{Average energies in the patterned state}

\begin{figure}
    \centering
\subfloat[]{ \includegraphics[width=0.5\columnwidth]{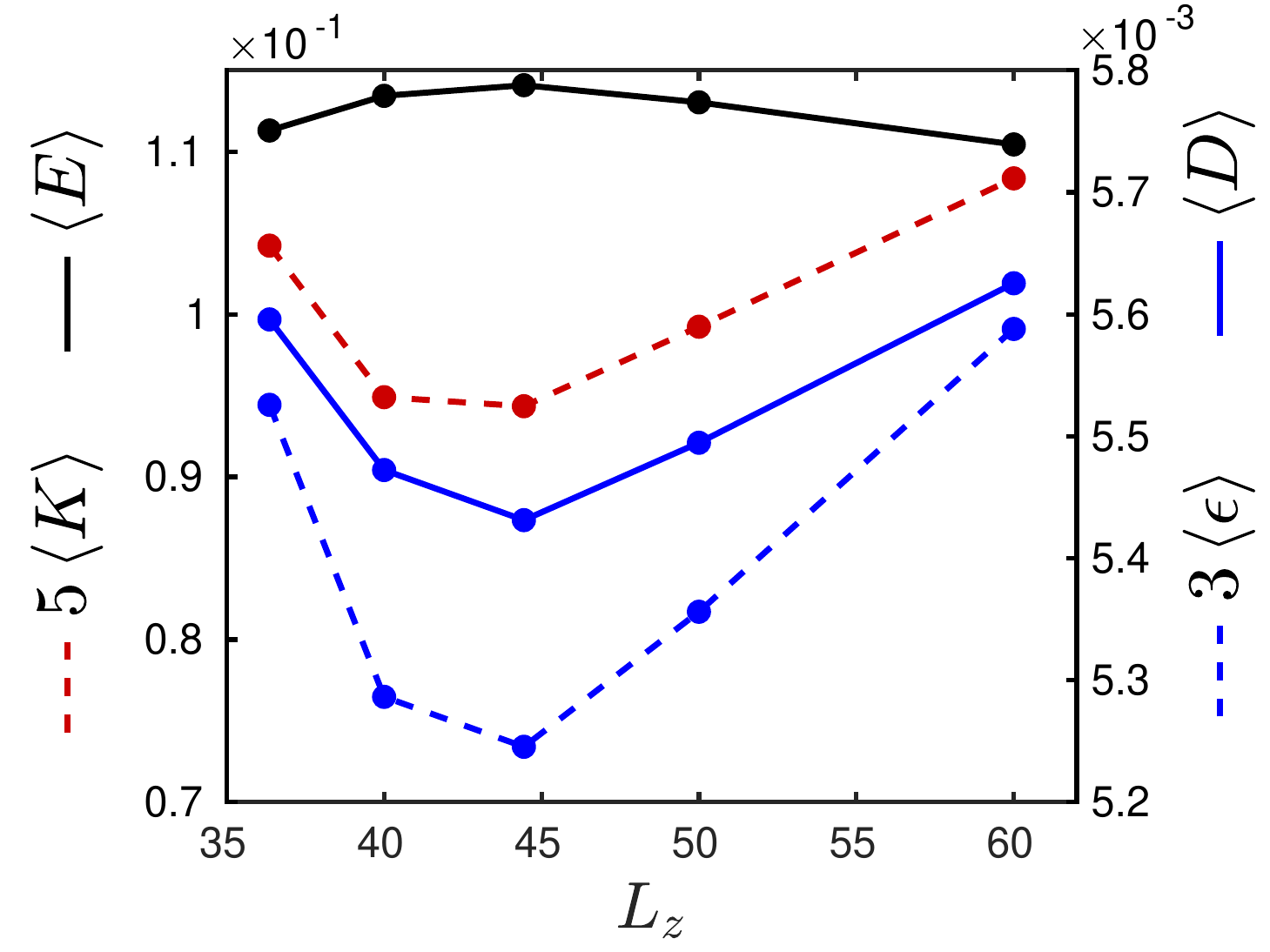}     \label{fig:E_Lz}}~
\subfloat[]{ \includegraphics[width=0.5\columnwidth]{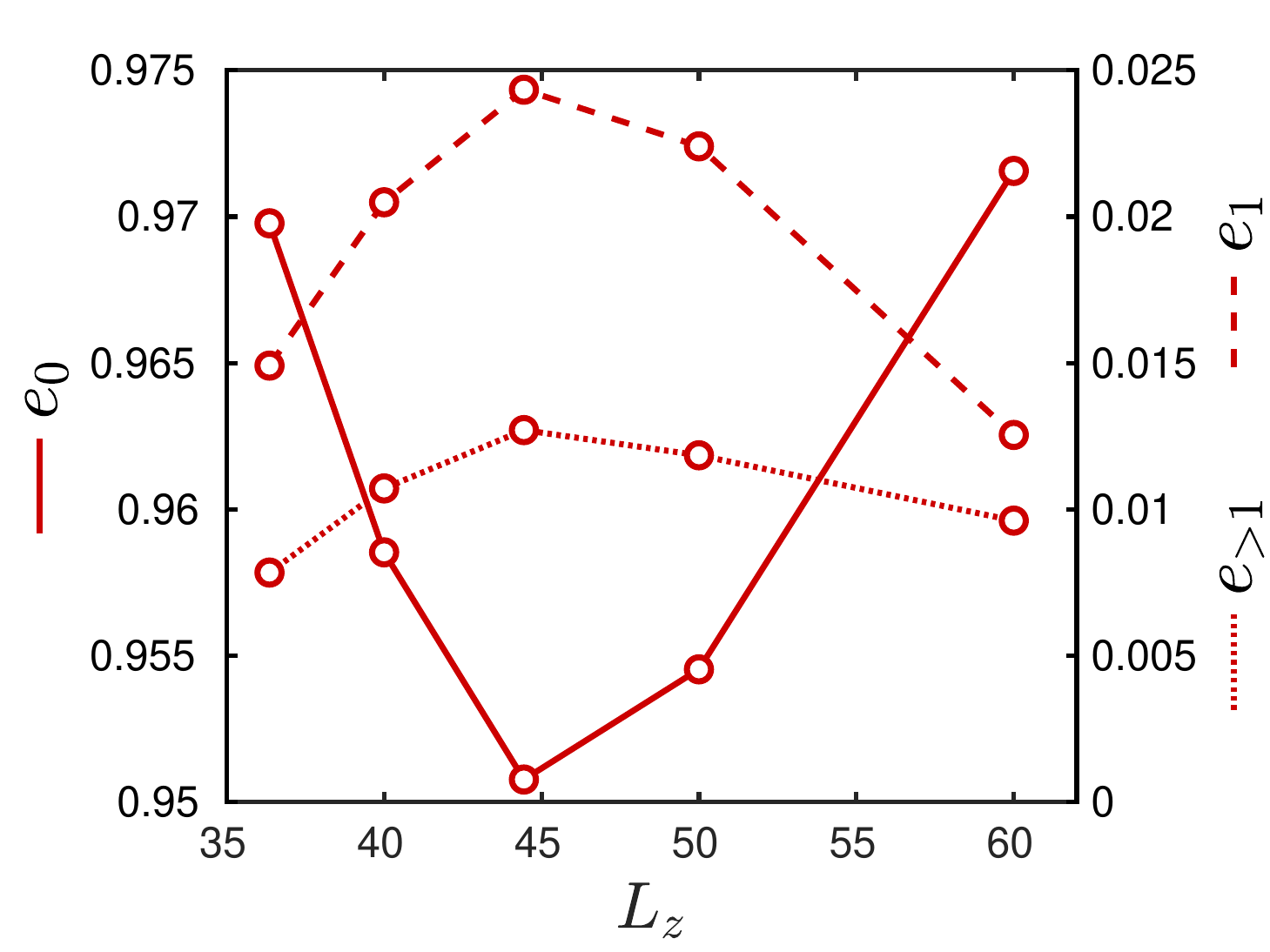}    \label{fig:trigo_Lz}} 
\caption{
Energy analysis for the patterned state at $Re=400$ as a function of the size $L_z$ of a Minimal Band Unit. (a) Spatially-averaged total energy $\left<E\right>$, mean TKE $\left<K\right>$ ($\times 5$), mean total dissipation $\left<D\right>$, mean turbulent dissipation $\left<\epsilon\right>$ ($\times 3$), for the patterned state at $Re=400$ as a function of $L_z$. 
(b) Energy in each of the $z$-Fourier components of the mean flow (equations \eqref{eq:trig} and \eqref{eq:trig_ratio}).}
\end{figure}

We first decompose the flow into a mean and fluctuations, $\utot = \ubar + \uprime$, where the mean (overbar) is taken over the statistically homogeneous directions $x$ and $t$. We compute energies of the total flow $\left<E\right> \equiv \left<\utot \cdot \utot\right>/2$ and of the fluctuations (turbulent kinetic energy) $\left<K\right> \equiv \left< \uprime \cdot \uprime \right>/2$, where $\left<\cdot\right>$ is the $(x,y,z,t)$ average.
Figure \ref{fig:E_Lz} shows these quantities as a function of 
$L_z$ for the patterned state at $Re=400$. At $L_z=44$, $\left<E\right>$ is maximal and $\left<K\right>$ is minimal. As a consequence, the mean-flow energy $\frac{1}{2} \left<\ubar \cdot \ubar \right>= \left<E\right> - \left<K\right>$ is also maximal at $L_z=44$. 
Figure \ref{fig:E_Lz} additionally shows average dissipation of the total flow $\left<D\right> \equiv \left< |\nabla \times \utot |^2\right>/Re$ and average dissipation of turbulent kinetic energy $\left<\epsilon\right>  \equiv \left< |\nabla \times \uprime |^2\right>/Re$, both of which are minimal at $L_z=44$. \resub{Note that these total energy and dissipation terms change very weakly with $L_z$, with a variation of less than $6\%$.}

%Hence there is a maximisation of the energy in the mean flow, and a corresponding minimization of energy in the fluctuations, at the wavelength observed to be selected in figure \ref{fig:lifetimes_pattern}, upper-most curves corresponding the $Re=400$. 

%The differences observed in both spectra and mean flow when $L_z$ is modified are quantified in figure \ref{fig:E_Lz}, where the evolution of various space-averaged energies is presented for different $L_z$. The total energy $\left<E\right>= \left<\utot \cdot \utot\right>/2$ and the turbulent kinetic energy $\left<K\right> = \left< \uprime \cdot \uprime \right>/2$ are computed for the patterned state at $Re=400$. (Here $\left<\cdot\right>$ is the $(x,y,z)$ average.)
%
%At $L_z=44$, $\left<E\right>$ is maximal, whereas $\left<K\right>$ is minimal. As a consequence, the mean-flow energy $\frac{1}{2} \left<\ubar \cdot \ubar \right>= \left<E\right> - \left<K\right>$ is maximal at  $L_z=44$.
%
%Space-averaged total dissipation $\left<D\right> \equiv \nu\left< |\nabla \times \utot |^2\right>$ (blue solid line) and kinetic dissipation $\left<\epsilon\right>  \equiv \left< |\nabla \times \uprime |^2\right>/Re$ (blue dashed line) are also shown and behave similarly to $\left<K\right>$. We recall from Section \ref{sec:existence} that at $Re=400$, $L_z=44$ is the most favourable wavelength (red curve of figure \ref{fig:lifetimes_pattern}), and this is correlated to a maximisation of the energy of the mean flow, and a minimisation of $\left<K\right>$ and dissipation. 

The mean flow is further analysed by computing the energy of each spectral component of the mean flow. For this, the $x$,  $t$ averaged flow $\ubar$ is decomposed into Fourier modes in $z$:
\begin{equation}
    \label{eq:trig}
    \ubar(y, z) = \ubar_0(y) + 2\mathcal{R} \left(\ubar_1(y) e^{i 2\pi z/L_z } \right) + \ubar_{>1} (y,z)
\end{equation}
where $\ubar_0$ is the uniform component of the mean flow,
$\ubar_1$ is the trigonometric Fourier coefficient corresponding to $k_z= 2\pi/L_z$ and $\ubar_{>1}$ is the remainder of the decomposition, for $k_z> 2\pi/L_z$. (We have omitted the hats on the $z$ Fourier components of $\ubar$.)
The energies of the spectral components relative to the total mean energy are 
\begin{equation}
    \label{eq:trig_ratio}
    e_0 = \frac{\left<\ubar_0\cdot\ubar_0\right>}{\left<\ubar\cdot\ubar\right>}, ~~~ e_1 =  \frac{\left<\ubar_1\cdot\ubar_1\right>}{\left<\ubar\cdot\ubar\right>}, ~~~ e_{>1} =  \frac{\left<\ubar_{>1}\cdot\ubar_{>1}\right>}{\left<\ubar\cdot\ubar\right>}
\end{equation}
%
%YD: it would be good to decompose <\ubar \cdot \ubar> into e_0, e_1 and e_>1 first 
%
These are presented in figure~\ref{fig:trigo_Lz}.
% at $Re=400$ as a function of $L_z$.
It can be seen that $e_0 \gg e_1 > e_{>1}$ and also that all %of these quantities 
have an extremum at $L_z=44$.
In particular, $L_z=44$ minimizes $e_0$ ($e_0=0.95$) while maximising the trigonometric component ($e_1=0.025$) along with the remaining components ($e_{>1} \simeq 0.011$).  
Note that for a banded state at $Re=350$, $L_z=40$, \citet{barkley2007mean} found that $e_0\simeq0.70$, $e_1\simeq0.30$ and $e_{>1}\simeq 0.004$, consistent with a strengthening of the bands as $Re$ is decreased. 

\subsection{Mean flow spectral balance}

%We now investigate how the various contributions to the spectral energetic content behave as a function of domain size. The equation for the time-averaged energy balance of the mean flow is written for its two main components $\ubar_0$ and $\ubar_1$ as \citep[Part 1]{gome1}:

We now investigate the spectral contributions to the budget of the mean flow $\ubar$, dominated by the mean flow's two main spectral components $\ubar_0$ and $\ubar_1$. The balances can be expressed as in \citet[Part 1]{gome1}:
\begin{align}
    \Advms_0 -\Prodms_0 - \Dissipms_0 + I = 0 \text{ for } \ubar_0  ~~~ \text{and} ~~~    \Advms_1 - \Prodms_1 - \Dissipms_1 = 0 \text{ for } \ubar_1  
    \label{eq:mf_bal}
\end{align}
where $I$ is the rate of energy injection by the viscous shear, and $\Prodms_0$, $\Dissipms_0$ and $\Advms_0$ stand for, respectively, production, dissipation and advection 
(i.e.\ non-linear interaction) contributions to the energy balance of mode $\ubar_0$ and similarly for $\ubar_1$. These are defined by
\begin{subequations}\begin{align}
    I &= \frac{2}{Re} \mathcal{R}\left\{ \int_{-1}^1 \frac{\partial}{\partial y} \left(\ujmhatc (k_z=0) \widehat{\overline{s}}_{yj} (k_z=0) \right) \, \text{d} y \right\}
    = \frac{1}{Re}\left( \frac{\partial \overline{u}_{\strm}}{\partial y}\bigg \rvert_{1} + \frac{\partial \overline{u}_{\strm}}{\partial y}\bigg \rvert_{-1}  \right) \label{eq:I}
    \\
    \Prodms_0 &= \mathcal{R}\left\{\int_{-1}^{1}  \frac{\partial \ujmhatc}{\partial x_i} (k_z=0) ~\widehat{\overline{u_i^\prime u_j^\prime}} (k_z=0) ~\text{d} y \right\}  \label{eq:Pi0}\\
    \Dissipms_0 &= \frac{2}{Re} \mathcal{R}\left\{\int_{-1}^{1}  \widehat{\overline{s}}_{ij} (k_z=0) ~ \widehat{\overline{s}}_{ij}^* (k_z=0) ~\text{d} y \right\} \\ 
     \Advms_0 &= - \mathcal{R}\left\{\int_{-1}^{1}\ujmhatc (k_z=0) ~ \widehat{ \overline{u}_i \frac{\partial \overline{u}_j}{\partial x_i}}(k_z=0)  ~\text{d} y \right\}
    \label{eq:A0}
\end{align}\end{subequations}
where $\mathcal{R}$ denotes the real part. We define $\Prodms_1$, $\Dissipms_1$ and $\Advms_1$ similarly by replacing $k_z=0$ by $k_z=2\pi/L_z$ in \eqref{eq:I}-\eqref{eq:A0}.

We recall two main results from \citet[Part 1]{gome1}: first, $\Advms_1 \simeq - \Advms_0$. This term represents the energetic transfer between modes $\ubar_0$ and $\ubar_1$ via the self-advection of the mean flow (the energetic spectral influx from $(\ubar \cdot \nabla ) \ubar$). 
Second, $\Prodms_1 <0$, and this term approximately balances the negative part of TKE production. This is an energy transfer from turbulent fluctuations to the component $\ubar_1$ of the mean flow. 
%On average, the patterned state absorbs energy from fluctuations.

\begin{figure}
    \centering
\subfloat[]{ \includegraphics[width=0.5\columnwidth]{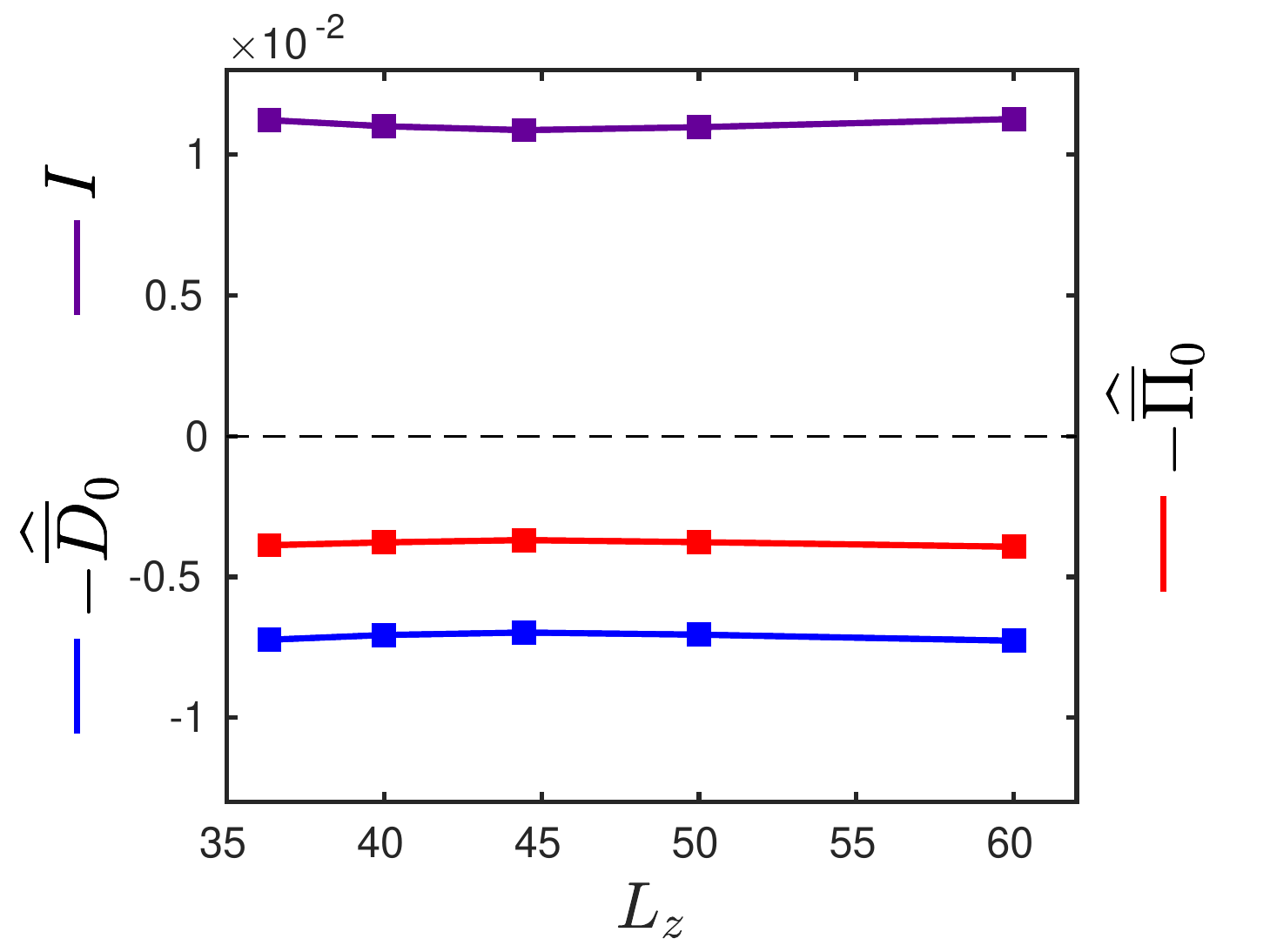} \label{fig:mean_spec_Lz0}} 
~
\subfloat[]{ \includegraphics[width=0.5\columnwidth]{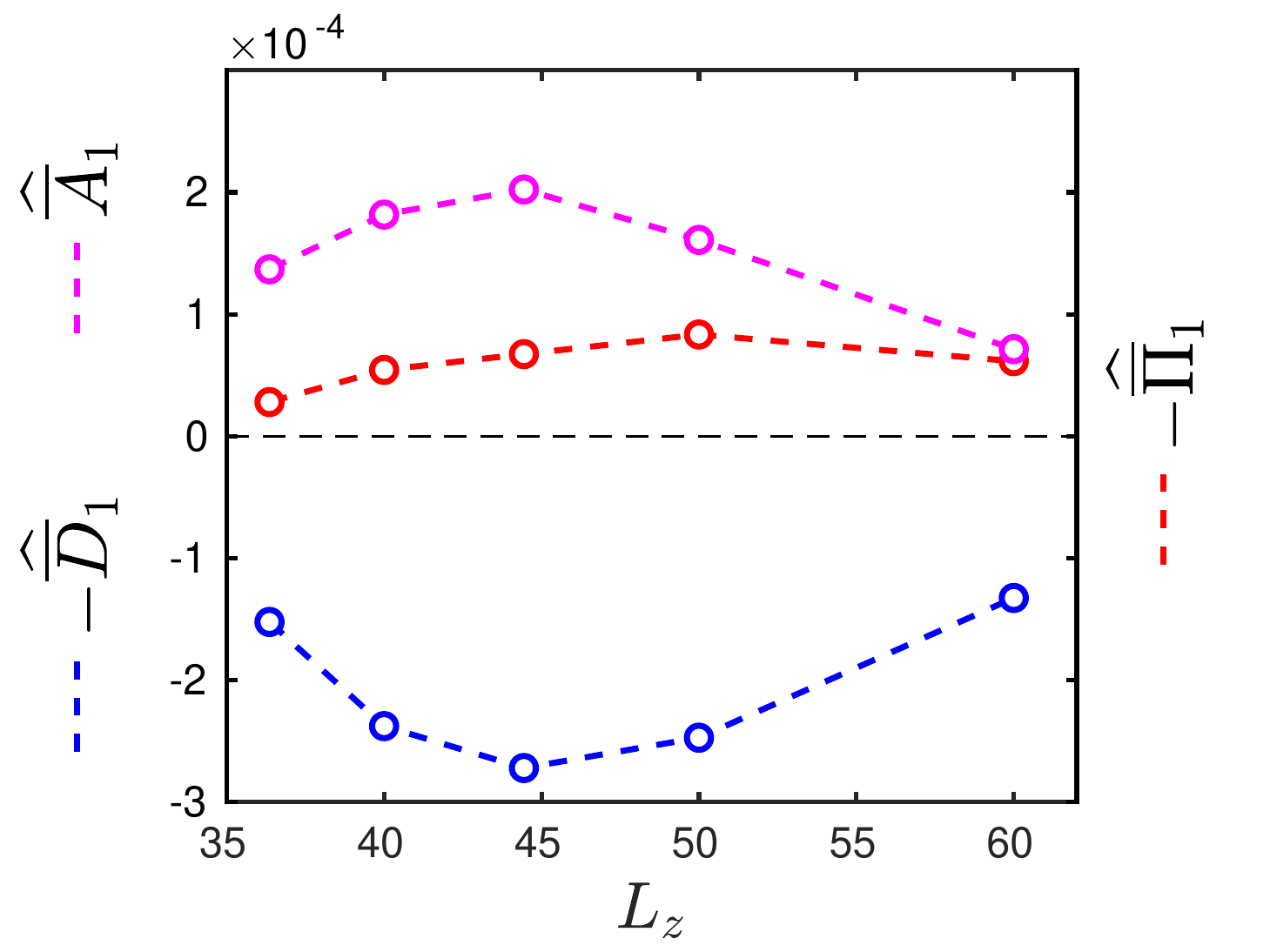}     \label{fig:mean_spec_Lz1}}    \caption{Spectral energy balance of the mean flow components (a) $\ubar_0$ (uniform component) and (b) $\ubar_1$ (large-scale flow along the laminar-turbulent interface). See equation
\eqref{eq:mf_bal}. \resub{Advection and dissipation of the large-scale flow, $\Advms_1$ and $\Dissipms_1$, show the strongest variations with $L_z$ and are optimal at the preferred wavelength $L_z\simeq 44$.}
}
\end{figure}

Each term contributing to the balance of $\ubar_0$ and $\ubar_1$ is shown as a function of $L_z$ in figures \ref{fig:mean_spec_Lz0} and \ref{fig:mean_spec_Lz1}. 
We do not show $\Advms_0$ because $\Advms_0\simeq -\Advms_1$.

We obtain the following results:
\begin{enumerate}[leftmargin=*,labelindent=8mm,labelsep=3mm, itemindent=0mm, label=(\arabic*)]
    \item Production $\Prodms_0$, dissipation $\Dissipms_0$ and energy injection $I$ are nearly independent of $L_z$, varing by no more than 6\% over the range shown. These $k_z=0$ quantities correspond to uniform fields in $z$ and hence it is unsurprising that they depend very little on $L_z$.
    \item 
    The non-linear term $\Advms_1 \simeq - \Advms_0$, i.e.\ the transfer from $\ubar_0$ to $\ubar_1$ which is the principal source of energy of $\ubar_1$, \resub{varies strongly with $L_z$ and} has a maximum at $L_z\simeq 44$. 
    %Dissipation $\Dissipms_1$ has a minimum, at $L_z\simeq 44$. 
    %The non-linear term $\Advms_0 \simeq - \Advms_1$ has a maximum, and dissipation $\Dissipms_1$ a minimum, at $L_z\simeq 44$. 
    %In contrast, the non-linear term $\Advms_0 \simeq - \Advms_1$, displayed in figure \ref{fig:trigo_Lz}, is maximised by $L_z\simeq 44$.
    This is the reason for which $\ubar_0$ is minimised by $L_z\simeq 44$ (see figure \ref{fig:trigo_Lz}): more energy is transferred from $\ubar_0$ to $\ubar_1$.
    \item 
    %In the rest of the balance of $\ubar_1$, dissipation $\Dissipms_1$ is maximised at $L_z\simeq44$, whereas 
     Production $\Prodms_1$ increases with $L_z$ and does not show an extremum at $L_z\simeq 44$ (it instead has a weak maximum at $L_z\simeq50$). In all cases, $\Prodms_1< \Advms_1$: the TKE feedback on the mean flow, although present, is not dominant and not selective.
     
     \item
     Dissipation $\Dissipms_1$ accounts for the remaining budget and its extremum at $L_z\simeq 44$ corresponds to maximal dissipation.
\end{enumerate}

The turbulent kinetic energy balance is also 
modified with changing $L_z$. This is presented in Appendix \ref{app:tke_bal}. The impact of TKE is however
secondary,
%because the feedback on the mean flow, via $\Prodms_1$, is not the leading term that fuels $\ubar_1$, and does not participate in maximising the energy of $\ubar_1$ at the preferred wavelength.
\resub{because of the results established in item (3).}

\section{Conclusion and discussion}

We have 
explored the appearance of patterns from uniform turbulence in plane Couette flow at $Re\leq500$. 
We used numerical domains of different sizes to quantify the competition between 
featureless (or uniform) turbulence and (quasi-) laminar gaps. In Minimal Band Units, intermittency reduces to a random alternation between two states: uniform or patterned.
In large slender domains, however, gaps nucleate randomly and locally in space, and the transition to patterns takes place continuously via the regimes presented in Section \ref{sec:intermittency}: the uniform regime in which gaps are rare and short-lived
(above $Re \simeq470$), and another regime ($Re<470$) in which gaps are more numerous and long-lived.
Below $Re\simeq430$, the large-scale spacing of these gaps starts to dominate the energy spectrum, which is a possible demarcation of the patterned regime.
%In this latter regime, 
With further decrease in 
$Re$, gaps eventually fill the entire flow domain, forming regular patterns.
The distinction between these regimes is observed in both gap lifetime and friction factor.

Spatially isolated gaps were already observed by \citet[]{prigent2003long}, \citet[]{barkley2005computational} and \citet[]{rolland2011ginzburg}. (See also \citet[]{manneville2015transition, manneville2017laminar} and references therein.) 
Our results confirm that pattern emergence, mediated by randomly-nucleated gaps, is 
necessarily more complex than the supercritical Ginzburg-Landau framework initially proposed by \citet[]{prigent2003long} and later developed by \citet[]{rolland2011ginzburg}. 
However, this does not preclude linear processes in the appearance of patterns, such as those reported by \citet[]{kashyap2022linear} from an ensemble-averaged linear response analysis.
%SG: review this sentence!

The intermittency between uniform turbulence and gaps that we quantify here in the range $380\lesssim Re\lesssim 500$ is not comparable to that between laminar flow and bands present for $325 \lesssim Re \lesssim 340$. 
The latter is a continuous phase transition in which the laminar flow is absorbing: laminar regions cannot spontaneously develop into turbulence and can only become turbulent by contamination from neighbouring turbulent flow. 
%
%when both turbulent and laminar zones coexist, no perturbation within the laminar region can develop into a turbulent structure. 
%
This is connected to the existence of a critical point at which the correlation length diverges with a power-law scaling with $Re$, as characterised by important past studies \citep{shi,chantry_universal,lemoult2016directed} which demonstrated a connection to directed percolation.
The emergence of gaps from uniform turbulence is of a different nature. Neither uniform turbulence nor gaps are absorbing states, since gaps can always appear spontaneously and can also disappear, returning the flow locally to a turbulent state. 
While the lifetimes of quasi-laminar gaps do exhibit an abrupt change in behaviour at $Re = 470$ (figure~\ref{fig:taug}), we observe no evidence of critical phenomena associated with the emergence of gaps from uniform turbulence. Hence, the change in behaviour appears to be in fact smooth. 
This is also true in pipe flow where quasi-laminar gaps form, but not patterns \citep{avila2013nature, frishman2022dynamical}.

%Moreover, we find no trace of a critical point, at which lifetimes (or any other measurement) 
%Not any other measurment ! order param does not diverge with system size
%would diverge with system size. %Critical point implies 
%The presence or not of a critical point associated to the transition from uniform to patterned state is only speculative at this stage

We used the pattern wavelength as a control parameter, via either the domain size or the initial condition, to investigate the existence
of a preferred pattern wavelength.
We propose that the finite spacing between gaps, visible in both local gaps and patterned regimes, is selected by the preferred size
of their associated large-scale flow.
Once gaps are sufficiently numerous and patterns are established, their average wavelength increases 
with decreasing $Re$, with changes in wavelength in a similar vein to the Eckhaus picture. 

The influence of the large-scale flow in wavelength selection is analysed in  Section \ref{sec:spec_optim}, where we carried out a spectral analysis like that in \citet[Part 1]{gome1} for various sizes of the Minimal Band Unit.
In particular, we investigated the roles of the turbulent fluctuations and of the mean flow, which is in turn decomposed into its uniform component $\ubar_0$ and its trigonometric component $\ubar_1$, 
associated to
%which represents 
the large-scale flow along the laminar-turbulent interface.
Our results demonstrate
a maximisation of the energy and dissipation of $\ubar_1$ by the wavelength naturally preferred by the flow, and this is primarily associated to an optimisation of the advective term $(\ubar \cdot \nabla) \ubar$ in the mean flow equation.
This term redistributes energy between modes $\ubar_0$ and $\ubar_1$ %(via $\Advms_0$) 
and is mostly responsible for energising the large-scale along-band flow. 
%As a result of this improved influx of energy, the mean flow is stronger at the preferred wavelength, contrary to turbulent kinetic energy.
Turbulent fluctuations are of secondary importance in driving the large-scale flow and do not play a significant role in the wavelength selection.
\resub{Our results of maximal transport of momentum and dissipation of the large-scale flow are therefore
analogous to the principles mentioned by \cite{malkus1956outline} and \citet{busse1981transition}.
Explaining this observation from 
%a guiding 
first principles
remains a  
prodigious
task.}

It is essential to understand the creation of the large-scale flow around a randomly emerging laminar hole. 
The statistics obtained in our tilted configuration 
%must be 
should be 
extended to 
large streamwise-spanwise domains, where short-lived and randomly-nucleated holes might align in the streamwise direction \citep[Fig. 5]{manneville2017laminar}. 
%presumably before the regime of 
\resub{This presumably occurs at $Re$ above the 
long-lived-gap regime, in which the two gap orientations $\pm \theta$ compete}. 
\resub{
%LST
%The selection of the angle of the laminar-turbulent interface might also be related to a most-dissipative large-scale flow, 
The selected pattern angle might also maximise the dissipation of the large-scale flow,
similarly to what we found for the preferred wavelength.}
Furthermore, a more complete dynamical picture of gap creation is needed. 
The excitable model of \citet[]{barkley2011a} might provide a proper framework, as it accounts for both the emergence of anti-puffs \citep[]{frishman2022dynamical} and of periodic solutions \citep[]{barkley2011b}. Connecting this model to the Navier-Stokes equations is, however, a formidable challenge. 
\resub{Our work emphasises the necessity of including the effects of the advective large-scale flow \citep{barkley2007mean, duguet2013oblique, klotz2021experimental, marensi2022dynamics}
to adapt this model 
to the establishment of the turbulent-laminar patterns of both preferred wavelength and angle observed in planar shear flows.}

\section*{Acknowledgements}

The calculations for this work were performed using high performance computing resources
provided by the Grand Equipement National de Calcul Intensif at the Institut du D\'eveloppement
et des Ressources en Informatique Scientifique (IDRIS, CNRS) through Grant No. A0102A01119.
This work was supported by a grant from the Simons Foundation (Grant No. 662985).
The authors wish to thank Anna Frishman, Tobias Grafke and Yohann Duguet for fruitful discussions, as well as the referees for their useful suggestions.

\section*{Declaration of Interests}
The authors report no conflict of interest.

\appendix

\section{Laminar and turbulent distributions in pipe vs Couette flows.}
\label{app:comp_pipe}

\begin{figure}[h!]
    \centering
\subfloat[]{ \includegraphics[width=0.5\columnwidth]{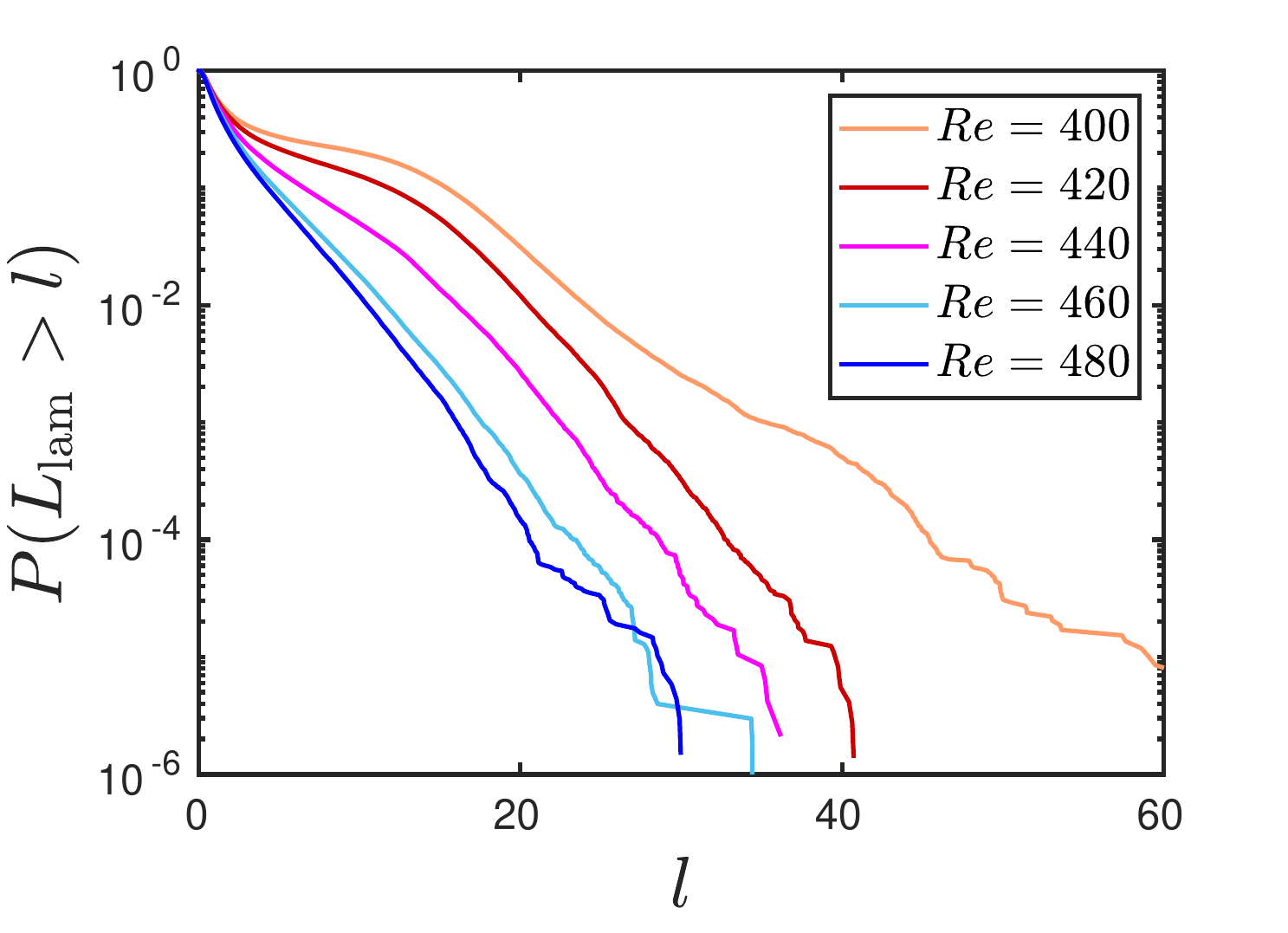} \label{fig:cdf_Llam}} 
~
\subfloat[]{ \includegraphics[width=0.5\columnwidth]{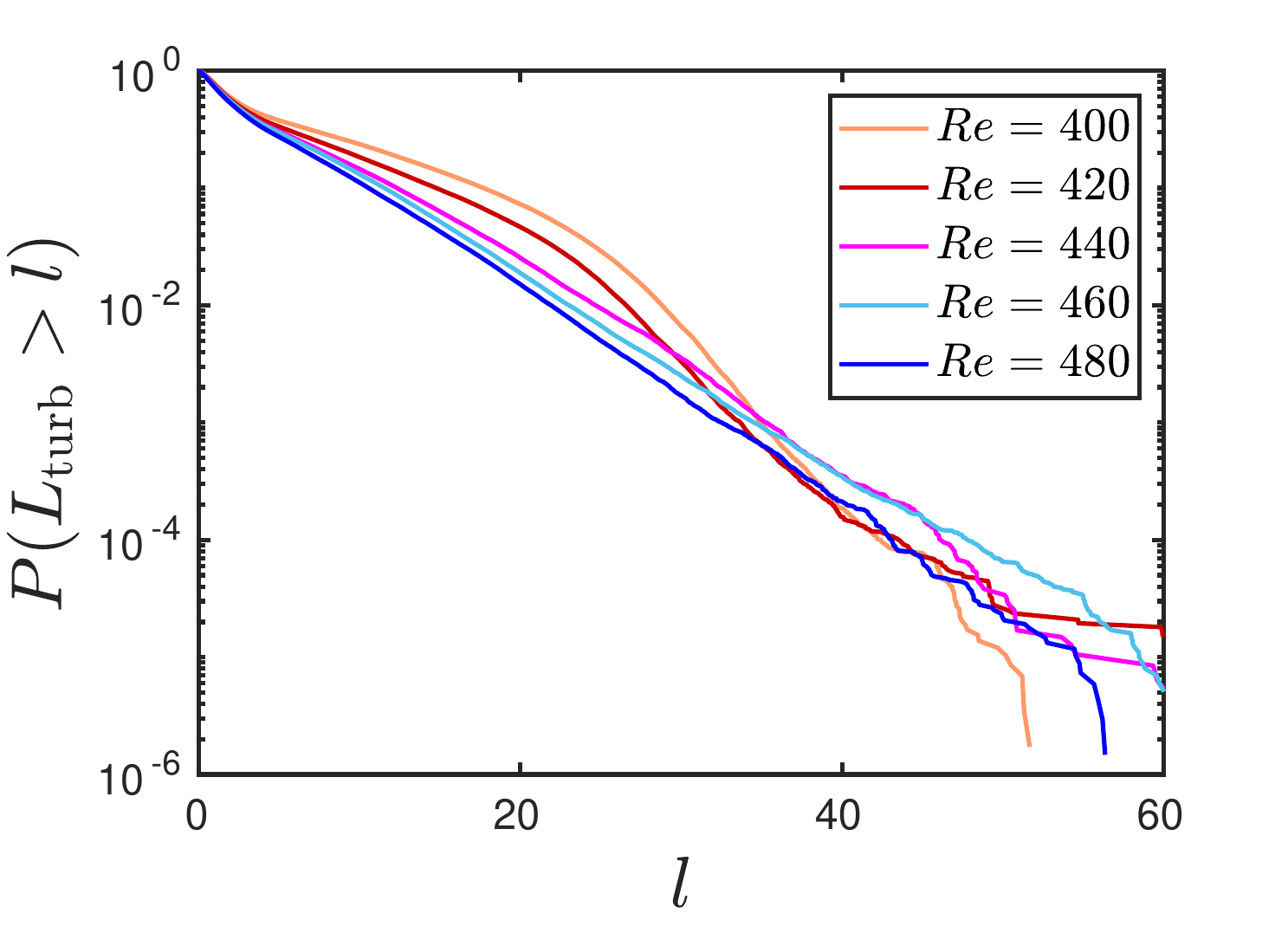}     \label{fig:cdf_Lturb}}  
\caption{Cumulative distribution of (a) laminar gaps and (b) turbulent zones, for various $Re$.}
\end{figure}

From figures \ref{fig:pdf_llam} and \ref{fig:pdf_lturb} of the main text, both distributions of laminar or turbulent lengths, $\Llam$ and $\Lturb$, are exponential for large enough lengths, similarly to pipe \citep[]{avila2013nature}. It is however striking that the distributions of $\Llam$ and $\Lturb$ have different shapes for $\Llam$ or $\Lturb>10$ in plane Couette flow: $\Llam$ shows a sharper distribution, whereas $\Lturb$ is more broadly distributed. We present on figures \ref{fig:cdf_Llam} and \ref{fig:cdf_Lturb} the cumulative distributions of $\Llam$ and $\Lturb$ for a complementary analysis.

We focus on the characteristic length $l_{\rm turb}^*$ or $l_{\rm lam}^*$ for which $P(\Llam>l_{\rm lam}^*) = P(\Lturb>l_{\rm turb}^*) = 10^{-2}$: for example, $l_{\rm lam}^*=15.5$ and $l_{\rm turb}^*=26.5$ at $Re=440$; $l_{\rm lam}^*=23.4$ and $l_{\rm turb}^*=30.3$ at $Re=400$.
We see that  $l_{\rm turb}^*$ and $l_{\rm lam}^*$ are of the same order of magnitude.
This differs from the same measurement in pipe flow, carried out by \citet[Fig. 2]{avila2013nature}: $l_{\rm lam}^*=6$ and $l_{\rm turb}^* \simeq 50$ at $Re=2800$; $l_{\rm lam}^*\simeq17$ and $l_{\rm turb}^* \simeq160$ at $Re=2500$ (as extracted from their figure 2.).
\resub{This confirms that turbulent and laminar spacings are of the same order of magnitude in plane Couette flow, contrary to pipe flow.}
% This contrasts with our measurement in plane Couette flow, where the turbulent and laminar spacings are of the same order of magnitude.

\section{Wavelet transform}
\label{app:wavelet}

\begin{figure}
    \centering
\subfloat[]{ %\includegraphics[width=0.5\columnwidth]
\includegraphics[width=0.455\columnwidth,height=5cm]
{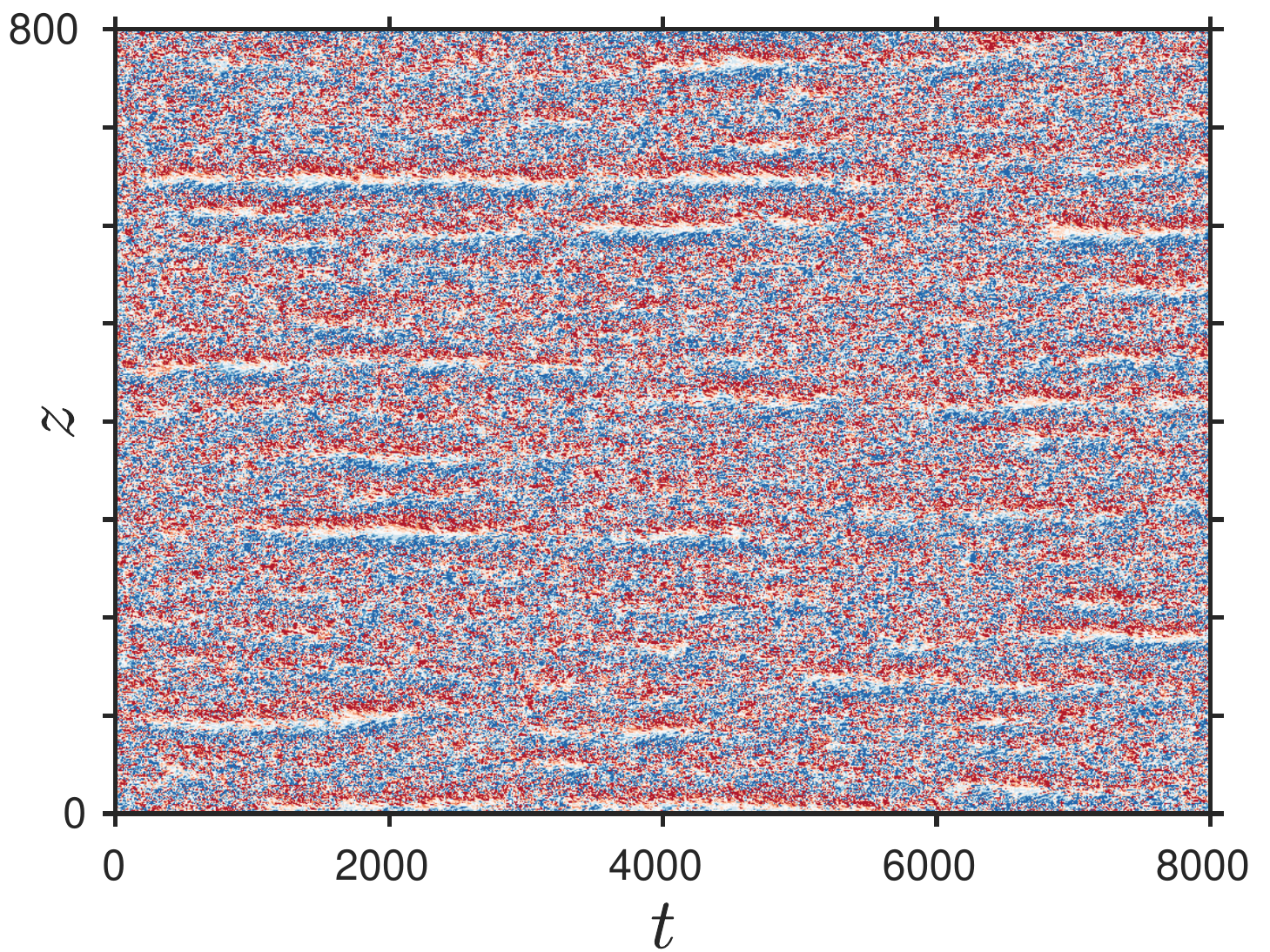} 
\label{fig:probes430}}
\subfloat[]{ %\includegraphics[width=0.5\columnwidth]
\includegraphics[width=0.545\columnwidth,height=5cm]
{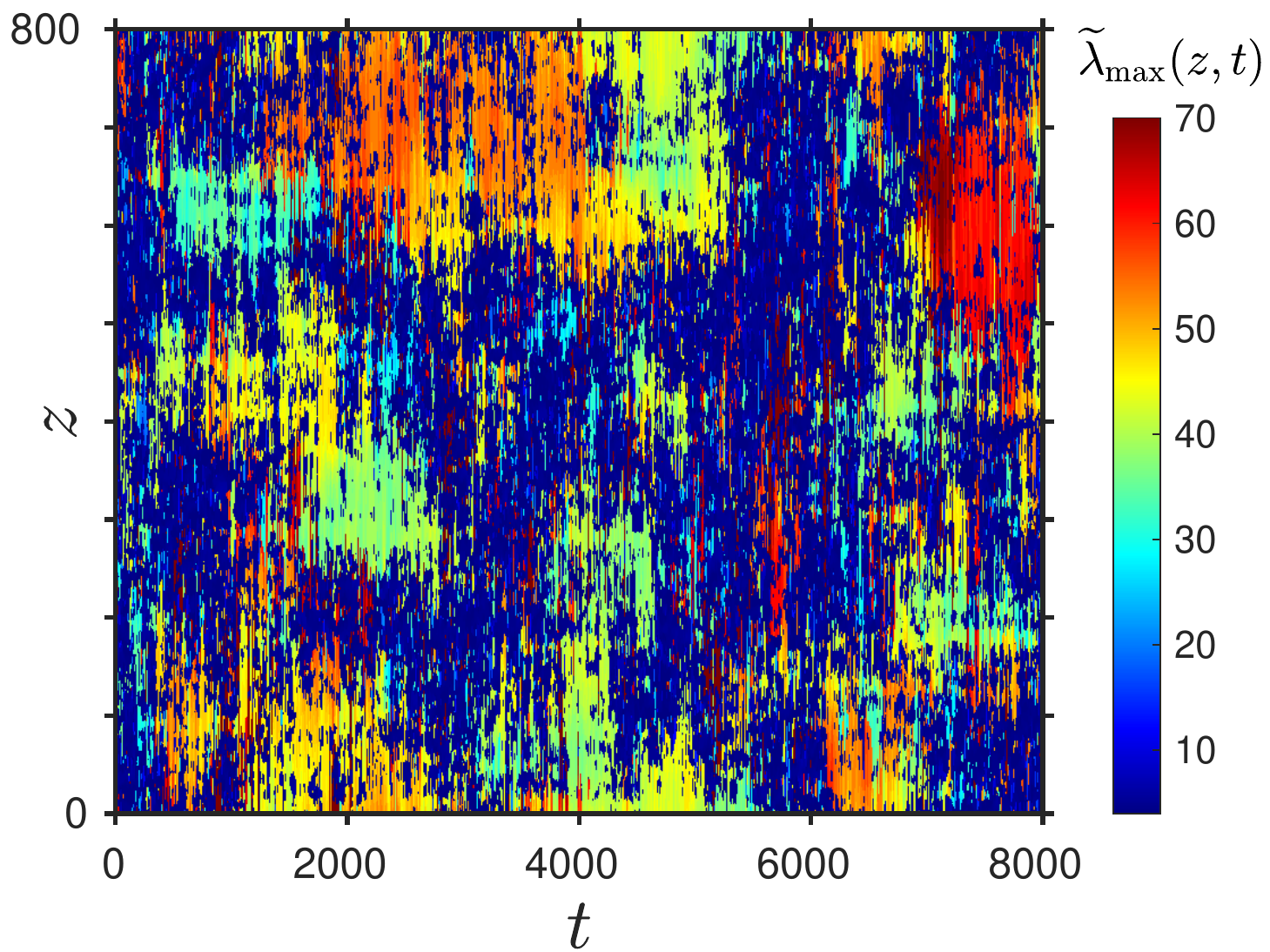} 
\label{fig:lambda_max}} ~
\caption{Space-time visualisation of a quench experiment at $Re=430$: (a) spanwise velocity (blue: $-0.2$, white: 0, red: 0.2), (b) $\lambdamax(z,t)$ defined by \eqref{eq:lambda_max}. 
$\lambdamax(z,t)$ (b) quantifies the presence of local large-scale modulations within the flow. Dark blue zones where $\lambdamax(z,t)<10$ correspond to locally featureless turbulence in (a). Large-scale modulation of gaps at different wavelengths are visible as the green-to-red spots in (b).
    }
\end{figure}

We introduce the one-dimensional continuous wavelet transform of the velocity
$\utot(z,t)$ taken along the line $(x,y)=(L_x/2, 0)$:
\begin{equation}
    \utilde(z, r, t) = C_\psi^{-1/2} r^{-1/2} \int_{0}^{L_z} \psi^* \left(\frac{z^\prime-z}{r}\right) \utot(z^\prime, t) dz^\prime
    \label{eq:wavelet_define}
\end{equation}
%LST * on \psi means complex conjugate?
Here $\psi$ is the Morlet basis function, defined in Fourier space as $\hat{\psi}(k) = \pi^{-1/4} e^{-(k - k_\psi )^2/2} $ for $k>0$.
Its central wavenumber is $k_\psi= 6 /\dz$, where $\dz$ is the grid spacing.  The scale factor $r$ is related to wavelength via $\lambda \simeq  2\pi r/k_\psi$.
$C_\psi \equiv \int |k|^{-1}  |\hat{\psi}(k)|^2 \text{d}k$ is a normalization constant.
Tildes are used to designate wavelet transformed quantities. 
The inverse transform is:
\begin{equation}
    \utot(z, t) = C_\psi^{-1/2} \int_0^{\infty} \int_{-\infty}^{\infty} \: r^{-1/2} \psi\left(\frac{z-z^\prime}{r}\right)\utilde(z^\prime, r, t) \: \frac{dz^\prime~ dr}{r^2}
\end{equation}
The wavelet transform is related to the Fourier transform in $z$ by:
\begin{equation}
    \utilde(z, r, t)  = \frac{1}{2\pi} C_\psi^{-1/2}  r^{1/2}\int_{-\infty}^{\infty}  \widehat{\psi}(r \,k_z) \widehat{\utot} (k_z, t) e^{i k_z z} \text{d} k_z
\end{equation}
%More info on: http://www.ece.northwestern.edu/local-apps/matlabhelp/toolbox/wavelet/ch06_a26.html
%
%The analysis of the wavelet signal is presented in Appendix \ref{appB} and shows that a most energetic wavelength emerges in space and time. 
%
We then define the most energetic instantaneous wavelength as:
\begin{equation}
\label{eq:lambda_max}
\lambdamax(z,t)= \frac{2\pi}{k_\psi}~ \underset{r}{\text{argmax}} ~ |\utilde(z,r,t)|^2
\end{equation}
%LST Make sure this is correct. Maybe explain more about scale factor r and why it makes sense to divide it by k_psi?
%
The characteristic evolution of $\lambdamax(z,t)$ is illustrated in  figure~\ref{fig:lambda_max} for the flow case corresponding to figure \ref{fig:probes430}. 
Regions in which $\lambdamax$ is large $(>10)$ and dominated by a single value correspond to the local patterns observed in figure \ref{fig:probes430}. 
In contrast, in regions where $\lambdamax$ is small $(<10)$ and fluctuating, the turbulence is locally uniform.

\begin{figure}
    \centering
    \subfloat[]{\includegraphics[width=0.5\columnwidth]{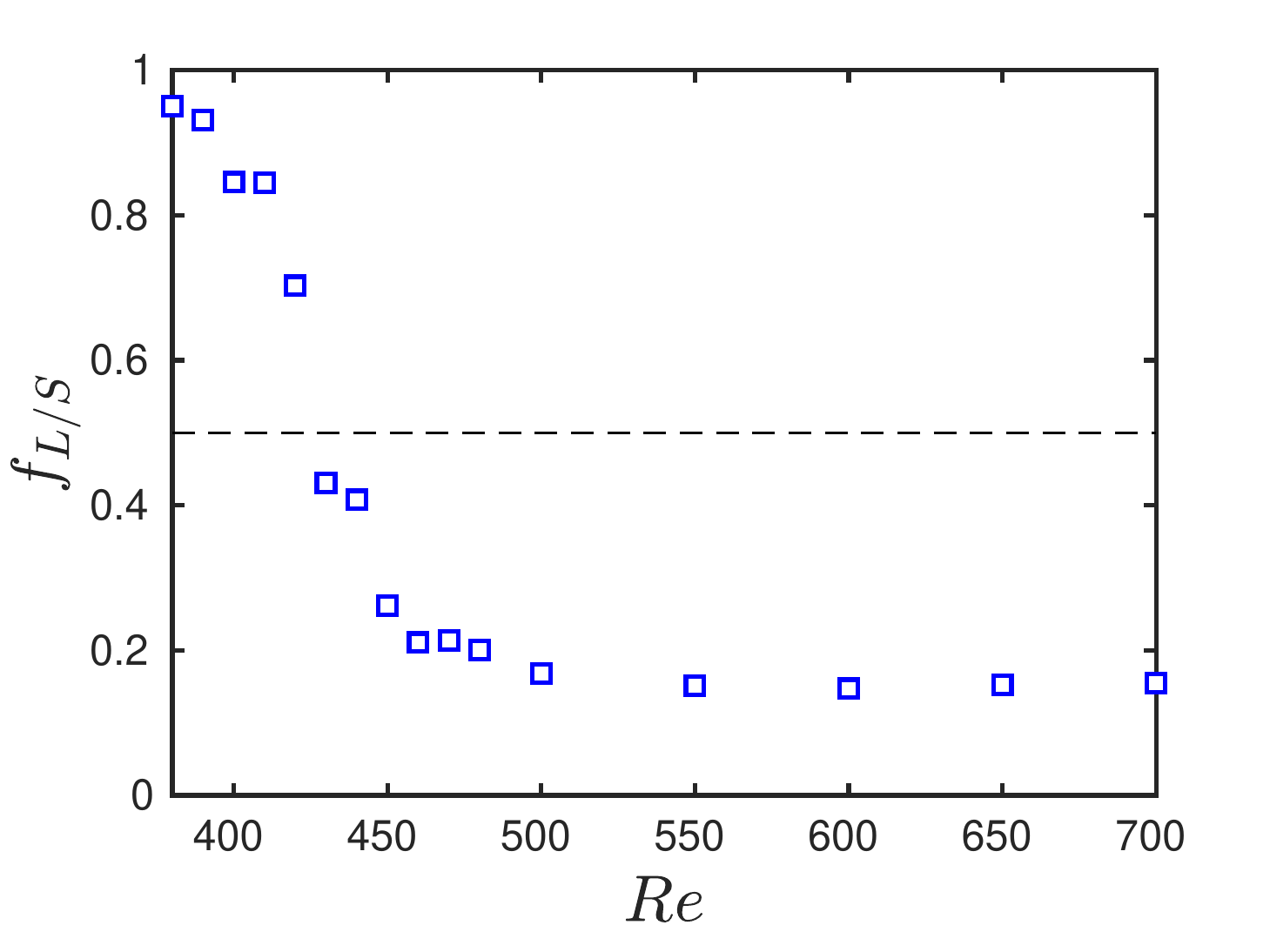}
        \label{fig:fLS}} ~
     \subfloat[]{\includegraphics[width=0.5\columnwidth]{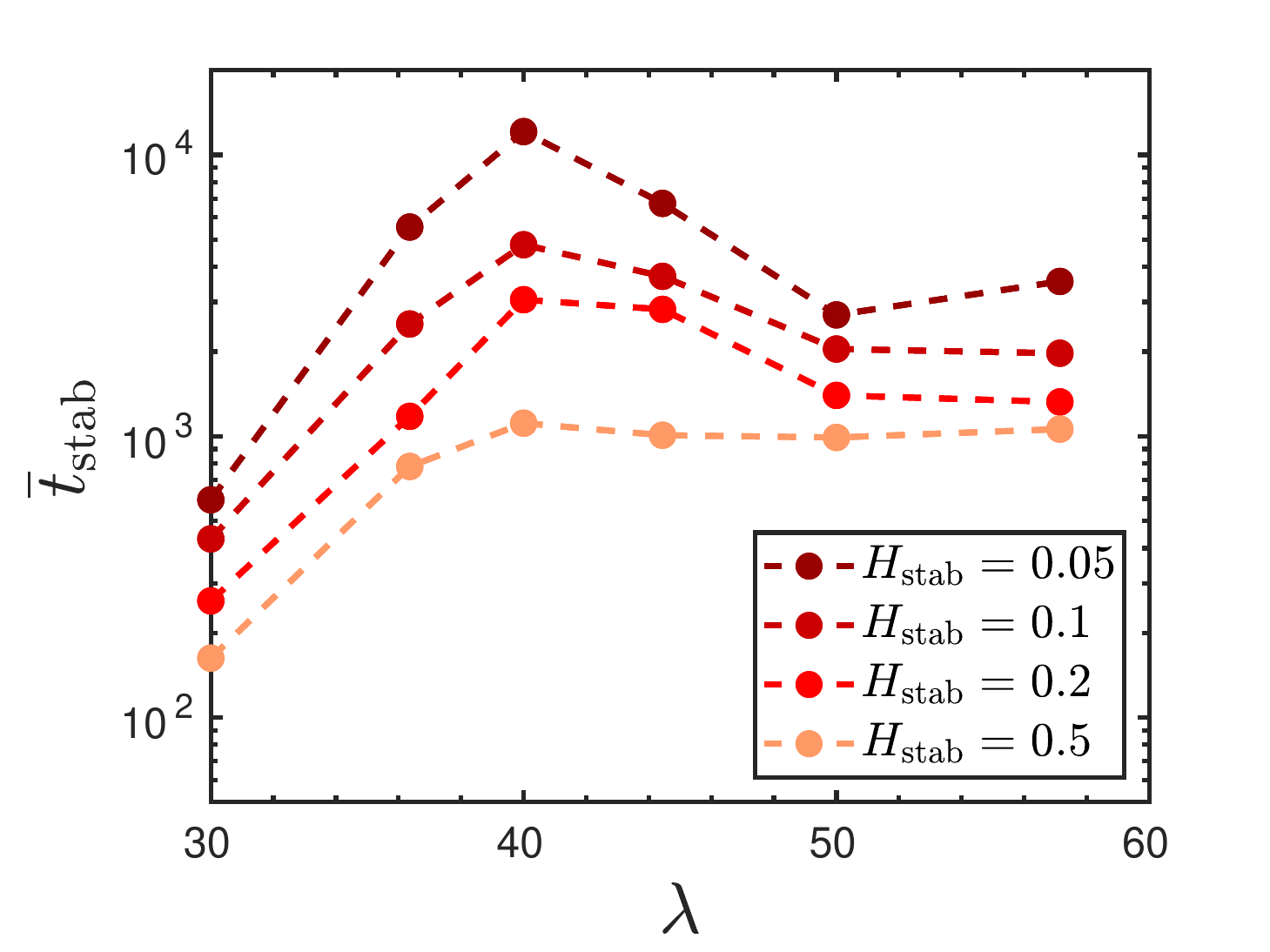} 
      \label{fig:stab_Hstab}}
\caption{(a) Space-time fraction of large to small wavelengths obtained by wavelet transform. 
$f_{L/S}$ crosses 0.5 at $Re \simeq 427 \simeq \Repg$.
(b) Sensitivity of the stability analysis in \ref{sec:stability} with regard to threshold $H_{\rm stab}$, at $Re=430$. 
    }
\end{figure}

This space-time intermittency of the patterns is quantified by measuring
\begin{equation}
    f_{L/S} = \left< \Theta (\lambdamax (z,t) - 10)  \right>_{z,t} 
\end{equation}
and is shown in figure \ref{fig:fLS} as a function of $Re$.

\section{Turbulent kinetic energy balance for various $L_z$}
\label{app:tke_bal}
%LST Define notation of y, replace nu by 1/Re
In this appendix, we address the balance of turbulent kinetic energy $\tke(k_z)$, written here in $y$-integrated form at a specific mode $k_z$ (see equation (5.3) of \citet[Part 1]{gome1} and the methodology in, e.g., \citet[]{bolotnov2010spectral, lee2015direct, mizuno2016spectra, cho2018scale}):
%
\begin{comment}
\begin{align}
    \label{eq:tke_bal_spec_y}
    0 = \Prodsy - \Dissipsy + \Advsy + \Transnlsy
\end{align}
with the $y$ subscripts designating $y$-integrated quantities:
\end{comment}
\begin{align}
    \label{eq:tke_bal_spec_y}
    0 = \Prods - \Dissips + \Advs + \Transnls
\end{align}
where the variables in \eqref{eq:tke_bal_spec_y} indicate $y$-integrated quantities:
\begin{align}
    \Prods (k_z)\equiv - \mathcal{R}\left\{ \int_{-1}^1 \overline{\ujphatc  \widehat{ \overline{u}_i\frac{\partial u_j^\prime}{\partial x_i} }} ~ \text{d}y  \right\}, \nonumber ~~~~
    \Dissips (k_z)\equiv \frac{2}{Re} \int_{-1}^1  \overline{ \widehat{s_{ij}^\prime} \widehat{s_{ij}^\prime}^*}~ \text{d}y, \\
    \Transnls (k_z)\equiv -\mathcal{R}\left\{  \int_{-1}^1  \overline{\ujphatc \widehat{  u_i^\prime \frac{\partial  u_j^\prime}{\partial x_i}  }}  ~ \text{d}y  \right\}, ~~~~
    \Advs (k_z)\equiv -\mathcal{R}\left\{ \int_{-1}^1 \overline{\ujphatc  \widehat{ \overline{u}_i\frac{\partial u_j^\prime}{\partial x_i} }} ~ \text{d}y \right\}
    \label{eq:tke_y_bal}
\end{align} 
respectively standing for production, dissipation, triadic interaction and advection terms. We recall that $\overline{(\cdot)}$ is an average in $(x,t)$.
The $y$ evolution of the energy balance was analysed in \citet[Part 1]{gome1}.
%but our current focus will be on the $y$-integrated balance. 

\citet[Part 1]{gome1} reported robust negative production at large scales, along with inverse non-linear transfers to large scales. If $\ksmax=1.41$ denotes the scale of rolls and streaks, this inverse transfer occurs  for $k_z < \kLS = 0.94$, while a downward transfer occurs for $k_z>\kSS= 3.6$ (We refer the reader to figure 5 of \citet[Part 1]{gome1}). %\ref{fig:tke_bal_spec_pattern}).
%In the transitional patterned regime, the negative production feeds $\ubar_1$ via the term $\Prodms_1$. 
%
This spectral organization of the energy balance will be quantified by the following transfer terms arising from \eqref{eq:tke_y_bal}:
\begin{align}
\widehat{T}_{LS}  \equiv    \sum_{k_z=0}^{\kLS} \Transnls (k_z), ~~~
\widehat{T}_{SS}  \equiv   \sum_{k_z=\kSS}^{\infty} \Transnls(k_z)
, ~~~
\widehat{D}_{LS}  \equiv   \sum_{k_z=0}^{\kLS} \Dissips (k_z),
~~~
\widehat{A}_{LS} \equiv 
\sum_{k_z=0}^{\kLS} \Advs (k_z)  
\label{eq:LS_trans}
\end{align}
\begin{comment}
\begin{align}
\widehat{T}_{LS}  =    \sum_{k_z=0}^{\kLS} \Transnlsy (k_z), ~~
\widehat{T}_{SS}  =   \sum_{k_z=\ksmax}^{\infty} \Transnlsy
, ~~
\widehat{D}_{LS}  =   \sum_{k_z=0}^{\ksmax} \Dissips_{y} (k_z), ~~
\widehat{A}_{LS} \equiv 
\sum_{k_z=0}^{\kLS} \Advs_{y} (k_z)  
\label{eq:LS_trans}
\end{align}
\end{comment}
%
$\widehat{T}_{LS} $ quantifies transfer to large scales, $\widehat{T}_{SS} $ the transfer to small scales, $\widehat{D}_{LS}$ the dissipation at large scales, and $\widehat{A}_{LS}$ is a transfer of energy from the mean flow to the large fluctuating scales. Large-scale production is not shown here, as we presented in figure \ref{fig:mean_spec_Lz1} a similar measurement of large-scale turbulent transfer to the mean flow, via $\Prodms_1$.

\begin{figure}
    \centering
\includegraphics[width=0.5\columnwidth]{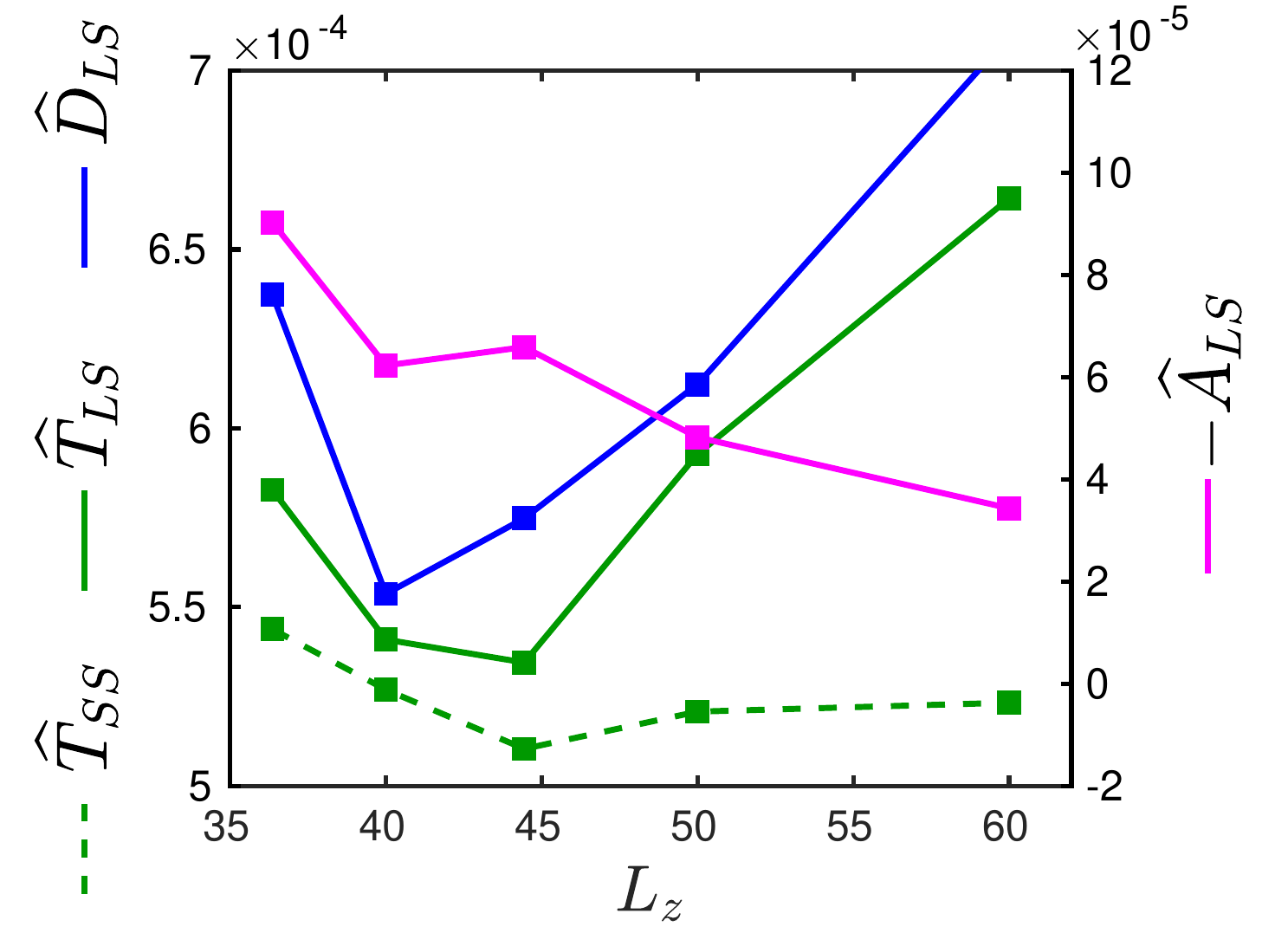}     
\caption{Evolution of the large-scale TKE balance with $L_z$ (\ref{eq:LS_trans}). }
\label{fig:tke_Lz}   
\end{figure}

The variables defined in \eqref{eq:LS_trans} are displayed in figure \ref{fig:tke_Lz} as a function of $L_z$. $\widehat{T}_{LS}  $ is minimal at $L_z \simeq 44$.
$\widehat{D}_{LS}  $ is minimal at $L_z \simeq 40$.
%whereas both large-scale dissipation and production are minimal near $L_z\simeq 40$.
%
%Over all large-scale quantities, production and advection vary the most with $L_z$ ($>60\% $ of variation). 
Contrary to $\widehat{T}_{LS} $, $\widehat{T}_{SS}$ is relatively constant with $L_z$ (green dashed line in figure \ref{fig:tke_Lz}), with a variation of around $6\%$. This demonstrates that transfers to small scales are unchanged with $L_z$. 
%This is consistent with the self-similarity in the small-scale spectrum observed in figure \ref{fig:spec_MBU}. 
Large-scale TKE advection decays with increasing $L_z$ hence it does not play a role in the preference of a  wavelength. Our results show that the balance at large-scale is minimised around $L_z\simeq 44$, confirming the less important role played by turbulent fluctuations in the wavelength selection, compared to that of the mean-flow advection reported in the main text.
%
%The optimisation of the mean flow, as presented in Section \ref{sec:spec_optim}, happens without perturbing the direct cascade to small scales.

%ADD NEGATIVE AND POSITIVE PROD?? 
%Why is TKE dissipation minimal 
% IS THE STREAK ENERGY DIFFERENT??

\FloatBarrier
\bibliographystyle{jfm}
\bibliography{bib}% Produces the bibliography via BibTeX.

\end{document}